\documentclass{aastex631}

\usepackage{amsmath}
\usepackage{verbatim}
\usepackage{hyperref}

\newcommand{\pivec}{\mbox{\boldmath $\pi$}}

\newcommand{\twosixtynine}{OGLE-2016-BLG-1635}       % KMT-2016-BLG-0269 = OGLE-2016-BLG-1635 / no MOA data
\newcommand{\fiveosix}{MOA-2016-BLG-532}             % KMT-2016-BLG-0506 = OGLE-2016-BLG-1749 = MOA-2016-BLG-532
\newcommand{\sixtwentyfive}{KMT-2016-BLG-0625}       % no OGLE / MOA avail. but, no data at the crucial part
\newcommand{\thirteenseven}{OGLE-2016-BLG-1850}      % KMT-2016-BLG-1307 = OGLE-2016-BLG-1850
\newcommand{\seventeenfiftyone}{KMT-2016-BLG-1751}   % MOA data available
\newcommand{\eighteenfiftyfive}{KMT-2016-BLG-1855}   % no OGLE / no MOA 

\newcommand{\doubleOtwenty}{OGLE-2016-BLG-0987}      % KMT-2016-BLG-0020 = OGLE-2016-BLG-0987 / MOA avail. but, KMT+OGLE coverage is enough
\newcommand{\oneOsix}{MOA-2016-BLG-123}              % KMT-2016-BLG-0106 = MOA-2016-BLG-123
\newcommand{\onefiftyseven}{OGLE-2016-BLG-0558}      % KMT-2016-BLG-0157 = OGLE-2016-BLG-0558
\newcommand{\threeseventyfour}{KMT-2016-BLG-0374}    % no OGLE / MOA avail.
\newcommand{\fourfortysix}{KMT-2016-BLG-0446}        % No OGLE / no MOA
\newcommand{\seventeensixteen}{OGLE-2016-BLG-1722}   % KMT-2016-BLG-1716 = OGLE-2016-BLG-1722 = MOA-2016-BLG-555
\newcommand{\eighteensixtythree}{OGLE-2016-BLG-0974} % KMT-2016-BLG-1863 = OGLE-2016-BLG-0974 = MOA-2016-BLG-351
\newcommand{\fourtwentyfive}{OGLE-2016-BLG-0185}     % KMT-2016-BLG-0425 = OGLE-2016-BLG-0185
\newcommand{\twotwoninefour}{KMT-2021-BLG-2294}      % KMT-2021-BLG-2294

\begin{document}

\title{Systematic KMTNet Planetary Anomaly Search. IX. Complete Sample of 2016 Prime-Field Planets
}

% Author List ------------------------------------------------------------------------------------------------------------
% 1
\author{In-Gu Shin} 
\affiliation{Center for Astrophysics $|$ Harvard \& Smithsonian 60 Garden St., Cambridge, MA 02138, USA}
%\correspondingauthor{In-Gu~Shin} \email{ingushin@gmail.com}
% 2
\author{Jennifer C. Yee}
\affiliation{Center for Astrophysics $|$ Harvard \& Smithsonian 60 Garden St., Cambridge, MA 02138, USA}
% 3
\author{Weicheng Zang}
\affiliation{Center for Astrophysics $|$ Harvard \& Smithsonian 60 Garden St., Cambridge, MA 02138, USA}
\affiliation{Department of Astronomy, Tsinghua University, Beijing 100084, China}
% 4
\author{Hongjing Yang}
\affiliation{Department of Astronomy, Tsinghua University, Beijing 100084, China}
% 5
\author{Kyu-Ha Hwang}
\affiliation{Korea Astronomy and Space Science Institute, Daejon 34055, Republic of Korea}
% 6
\author{Cheongho Han}
\affiliation{Department of Physics, Chungbuk National University, Cheongju 28644, Republic of Korea}
% 7
\author{Andrew Gould}
\affiliation{Max Planck Institute for Astronomy, K\"onigstuhl 17, D-69117 Heidelberg, Germany}
\affiliation{Department of Astronomy, The Ohio State University, 140 W. 18th Ave., Columbus, OH 43210, USA}
% 8
\author{Andrzej~Udalski}
\affiliation{Astronomical Observatory, University of Warsaw, Al.~Ujazdowskie 4, 00-478 Warszawa, Poland}
% 9
\author{Ian A. Bond}
\affiliation{Institute of Natural and Mathematical Sciences, Massey University, Auckland 0745, New Zealand}
\collaboration{10}{(Leading authors),}
%
% KMTNet ---------------------------
% Science Team
% 1
\author{Michael D. Albrow} 
\affiliation{University of Canterbury, Department of Physics and Astronomy, Private Bag 4800, Christchurch 8020, New Zealand}
% 2
\author{Sun-Ju Chung}
\affiliation{Korea Astronomy and Space Science Institute, Daejon 34055, Republic of Korea}
% 3
\author{Youn Kil Jung}
\affiliation{Korea Astronomy and Space Science Institute, Daejon 34055, Republic of Korea}
\affiliation{University of Science and Technology, Korea, (UST), 217 Gajeong-ro, Yuseong-gu, Daejeon 34113, Republic of Korea}
% 4
\author{Yoon-Hyun Ryu}
\affiliation{Korea Astronomy and Space Science Institute, Daejon 34055, Republic of Korea}
% 5
\author{Yossi Shvartzvald}
\affiliation{Department of Particle Physics and Astrophysics, Weizmann Institute of Science, Rehovot 76100, Israel}
% 
% Operation Team
% 6
\author{Sang-Mok Cha}
\affiliation{Korea Astronomy and Space Science Institute, Daejon 34055, Republic of Korea}
\affiliation{School of Space Research, Kyung Hee University, Yongin, Kyeonggi 17104, Republic of Korea}
% 7 
\author{Dong-Jin Kim}
\affiliation{Korea Astronomy and Space Science Institute, Daejon 34055, Republic of Korea}
% 8
\author{Seung-Lee Kim} 
\affiliation{Korea Astronomy and Space Science Institute, Daejon 34055, Republic of Korea}
% 9
\author{Chung-Uk Lee}
\affiliation{Korea Astronomy and Space Science Institute, Daejon 34055, Republic of Korea}
% 10
\author{Dong-Joo Lee}
\affiliation{Korea Astronomy and Space Science Institute, Daejon 34055, Republic of Korea}
% 11
\author{Yongseok Lee}
\affiliation{Korea Astronomy and Space Science Institute, Daejon 34055, Republic of Korea}
\affiliation{School of Space Research, Kyung Hee University, Yongin, Kyeonggi 17104, Republic of Korea}
% 12
\author{Byeong-Gon Park}
\affiliation{Korea Astronomy and Space Science Institute, Daejon 34055, Republic of Korea}
% 13
\author{Richard W. Pogge}
\affiliation{Department of Astronomy, The Ohio State University, 140 W. 18th Ave., Columbus, OH 43210, USA}
\collaboration{14}{(The KMTNet Collaboration),}
% OGLE -------------------------------
% 1
\author{Przemek~Mr{\'o}z}
\affiliation{Astronomical Observatory, University of Warsaw, Al.~Ujazdowskie 4, 00-478 Warszawa, Poland}
% 2
\author{Micha{\l}~K.~Szyma{\'n}ski}
\affiliation{Astronomical Observatory, University of Warsaw, Al.~Ujazdowskie 4, 00-478 Warszawa, Poland}
% 3
\author{Jan~Skowron}
\affiliation{Astronomical Observatory, University of Warsaw, Al.~Ujazdowskie 4, 00-478 Warszawa, Poland}
% 4
\author{Rados{\l}aw~Poleski}
\affiliation{Astronomical Observatory, University of Warsaw, Al.~Ujazdowskie 4, 00-478 Warszawa, Poland}
% 5
\author{Igor~Soszy{\'n}ski}
\affiliation{Astronomical Observatory, University of Warsaw, Al.~Ujazdowskie 4, 00-478 Warszawa, Poland}
% 6
\author{Pawe{\l}~Pietrukowicz}
\affiliation{Astronomical Observatory, University of Warsaw, Al.~Ujazdowskie 4, 00-478 Warszawa, Poland}
% 7
\author{Szymon~Koz{\l}owski}
\affiliation{Astronomical Observatory, University of Warsaw, Al.~Ujazdowskie 4, 00-478 Warszawa, Poland}
% 8
\author{Krzysztof~A.~Rybicki}
\affiliation{Astronomical Observatory, University of Warsaw, Al.~Ujazdowskie 4, 00-478 Warszawa, Poland}
\affiliation{Department of Particle Physics and Astrophysics, Weizmann Institute of Science, Rehovot 76100, Israel}
% 9
\author{Patryk~Iwanek}
\affiliation{Astronomical Observatory, University of Warsaw, Al.~Ujazdowskie 4, 00-478 Warszawa, Poland}
% 10
\author{Krzysztof~Ulaczyk}
\affiliation{Department of Physics, University of Warwick, Gibbet Hill Road, Coventry, CV4 7AL, UK}
% 11
\author{Marcin~Wrona}
\affiliation{Astronomical Observatory, University of Warsaw, Al.~Ujazdowskie 4, 00-478 Warszawa, Poland}
% 12
\author{Mariusz~Gromadzki}
\affiliation{Astronomical Observatory, University of Warsaw, Al.~Ujazdowskie 4, 00-478 Warszawa, Poland}
\collaboration{13}{(The OGLE Collaboration)}
% MOA --------------------------------
% 1
\author{Fumio Abe}
\affiliation{Institute for Space-Earth Environmental Research, Nagoya University, Nagoya 464-8601, Japan}
% 2
\author{Richard Barry}
\affiliation{Code 667, NASA Goddard Space Flight Center, Greenbelt, MD 20771, USA}
% 3
\author{David P.~Bennett}
\affiliation{Code 667, NASA Goddard Space Flight Center, Greenbelt, MD 20771, USA}
\affiliation{Department of Astronomy, University of Maryland, College Park, MD 20742, USA}
% 4
\author{Aparna Bhattacharya}
\affiliation{Code 667, NASA Goddard Space Flight Center, Greenbelt, MD 20771, USA}
\affiliation{Department of Astronomy, University of Maryland, College Park, MD 20742, USA}
% 5
\author{Hirosane Fujii}
\affiliation{Institute for Space-Earth Environmental Research, Nagoya University, Nagoya 464-8601, Japan}
% 6
\author{Akihiko Fukui}
\affiliation{Department of Earth and Planetary Science, Graduate School of Science, The University of Tokyo, 7-3-1 Hongo, Bunkyo-ku, Tokyo 113-0033, Japan}
\affiliation{Instituto de Astrof\'isica de Canarias, V\'ia L\'actea s/n, E-38205 La Laguna, Tenerife, Spain}
% 7
\author{Ryusei Hamada}
\affiliation{Department of Earth and Space Science, Graduate School of Science, Osaka University, Toyonaka, Osaka 560-0043, Japan}
% 8
\author{Yuki Hirao}
\affiliation{Department of Earth and Space Science, Graduate School of Science, Osaka University, Toyonaka, Osaka 560-0043, Japan}
% 9 
\author{Stela Ishitani Silva}
\affiliation{Department of Physics, The Catholic University of America, Washington, DC 20064, USA}
\affiliation{Code 667, NASA Goddard Space Flight Center, Greenbelt, MD 20771, USA}
% 10
\author{Yoshitaka Itow}
\affiliation{Institute for Space-Earth Environmental Research, Nagoya University, Nagoya 464-8601, Japan}
% 11
\author{Rintaro Kirikawa}
\affiliation{Department of Earth and Space Science, Graduate School of Science, Osaka University, Toyonaka, Osaka 560-0043, Japan}
% 12
\author{Iona Kondo}
\affiliation{Department of Earth and Space Science, Graduate School of Science, Osaka University, Toyonaka, Osaka 560-0043, Japan}
% 13
\author{Naoki Koshimoto}
\affiliation{Department of Astronomy, Graduate School of Science, The University of Tokyo, 7-3-1 Hongo, Bunkyo-ku, Tokyo 113-0033, Japan}
% 14
\author{Yutaka Matsubara}
\affiliation{Institute for Space-Earth Environmental Research, Nagoya University, Nagoya 464-8601, Japan}
% 15
\author{Shota Miyazaki}
\affiliation{Department of Earth and Space Science, Graduate School of Science, Osaka University, Toyonaka, Osaka 560-0043, Japan}
% 16
\author{Yasushi Muraki}
\affiliation{Institute for Space-Earth Environmental Research, Nagoya University, Nagoya 464-8601, Japan}
% 17
\author{Greg Olmschenk}
\affiliation{Code 667, NASA Goddard Space Flight Center, Greenbelt, MD 20771, USA}
% 18
\author{Cl\'ement Ranc}
\affiliation{Sorbonne Universit\'e, CNRS, UMR 7095, Institut d'Astrophysique de Paris, 98 bis bd Arago, 75014 Paris, France}
% 19
\author{Nicholas J. Rattenbury}
\affiliation{Department of Physics, University of Auckland, Private Bag 92019, Auckland, New Zealand}
% 20
\author{Yuki Satoh}
\affiliation{Department of Earth and Space Science, Graduate School of Science, Osaka University, Toyonaka, Osaka 560-0043, Japan}
% 21
\author{Takahiro Sumi}
\affiliation{Department of Earth and Space Science, Graduate School of Science, Osaka University, Toyonaka, Osaka 560-0043, Japan}
% 22
\author{Daisuke Suzuki}
\affiliation{Department of Earth and Space Science, Graduate School of Science, Osaka University, Toyonaka, Osaka 560-0043, Japan}
% 23
\author{Mio Tomoyoshi}
\affiliation{Department of Earth and Space Science, Graduate School of Science, Osaka University, Toyonaka, Osaka 560-0043, Japan}
% 24
\author{Paul . J. Tristram}
\affiliation{University of Canterbury Mt.\ John Observatory, P.O. Box 56, Lake Tekapo 8770, New Zealand}
% 25
\author{Aikaterini Vandorou}
\affiliation{Code 667, NASA Goddard Space Flight Center, Greenbelt, MD 20771, USA}
\affiliation{Department of Astronomy, University of Maryland, College Park, MD 20742, USA}
% 26
\author{Hibiki Yama}
\affiliation{Department of Earth and Space Science, Graduate School of Science, Osaka University, Toyonaka, Osaka 560-0043, Japan}
% 27
\author{Kansuke Yamashita}
\affiliation{Department of Earth and Space Science, Graduate School of Science, Osaka University, Toyonaka, Osaka 560-0043, Japan}
\collaboration{28}{(the MOA Collaboration)}
% ------------------------------------------------------------------------------------------------------------------------

\begin{abstract}
As a part of the ``Systematic KMTNet Planetary Anomaly Search" series, we report five new planets (namely, \twosixtynine Lb, \fiveosix Lb, \sixtwentyfive Lb, \thirteenseven Lb, and \seventeenfiftyone Lb) and one planet candidate (\eighteenfiftyfive), which were found by searching $2016$ KMTNet prime fields. These {\it buried} planets show a wide range of masses from Earth--class to Super--Jupiter--class, and are located in both the disk and the bulge.  The ultimate goal of this series is to build a complete planet sample. Because our work provides a complementary sample to other planet detection methods, which have different detection sensitivities, our complete sample will help us to obtain a better understanding of planet demographics in our Galaxy.
\end{abstract}

\section{Introduction}

To build a complete microlensing planet sample, we conduct a series of works called ``Systematic KMTNet Planetary Anomaly Search" based on a large microlensing survey archive obtained by the Korea Microlensing Telescope Network \citep[KMTNet:][]{kim16}. We identify planet--like anomalies using the ``AnomalyFinder" algorithm \citep{zang21, zang22a} instead of a traditional ``by--eye" method, which can systematically identify almost all candidates showing anomalies on the light curve\footnote{Although the AnomalyFinder (AF) detects anomalies using criteria optimized to the KMTNet data, some anomalous events can be omitted because the criteria are not perfect, yet. For example, AF missed KMT-2021-BLG-2294Lb \citep{shin23}. Thus, the ``by-eye" method can help us to improve the criteria and understand the completeness of the final planet sample.}. However, to reveal the origin of the anomaly requires (preliminary) models including possible degenerate solutions to figure out the mass ratio of the lens component (i.e., $q$). Also, it requires investigating the data for the anomaly to check whether or not the anomaly is caused by a false--positive signal. Thus, detailed analyses for all anomalous events found by the AnomalyFinder require significant resources and human efforts. 

Hence, for the KMTNet data obtained from $2016$ to $2021$, we conduct the work separately for each bulge season and for observing fields with different cadences, which are divided into the prime fields (high cadence: $\Gamma = 2.0$--$4.0\, {\rm hr^{-1}}$) and sub--prime fields (low cadence: $\Gamma = 0.2$--$1.0\, {\rm hr^{-1}}$). The KMTNet field information is described in \citet{kim18}. We have already done the systematic searches for the $2018$ prime field \citep{wang22, hwang22, gould22}, $2018$ sub--prime fields \citep{jung22}, $2019$ prime fields \citep{zang21, hwang22, zang22a}, and $2019$ sub--prime fields \citep{jung23}. In addition, \citet{zang23} present a complete sample of planets with the mass ratio $q < 10^{-4}$ discovered from all candidate events observed from $2016$ to $2019$.   

This is the 9th work to build the complete sample, which is conducted for the $2016$ prime--fields (i.e., BLG$01$, BLG$41$, BLG$02$, BLG$42$, BLG$03$, BLG$43$). The AnomalyFinder algorithm and candidate review identified $106$ anomalous events (plus $14$ events that were already published). Based on visual inspection and/or preliminary modeling, $79$ were eliminated as binaries. For the remaining $13$ new candidates with at least one solution with $q < 0.06$, we re-reduce the photometry to check for/remove the systematics in the data sets. Based on further analysis with the best quality data sets, $7$ were eliminated because they had no reliable planetary solutions (i.e., $q < 0.03$) with $\Delta\chi^{2} < 10.0$. We also investigate one additional $2016$ prime--field event for the detailed analysis, which was identified using ``by-eye" method and reported as a planet-like event, but was not in the final AnomalyFinder candidate list (see Appendix \ref{sec:appendixB}). Then, we find $5$ new planets and one planet candidate based on detailed analyses, which are \twosixtynine Lb, \fiveosix Lb, \sixtwentyfive Lb, \thirteenseven Lb, \seventeenfiftyone Lb, and \eighteenfiftyfive. We note that these planetary systems are designated by the survey projects that first announced the events as is traditional, even though the planetary systems were discovered based on the systematic search using the KMTNet data archive. We describe observations of each survey in Section \ref{sec:obs}. Then, we describe the light curve analysis for the planet candidates in Section \ref{sec:lc_analysis}. We note that, for the $9$ non--planetary events, we report the analysis results in Appendix \ref{sec:appendixA} for the record. In Section \ref{sec:cmd_analysis}, we present analyses for color-magnitude diagrams of the $5$ planetary events. In Section \ref{sec:bayes_analysis}, we present properties of the planetary systems determined based on the Bayesian analyses. Lastly, we summarize the results of this work in Section \ref{sec:summary}.

\section{Observations} \label{sec:obs}

In Tables \ref{table:obs_planet} and \ref{table:obs_binary} (see Appendix \ref{sec:appendixA}), we present observational information for the anomalous events, which have at least one solution with $q < 0.06$ found from preliminary modeling. For the anomalous events, we gather all available data taken from microlensing surveys for preliminary modeling. The KMTNet pipeline data are available from the KMTNet Alert System \citep[\url{https://kmtnet.kasi.re.kr/~ulens/}]{kim18}. They were obtained using three identical $1.6$ m telescopes equipped with wide--field ($4$ square degree) cameras. The telescopes are located at the Cerro Tololo Inter-American Observatory in Chile (KMTC), the South African Astronomical Observatory in South Africa (KMTS), and the Siding Spring Observatory in Australia (KMTA), which are in well--separated time zones to achieve near--continuous observations. Thus, the ``prime--fields" of the KMTNet have high--cadences ($\Gamma \geq 2 \,{\rm hr^{-1}}$) in $I$--band (Johnson--Cousins {\it BVRI} filter system). Also, for the KMTC observations, KMTNet regularly takes one observation in $V$--band for every $10$th $I$--band observation. We note that, for the KMTS observations, it takes one $V$--band observation for every $20$th $I$--band observation.

The OGLE \citep[Optical Gravitational Lensing Experiment:][]{udalski03, udalski15} data are available from the OGLE Early Warning System \citep[\url{http://ogle.astrouw.edu.pl/ogle4/ews/ews.html}]{udalski94} and were obtained using the $1.3$ m Warsaw telescope with a $1.4\, {\rm deg}^{2}$ camera located at Las Campanas Observatory in Chile. The OGLE observations were mainly made in $I$ band. Also, they periodically observe in $V$ band. 

The MOA \citep[Microlensing Observations in Astrophysics:][]{bond01, sumi03} data are available on their alert system website (\url{http://www.massey.ac.nz/~iabond/moa/alerts/}), and were obtained using a $1.8$ m telescope located at Mt. John University Observatory in New Zealand. The MOA observations were taken using the MOA--Red filter (hereafter, referred to $R$ band), which is roughly the sum of the Cousins $R$ and $I$ bands (wavelength ranges: $609$--$1109$ nm, transmission ranges: $0.0$--$0.978$).

The data of each survey were reduced by their own pipelines (KMTNet: \citealt{albrow09}, OGLE: \citealt{wozniak00},  and MOA: \citealt{bond01}), which adopt/modify the difference image analysis technique \citep{tomaney96, alard98}. We note that, for planet-like events listed in Tables \ref{table:obs_planet} and \ref{table:obs_binary}, the KMTNet data are re-reduced using an optimized version of pySIS (Yang et al. in prep) to obtain the best quality of data sets (hereafter, ``TLC (tender loving care)" reductions) for the analyses. Also, we reduce $V$--band data to determine the source color using the pyDIA package \citep{albrow17, bramich13}. We also note that some events require re-reduced data obtained from the OGLE and MOA surveys for the detailed analyses. For the \seventeenfiftyone\ and \threeseventyfour\ events, MOA did not alert these events. However, the events are located in the MOA fields. Therefore, the MOA team provided re-reduced data for these two events. \thirteenseven\ has a long baseline extending to $2017$ season. The OGLE team provided re-reduced data for this event including the long baseline.

\section{Light Curve Analysis} \label{sec:lc_analysis}

\subsection{Basics of the Analysis}

We conduct detailed analysis for $13$ candidates with re-reduced data sets using the optimized pySIS package (Yang et al. in prep). The analysis of the $5$ planetary events and one planet candidate is presented in this section and remaining $7$ events are briefly presented in Appendix \ref{sec:appendixA}. We follow the methodology of the light curve analysis described in \citet{shin23}. We briefly describe the analysis process, which consists of two steps, to present terminology used in this work. 

First, we conduct a grid search to find all possible solutions, in particular, local minima having planetary mass ratios (i.e., $q \lesssim 0.03$). For the grid search, we start from the static 2L1S case, i.e., without motions of the lenses or source (STD), where $n$L$m$S indicates number of lenses ($n$) and sources ($m$), respectively. To describe a microlensing light curve, the STD model requires seven parameters: ($t_{0}, u_{0}, t_{\rm E}, s, q, \alpha, \rho_{\ast}$), which are respectively defined as the time at the peak of the light curve, impact parameter, Einstein timescale, projected separation between binary lens components in units of the angular Einstein radius ($\theta_{\rm E}$), mass ratio of the lens components (i.e., $q \equiv M_{\rm secondary}/M_{\rm primary}$), angle between the source trajectory and binary axis, and angular source radius ($\theta_{\ast}$) scaled by $\theta_{\rm E}$ (i.e., $\rho_{\ast} \equiv \theta_{\ast}/\theta_{\rm E}$). We set $(s, q)$ as grid parameters for the grid search, which are most sensitive to describe anomalies on the light curve. The ranges of $(s, q)$ are $\log_{10}(s) \in [-1.0, 1.0]$ and $\log_{10}(q) \in [-5.5, 1.0]$ with $100$ grid points for each range. For five remaining parameters, we optimize them using $\chi^{2}$ minimizing method called the Markov Chain Monte Carlo (MCMC) algorithm \citep{doran04}. We note that $\alpha$ is treated as a semi-grid parameter because it is also sensitive to describing the anomalies: we start $21$ seeds for $\alpha$ parameter within the range of $\alpha \in [0, 2\pi]$. 

Second, once we find all plausible models, we refine model parameters for all cases by allowing all parameters can be freely vary within physically possible ranges. During this second process, we re--scale the errors of the data sets based on the best--fit model to make each data point contribute $\chi^{2} \sim 1.0$. The error re--scaling procedure is described in \citet{yee12}. Briefly, $e_{\rm R} = k\sqrt{e_{\rm O}^{2} + e_{\rm S}^{2}}$ where $e_{\rm R}$ is re--scaled error, $k$ is the re-scaling factor, $e_{\rm O}$ is original error, and $e_{\rm S}$ is systematics term. 

Based on the STD models, we consider higher--order effects if the solutions have a high chance to detect the effects. Specifically, we firstly consider the annual microlens parallax (APRX) effect \citep{gould92} if the models show relatively long timescale (i.e., $t_{\rm E} \gtrsim 15$ days at least). Once we find the APRX effect, we also consider the lens--orbital (OBT) effect because the OBT may affect the APRX measurements. Lastly, in the cases of the APRX effect detected, we test the ``xallarap" effect \citep[which is spelled backward of ``parallax":][]{griest92,han97,paczynski97,poindexter05}, which reflects the accelerating orbital motion of secondary source without brightness contribution of the secondary source. Because the xallarap effect can mimic the APRX effect, the xallarap test is required to confirm the APRX measurements. 

From the detailed analysis, we claim the detection of planetary systems if the fiducial solutions satisfy both detection criteria: (1) the solution(s) should have $q \lesssim 0.03$ and (2) the solution(s) should have $\Delta\chi^{2} \lesssim 10$ compared to other non--planetary solution(s).

Lastly, we note that, to indicate the degenerate solutions that we found, we follow the unified notation of the $s^{\dagger}$ formalism described in \citet{hwang22} and \citet{ryu22}. Also, we can check our solutions using the formalism for validation. Here, we briefly present the $s^{\dagger}$ formalism for  the description of each event in the following sections. 

The separations ($s^{\dagger}_{\pm}$) caused by major and minor images \citep{gouldloeb92} are expected as
\begin{equation}
s_{\pm}^{\dagger} \equiv \frac{\sqrt{u_{\rm anom}^{2} + 4} \pm u_{\rm anom}}{2}, 
\end{equation}
where $u_{\rm anom} = ({\tau_{\rm anom}^{2} + u_{0}^{2}})^{1/2}$ is the offset of the source position from the host obtained from the scaled time offset from the peak of the light curve, $\tau_{\rm anom} \equiv (t_{\rm anom}-t_{0})/t_{\rm E}$. The expected $s_{\pm}^{\dagger}$ can be compared to the empirical results. The comparison depends on the type of anomaly shape and the number of solutions. In general, the ``bump"--shaped anomaly caused by the major image perturbation should correspond to the $s_{+}^{\dagger}$ expectation, while the ``dip"--shaped anomaly caused by the minor image perturbation should correspond to the $s_{-}^{\dagger}$ expectation. For the number of solutions (i.e., degenerate cases), if we have unique solution, the empirical $s$ should correspond to one of $s_{\pm}^{\dagger}$. If we have two degenerate solutions such as $s_{\pm}$, the empirical solutions have a relation of  
\begin{equation}
s^{\dagger} = \sqrt{s_{+}s_{-}},
\end{equation}
which should correspond to one of $s_{\pm}^{\dagger}$ values. The $\alpha$ can be also predicted as
\begin{equation}
\tan{\alpha} = \frac{u_0}{\tau_{\rm anom}}.
\end{equation}
More specifically, $\alpha = \tan^{-1}(u_0/\tau_{\rm anom}) + j\pi$, where $j=(0,1)$ for (major, minor) images, and where the range of $\tan^{-1}$ is defined as $[0, \pi]$. We note that the $\alpha$ expectation depends on the coordinate system of the modeling. Lastly, in the case of the dip--shaped anomaly, we can obtain the first order approximation of the $q$ values. That is, 
\begin{equation}
q = \left( \frac{\Delta t_{\rm dip}}{4 t_{\rm E}} \right)^{2} \frac{s \sin^{2} \alpha}{u_{\rm anom}} 
  = \left( \frac{\Delta t_{\rm dip}}{4 t_{\rm E}} \right)^{2} \frac{s}{|u_{0}|} |\sin^{3} \alpha|.
\end{equation}
We note that the predicted $q$ generally matches the empirical $q$ value within a factor of $\sim 2$. This expectation is useful for judging how valuable an event is to conduct a detailed analysis (i.e., whether or not it is a planetary event), even if the expectation could not be very accurate. The theoretical origins of the heuristic analysis and such degeneracies are described in \citet{gaudi97}, \citet{griest98}, and \citet{zhang22a}.

\subsection{\twosixtynine} % KMT-2016-BLG-0269

\begin{figure}[htb!]
\epsscale{1.00}
\plotone{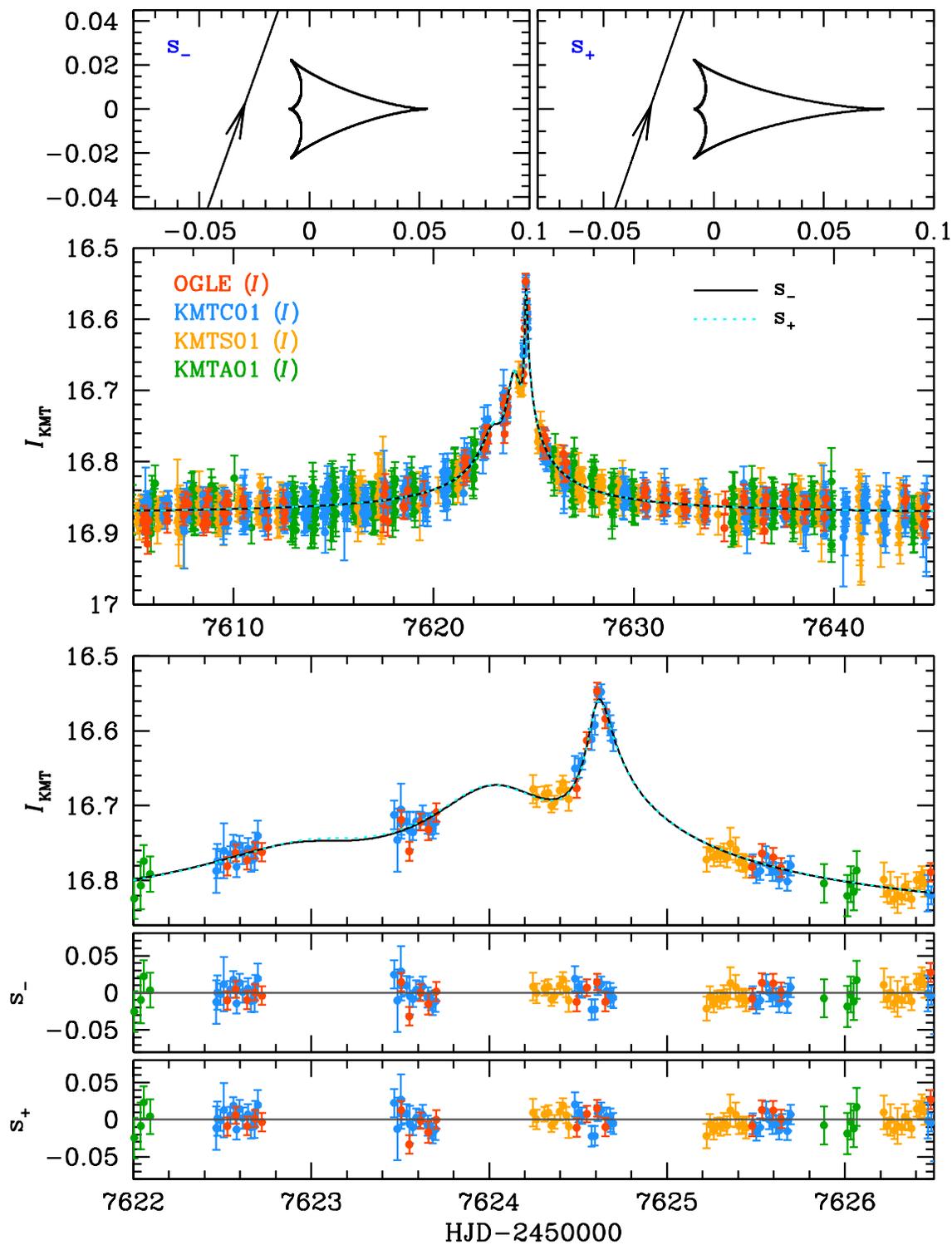}
\caption{Light curve of \twosixtynine\ with $s_{\pm}$ models and their caustic geometries.
\label{fig:lc_0269}}
\end{figure}

The light curve of \twosixtynine\ (which we identified as KMT-2016-BLG-0269) exhibits a bump-shaped anomaly at HJD$^{\prime} = 7624.6$. In Figure \ref{fig:lc_0269}, we present the light curve with degenerate (i.e., $s_{\pm}$) models. We also present the model parameters in Table \ref{table:model_0269}. From a heuristic analysis, we find $\tau_{\rm anom} = 0.023$ and $u_{\rm anom} = 0.036$ based on $t_{\rm anom} = 7624.60$, $t_{0} = 7624.12$, $u_{0} = 0.028$, and $t_{\rm E} = 21$ days. As a result, we expect $s_{-}^{\dagger} = 0.98$ and $s_{+}^{\dagger} = 1.02$. The $s_{-}^{\dagger}$ is consistent with $s^{\dagger} = 0.99$ for our solutions. Although degenerate solutions exist, the mass ratios of both cases are less than $0.03$, which implies that the companion is a planet by our formal definition. 

The timescale of this event is about $21$ days, which implies that there is a possibility of measuring the APRX considering the empirical criterion $t_{\rm E} \gtrsim 15$ days. Thus, we test the APRX models for this event. We find $\chi^{2}$ improvements $\Delta\chi^{2} = 7.24$ and $\Delta\chi^{2} = 9.96$ for the $s_{-}$ and $s_{+}$ cases, respectively. The $\Delta\chi^{2}$ values are too small to claim the APRX detection. Moreover, the APRX parameters are not converged for the $s_{-}$ case. For the $s_{+}$ case, the APRX model favors values of $|\pi_{\rm E}| > 10$ that are not reliable because they are caused by over fitting systematics at the baseline. Hence, we conclude that the STD models should be the fiducial solutions for \twosixtynine. We can only measure the upper limits of $\rho_{\ast}$ (i.e., $3\sigma$ ranges) because the source does not cross the caustic as shown in Figure \ref{fig:lc_0269}.

\subsection{\fiveosix} % KMT-2016-BLG-0506

\begin{figure}[htb!]
\epsscale{1.00}
\plotone{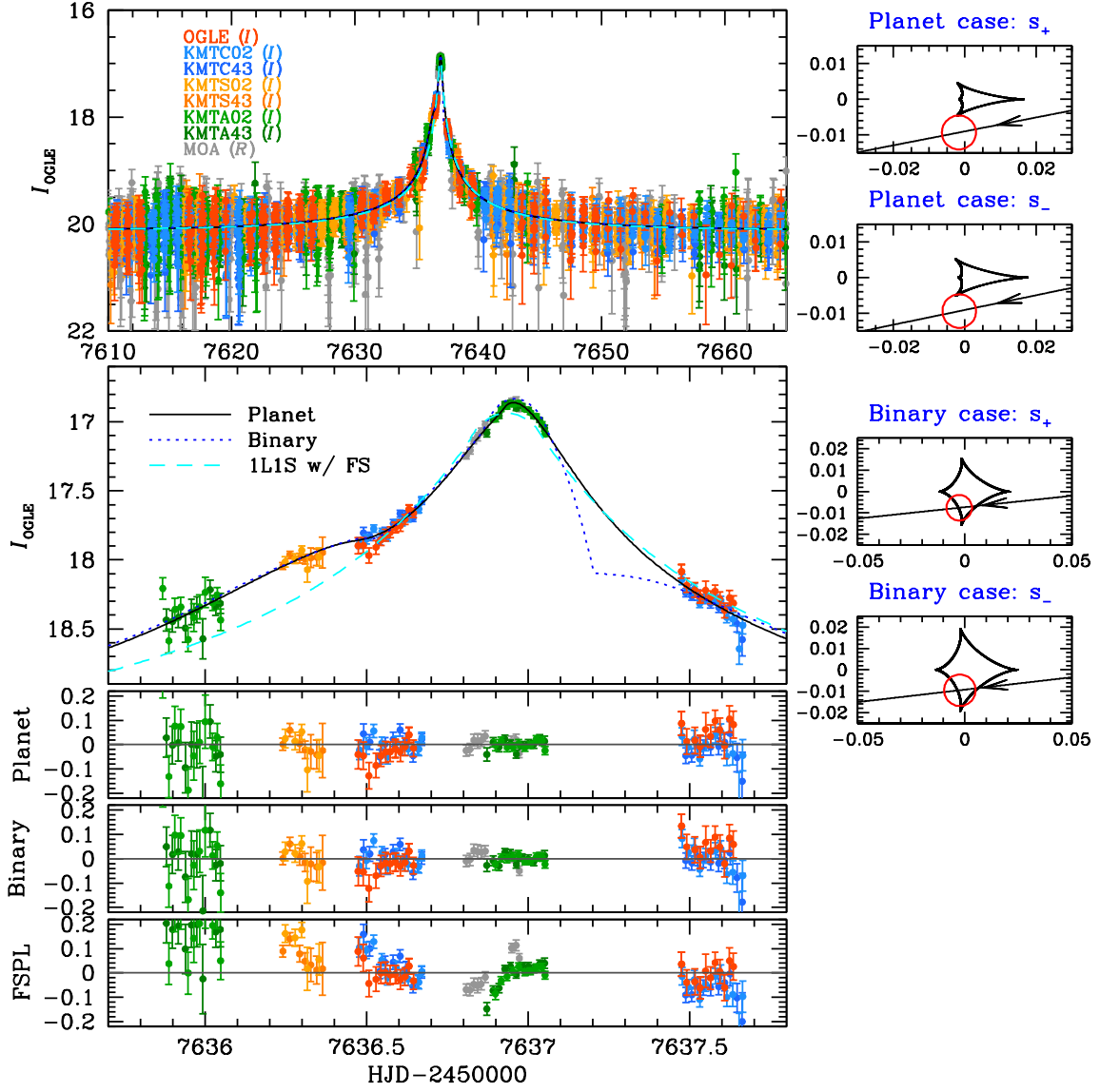}
\caption{Light curve of \fiveosix\ with the best-fit planetary model. We compare the planetary 
solution (black solid line) to binary (blue dotted line) and 1L1S with finite--source (cyan dashed line) 
models. We also present caustic geometries of 2L1S models on the right side for comparison.  
\label{fig:lc_0506}}
\end{figure}

In Figure \ref{fig:lc_0506}, we present the light curve of \fiveosix\ (which we identified as KMT-2016-BLG-0506), which shows a clear deviation from the 1L1S model with a finite source. Although the anomaly is neither obviously bump--shaped nor dip--shaped, we find that the heuristic analysis is valid. It yields $\tau_{\rm anom} = -0.032$ and $u_{\rm anom} = 0.034$ from $t_{\rm anom} = 7636.20$ and $t_{\rm E} =21$ days. Then, we expect $s^{\dagger}_{+} = 1.017$, which matches exactly with $s^{\dagger} = 1.017$ (derived from the modeling). The light curve can be well described by a 2L1S interpretation with both planet and binary cases (see light curves and geometries in Figure \ref{fig:lc_0506}). In Table \ref{table:model_0506}, we present the best-fit parameters. However, we find that the binary case shows worse fits by $\Delta\chi^{2} = 20.62$ and $22.32$ for the $s_{+}$ and $s_{-}$ cases, respectively. The $\Delta\chi^{2}$ amounts are larger than our criterion to claim the planet detection. Although we conclude that this event is caused by a planetary lens system, we report both cases because the crucial part of the light curve (HJD${^\prime} = 7637.1$--$7637.4$) for clearly distinguishing between planet and binary solutions is not covered.

For this event, $\rho_{\ast}$ is measured. The signal of the finite source effect comes from the peak of the light curve, which cannot be properly described by the 1L1S interpretation (see the residual of Figure \ref{fig:lc_0506}). The peak part can be described by 2L1S solutions (planet cases) by {\it touching} the cusp of the central caustic (see Figure \ref{fig:lc_0506}). As a result, $\rho_{\ast}$ is well measured.

We also test the APRX effect because of relatively long timescale of the event (i.e., $t_{\rm E} \sim 21$ days). We find that the $\chi^{2}$ improvements of $\Delta\chi^{2} = 21.57$ and $\Delta\chi^{2} = 12.73$ for the $s_{+}$ and $s_{-}$ cases, respectively. However, the APRX fits show values too big for both cases, i.e., $|\pi_{\rm E, {\it N}}| > 10$, which comes from over fitting systematics at the baseline. This fact implies that the APRX measurement is not reliable. Thus, we do not adopt the results of the APRX models.

\subsection{\sixtwentyfive} % KMT-2016-BLG-0625

\begin{figure}[htb!]
\epsscale{1.00}
\plotone{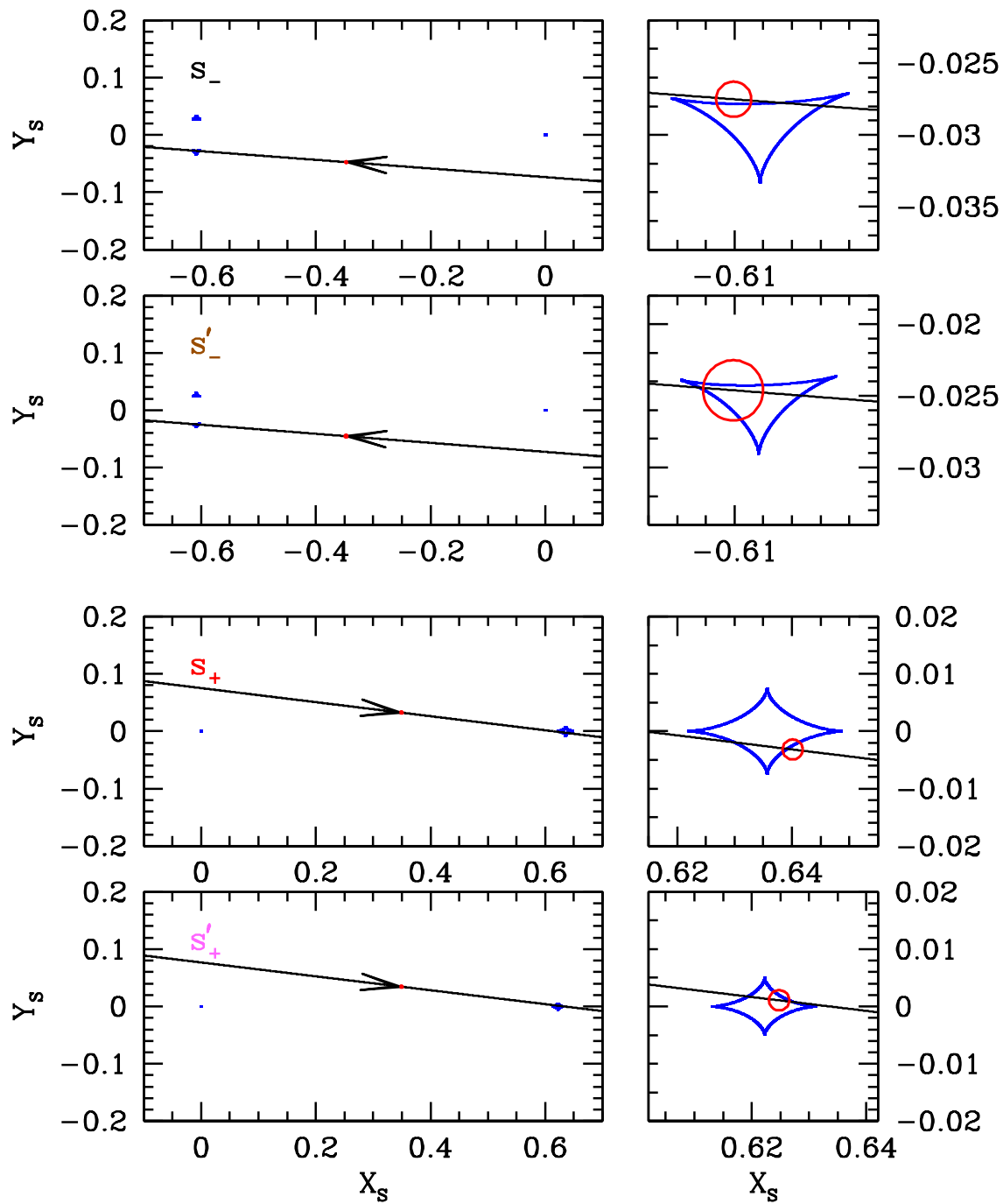}
\caption{Caustic geometries of \sixtwentyfive\ for the 2L1S models. 
\label{fig:geo_0625}}
\end{figure}

As shown in Figure \ref{fig:geo_0625}, the light curve of \sixtwentyfive\ shows a clear bump-shaped anomaly at HJD$^{\prime} \sim 7662.95$. Based on the heuristic analysis, we find $\tau_{\rm anom} = 0.609$ and $u_{\rm anom} = 0.613$ from $t_{\rm anom} = 7662.95$ and $t_{\rm E} = 11.5$ days. Then, we can expect that $s^{\dagger}_{-} = 0.739$ and $s^{\dagger}_{+} = 1.352$, which are consistent with $s_{-} = 0.741$ and $s_{+} = 1.367$, respectively, among the solutions presented in Table \ref{table:model_0625}. Also, we expect $\alpha = 0.12$ or $3.26$ radians, which are consistent with $\alpha = 0.12$ and $3.22$ of the $s_{+}$ and $s_{-}$ cases, respectively. 

As shown in Table \ref{table:model_0625}, we find four planetary solutions ($s_{\pm}$ and $s_{\pm}^{\prime}$) that can explain the anomaly. Because of the gaps near the anomaly, the $\Delta\chi^{2}$ values between the models (i.e., $\Delta\chi^{2} = 0.98$--$3.30$) are too small to distinguish between them, although the model light curves show quite different features caused by the different caustic geometries presented in Figure \ref{fig:geo_0625}. Although we cannot break the degeneracy of the planetary solutions, all cases indicate the companion of the lens system is a planet, i.e., $q = \mathcal{O}(10^{-4})$ (see Table \ref{table:model_0625}).

As shown in Figure \ref{fig:geo_0625}, all planetary solutions produce the anomaly by crossing the caustic(s). As a result, we can measure the $\rho_{\ast}$ despite the non-optimal coverage. We do not test for the APRX measurement because of the relatively short timescale (i.e., $t_{\rm E} \sim 11$ days).

Because the bump-type planetary anomaly can often lead to a 2L1S/1L2S degeneracy \citep{gaudi98}, we check the 1L2S case for this event. In Table \ref{table:model_0625}, we present the best-fit model of the 1L2S interpretation. We find that the 1L2S case is disfavored by $\Delta\chi^{2} = 7.35$. However, the $\Delta\chi^{2}$ amount is not enough to conclusively resolve the 2L1S/1L2S degeneracy. Nevertheless, because we measure the $\rho_{\ast}$ of the secondary source, we can measure the lens--source relative proper motion of the secondary source ($\mu_{\rm rel, S_{2}}$) to check the 1L2S model. We find (see Section \ref{sec:cmd_analysis}) that $\mu_{\rm rel, S_{2}} = 0.83\pm 0.22\, {\rm mas\, yr^{-1}}$. By comparison, \citet{masada} found that for observed microlensing events with planetary-type anomalies, low proper motions have probabilities
\begin{equation}
p(\leq\mu_{\rm rel}) = \frac{(\mu_{\rm rel}/2\sigma_\mu)^{\nu+1}}{[(\nu+1)/2]!}
\rightarrow \frac{\mu_{\rm rel}^{2}}{4\sigma_\mu^2}  
\rightarrow 2.8\times 10^{-2} \left( \frac{\mu_{\rm rel}}{1\, {\rm mas\, yr^{-1}}} \right)^{2},
\label{eqn:masada}
\end{equation}
where $\sigma_\mu=3\, {\rm mas\,yr^{-1}}$ and $\nu=1$.  See also Equation~(9) of \citet{jung23}. Applying this formula to the 1L2S solution, we find $p=1.9\%$.  This would, in itself, be a reasonably strong argument against the 1L2S solution. When combined with the fact that this solution is disfavored by $\Delta\chi^{2}=7.35$, we consider it to be decisive. Therefore, we reject the 1L2S solution and conclude that \sixtwentyfive\ is caused by a planetary lens system. However, we note that the mass ratio $q$ varies by a factor $\sim 3$ over the four degenerate solutions.

\subsection{\thirteenseven} % KMT-2016-BLG-1307

The light curve of \thirteenseven\ (which we identified as KMT-2016-BLG-1307) shows a dip--shaped anomaly at HJD$^{\prime} \sim 7663$. Based on the heuristic analysis, we can expect $s^{\dagger}_{-} = 0.812$ and $q = 0.9 \times 10^{-4}$ (based on $\tau_{\rm anom} = 0.126$ and $u_{\rm anom} = 0.419$ that are found from $t_{\rm anom} = 7663.15$ and $t_{\rm E} = 63.0$ days), which corresponds well with the empirical values: $s^{\dagger} = 0.813$ and $q \sim 1.0\times 10^{-4}$.

\begin{figure}[htb!]
\epsscale{1.00}
\plotone{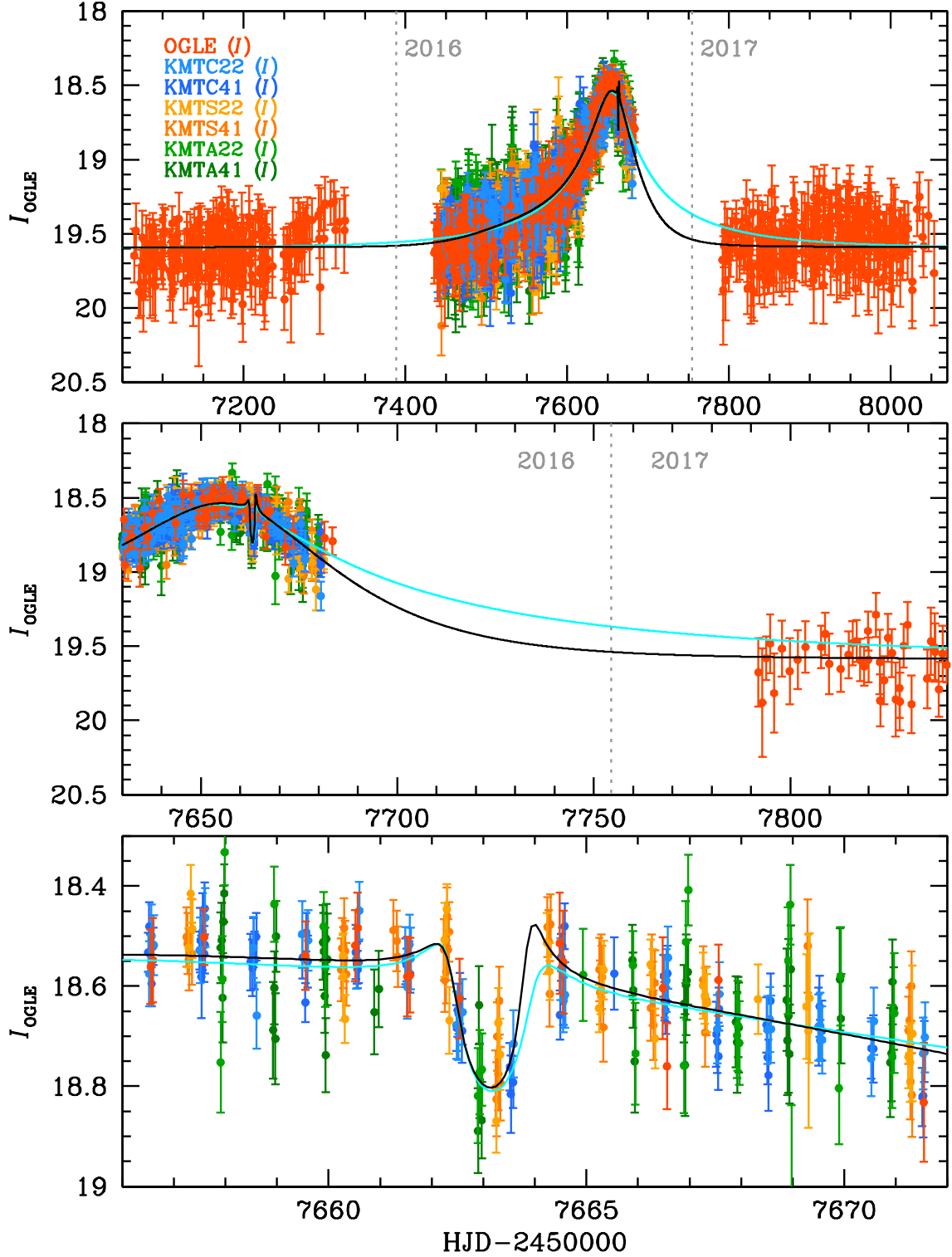}
\caption{Light curve of \thirteenseven\ with STD (cyan) and APRX (black) models. 
\label{fig:lc_whole_1307}}
\end{figure}

\begin{figure}[htb!]
\epsscale{1.00}
\plotone{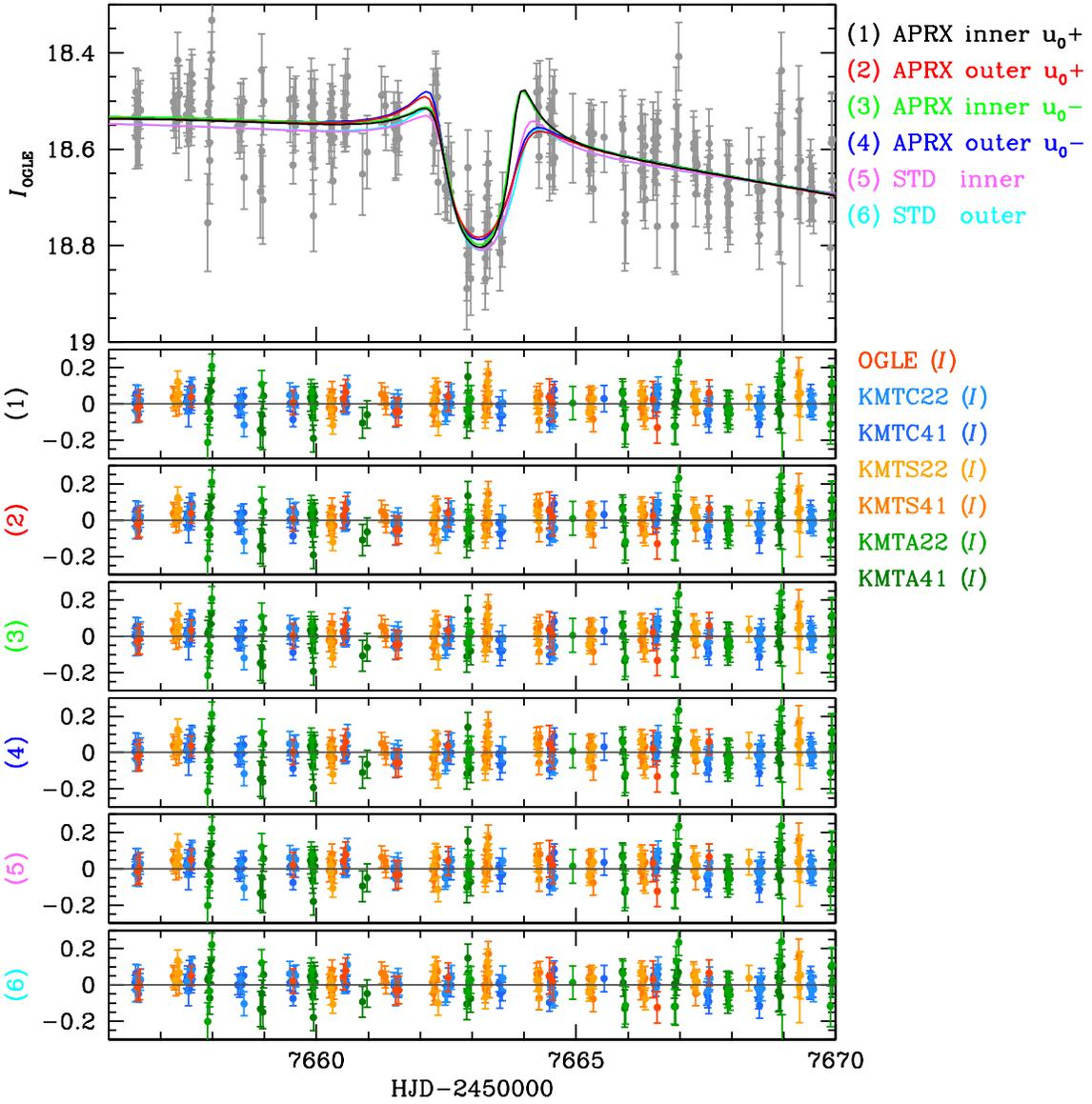}
\caption{Zoom-in of the anomaly part of \thirteenseven\ for comparing all models with their residuals.  
\label{fig:lc_zoom_1307}}
\end{figure}

\begin{figure}[htb!]
\epsscale{1.00}
\plotone{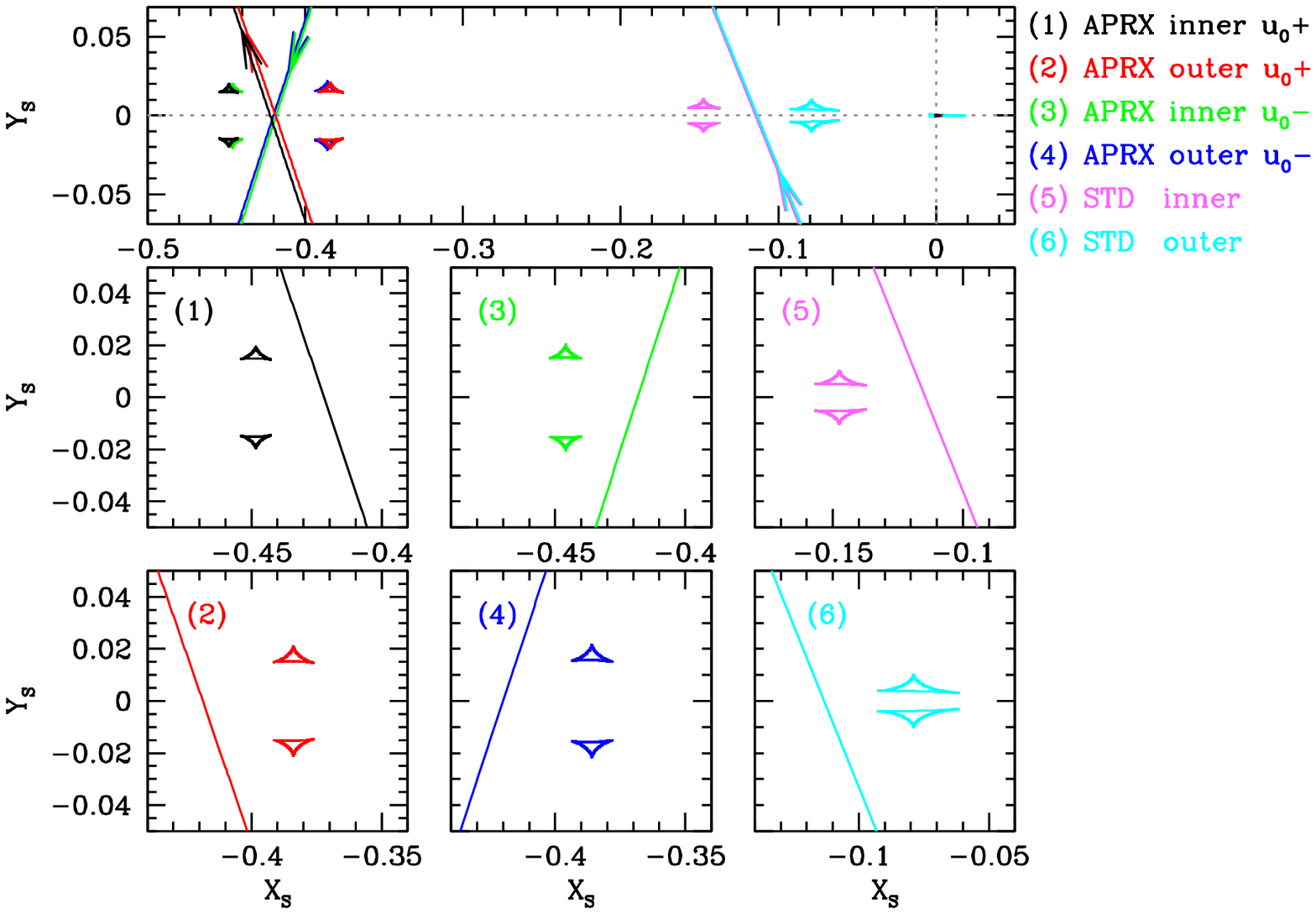}
\caption{Caustic geometries of \thirteenseven. 
\label{fig:geo_1307}}
\end{figure}

In Figure \ref{fig:lc_whole_1307}, we present the observed light curve with zoom--ins of the anomaly. We also present the best-fit model light curves of the STD and APRX cases shown in Table \ref{table:model_1307}. We find that both STD models (i.e., inner and outer cases) can describe the planetary anomaly as shown in Figure \ref{fig:lc_zoom_1307}. However, the STD cases show a very long timescale ($t_{\rm E} \sim 210$ days), which implies that the light curve is likely to be affected by a strong APRX effect. As expected, we find that the STD model cannot properly describe the $2017$ baseline. Thus, we consider the ARPX effect. Then, we find a substantial $\chi^{2}$ improvement of $\Delta\chi^{2} \gtrsim 100$, which mostly comes from the better fit of $2017$ baseline (see Figure \ref{fig:lc_whole_1307}). Also, all APRX solutions can well describe the planetary anomaly as shown in Figure \ref{fig:lc_zoom_1307}. In Figure \ref{fig:geo_1307}, we present caustic geometries of all cases for comparison. We note that \thirteenseven\ is a non--caustic--crossing event. As a result, we cannot precisely measure $\rho{\ast}$ (only upper limits are available).

\begin{figure}[htb!]
\epsscale{1.00}
\plotone{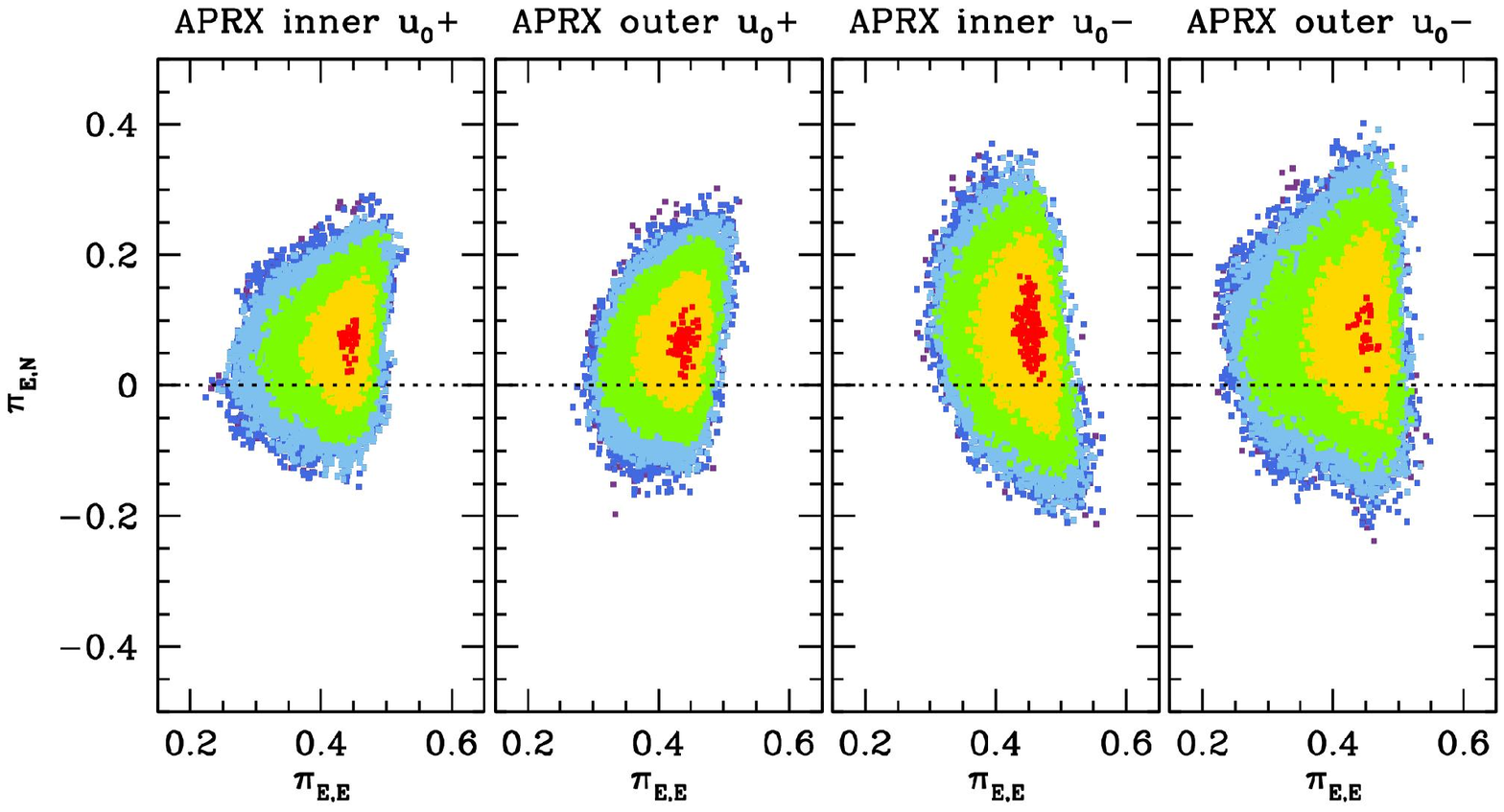}
\caption{The APRX distributions of \thirteenseven. 
\label{fig:APRX_1307}}
\end{figure}

In Figure \ref{fig:APRX_1307}, we present the distributions of the APRX measurements, which are well converged. However, tests are required before we conclude that the APRX models should be the fiducial solutions for this event. First, because the lens--orbital motion can affect the APRX measurements (especially, the uncertainty of the APRX measurement), we test the lens--orbital effect (OBT). We conduct OBT+APRX models for each APRX case. We find that the OBT+APRX models show negligible $\chi^{2}$ improvements, which are $\Delta\chi^{2} \lesssim 0.5$ for the inner cases and $\Delta\chi^{2} \lesssim 3.0$ for the outer cases, respectively. We also find that there is no effect on the uncertainties of the APRX measurements. 

\begin{figure}[htb!]
\epsscale{1.00}
\plotone{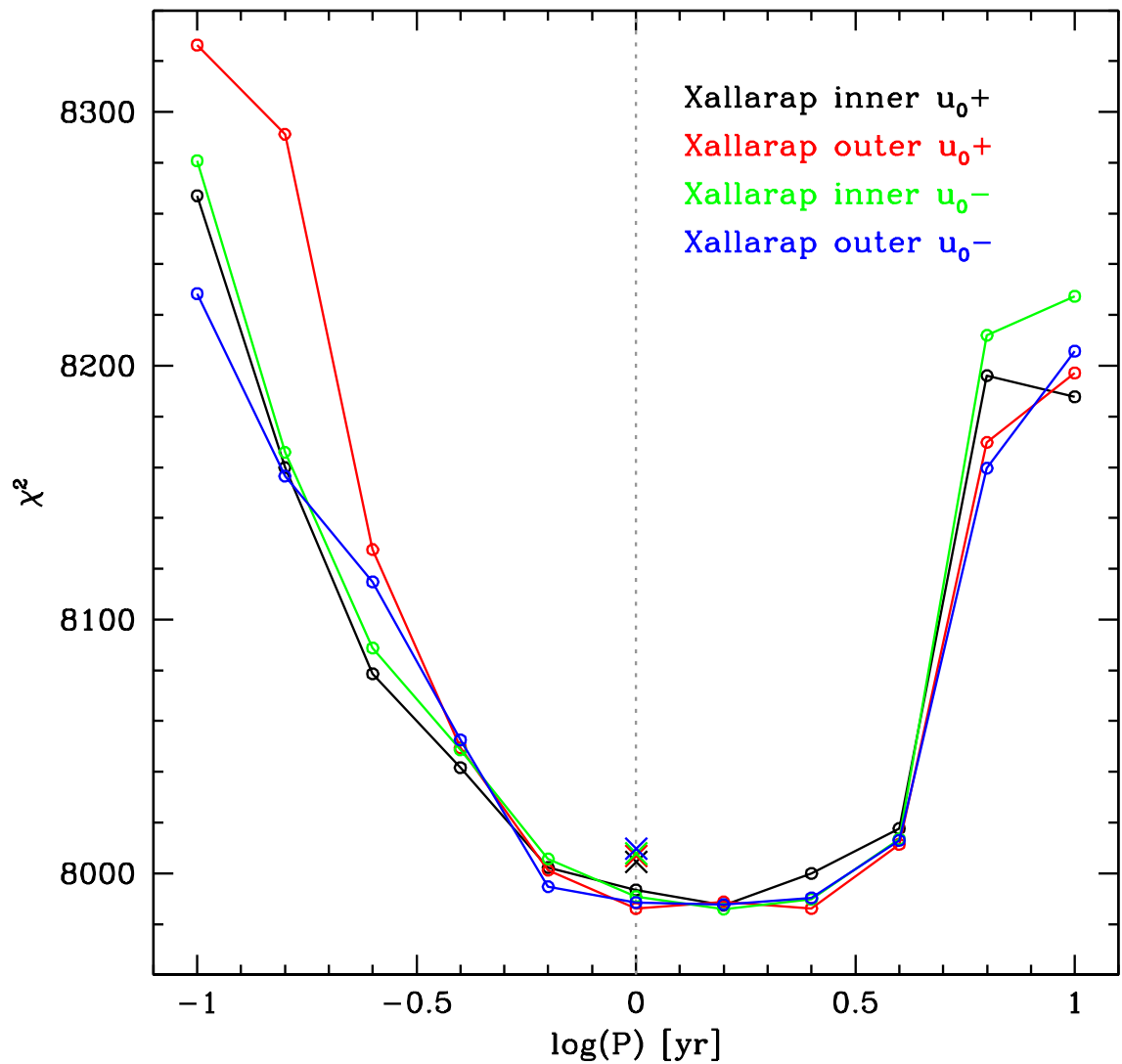}
\caption{$\log_{10}(P)$ and $\chi^{2}$ plots of xallarap models of \thirteenseven. 
The ``X" marks indicate $\chi^{2}$ of the APRX models.
\label{fig:P_chi2_1307}}
\end{figure}

Second, to check the APRX models, we add xallarap to the models by introducing five parameters, which are North and East components of the xallarap vector ($\xi_{\rm E, {\it N}}$, $\xi_{\rm E, {\it E}}$), the phase angle ($\phi$), the inclination of the orbit ($i$), and orbital period ($P$). We find that xallarap cases show $\chi^{2}$ improvements of $\Delta\chi^{2} = 17.0$--$22.1$ compared to the APRX cases, which are marginal $\Delta\chi^{2}$ amounts to firmly claim the xallarap models can be fiducial solutions for this event. Moreover, although the best--fit model favors $\log_{10}(P) = 0.2$ as shown in Figure \ref{fig:P_chi2_1307}, we find that the xallarap models at $\log_{10}(P) = 0.0$ show $\Delta\chi^{2} \lesssim 6.0$ compared to the best--fit xallarap model of each case. The clues imply that the asymmetry of the light curve is due to the APRX effect rather than xallarap effect. Thus, we conclude that the fiducial solutions for this event are the APRX models.

\subsection{\seventeenfiftyone} % KMT-2016-BLG-1751
\label{sec:lc_analysis_1751}

\begin{figure}[htb!]
\epsscale{1.00}
\plotone{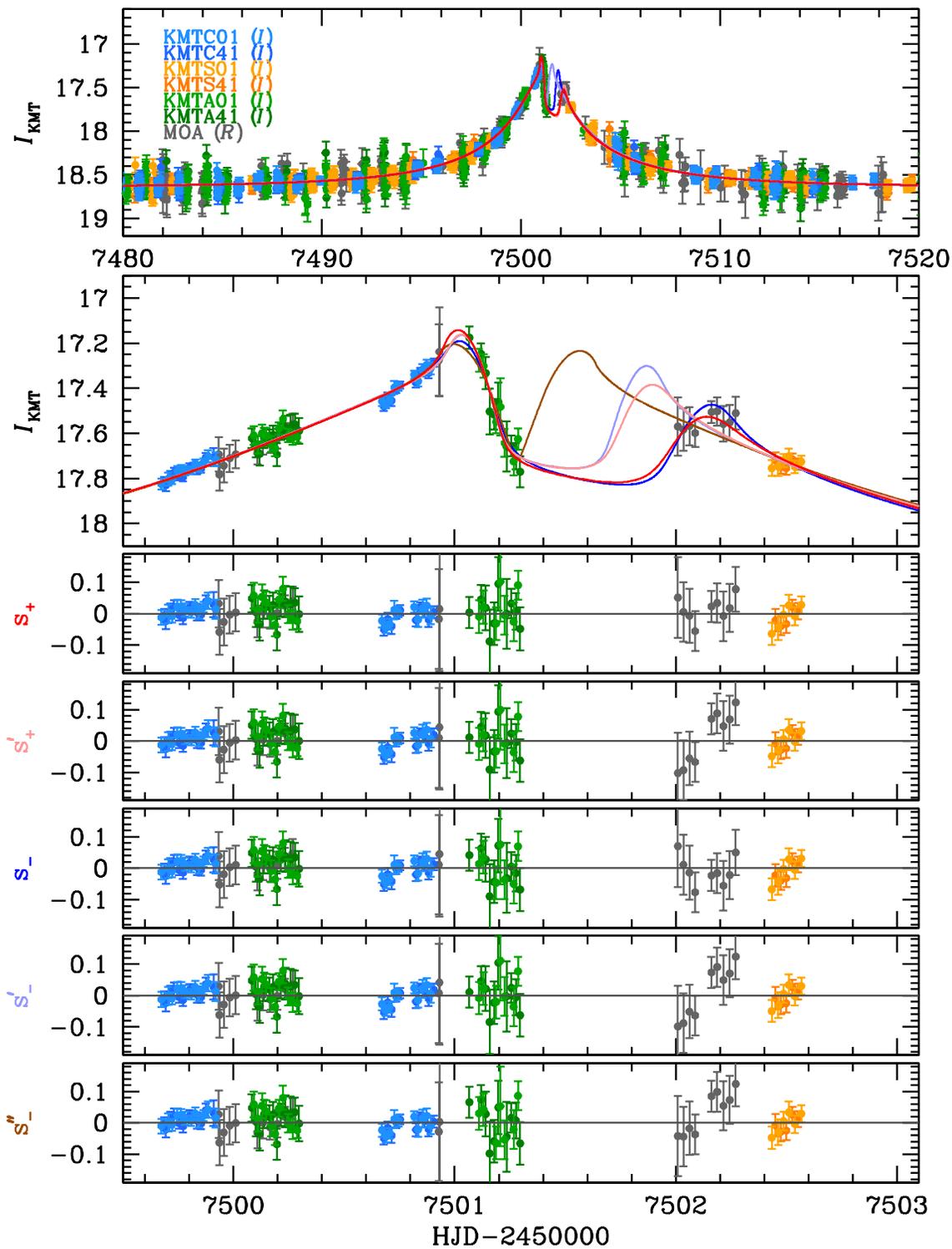}
\caption{Light curve of \seventeenfiftyone\ with degenerate models.
\label{fig:lc_1751}}
\end{figure}

In Figure \ref{fig:lc_1751}, we present the observed light curve of \seventeenfiftyone, which shows a clear planetary anomaly (i.e., dip--feature) at the peak of the light curve. Based on the heuristic analysis ($\tau_{\rm anom} \sim 0.00$ and $u_{\rm anom} = 0.11$ from $t_{\rm anom} = 7501.00$ and $t_{\rm E} = 10.0$ days), we expect $s^{\dagger}_{-} = 0.946$, which is well matched to both $s^{\dagger} = \sqrt{s_{-}s_{+}} = 0.944$ and $s^{\dagger} = \sqrt{s^{\prime}_{-}s^{\prime}_{+}} = 0.947$. We also expect $q \simeq 0.003$ (for both $s^{\dagger}$ cases), which agrees with the $q$ values presented in Table \ref{table:model_1751} to within a factor of $\sim 2$. 

\begin{figure}[htb!]
\epsscale{1.00}
\plotone{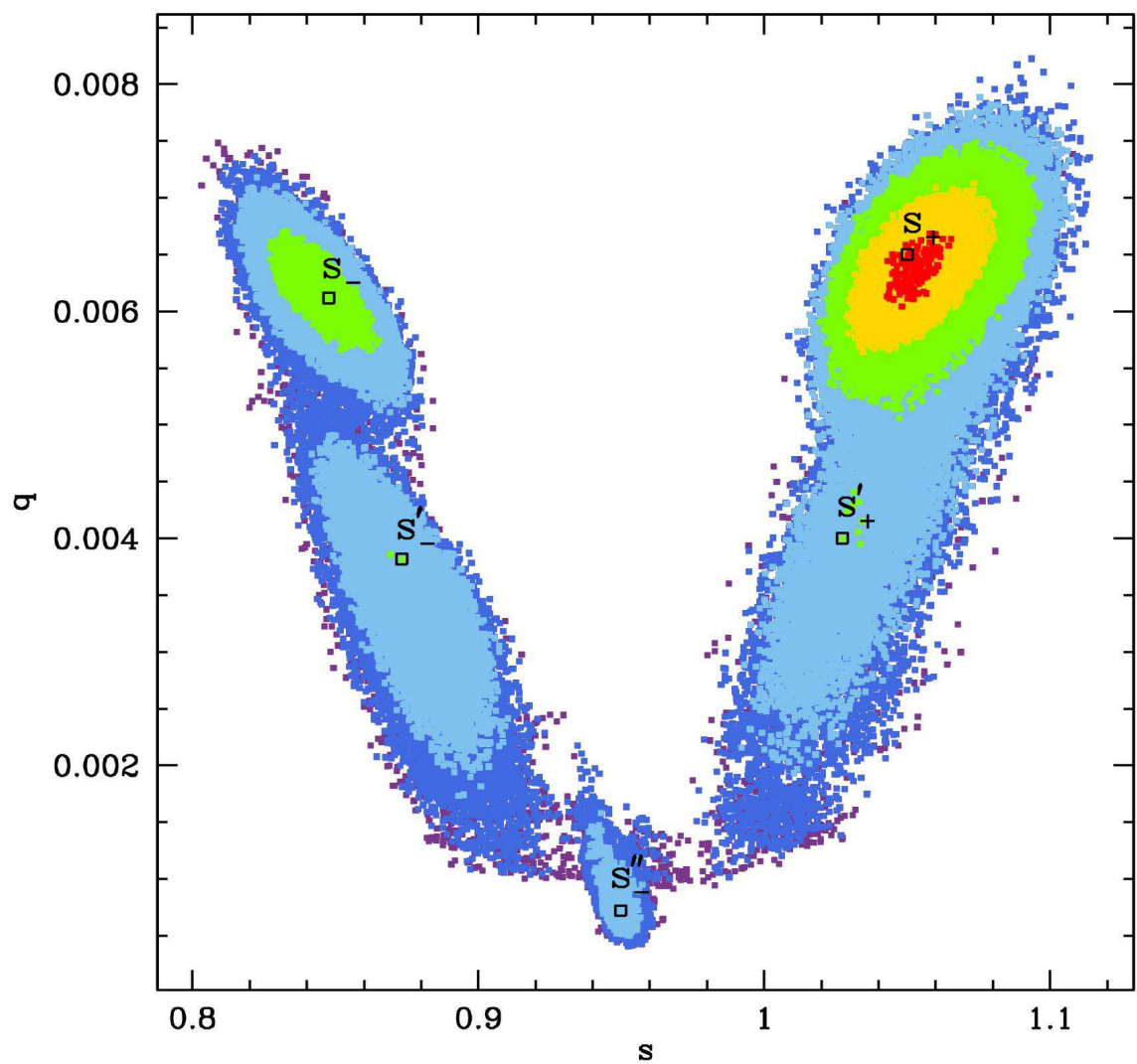}
\caption{$s$--$q$ parameter space of \seventeenfiftyone\ showing possible solutions.
\label{fig:sq_locals_1751}}
\end{figure}

\begin{figure}[htb!]
\epsscale{1.00}
\plotone{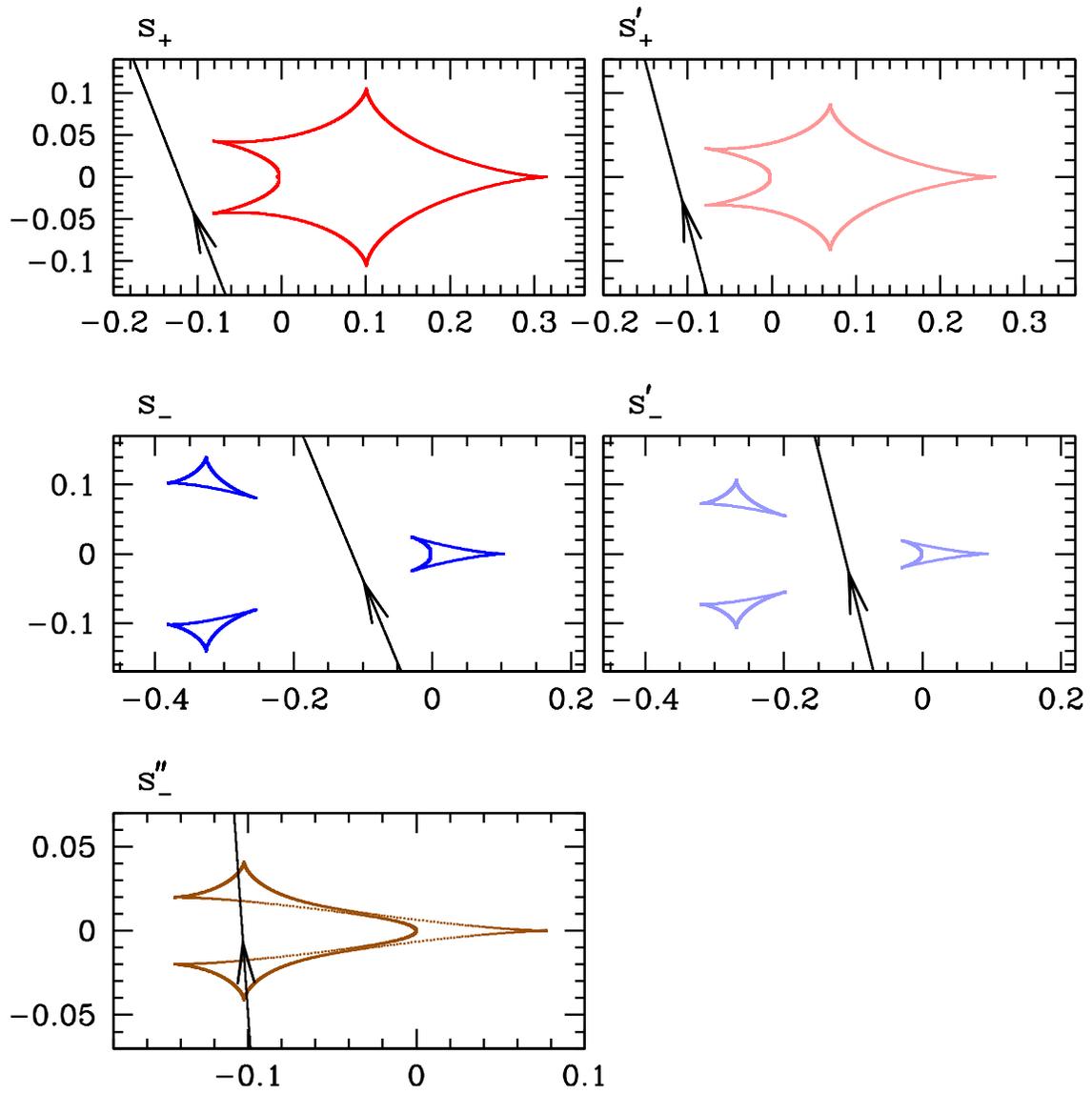}
\caption{Caustic geometry of each solution of \seventeenfiftyone. The color of caustic is identical 
to the color of the light curve shown in Figure \ref{fig:lc_1751}. 
\label{fig:geo_1751}}
\end{figure}

We find that several solutions can explain the anomaly because the coverage of the anomaly (HJD$^{\prime} = 7500.8$--$7502.4$) is non-optimal. Thus, despite including MOA data, the gap in the anomaly produces degenerate solutions. In Table \ref{table:model_1751}, we present model parameters of the solutions. In Figures \ref{fig:sq_locals_1751} and \ref{fig:geo_1751}, we also present the $s$--$q$ parameter space with the locations of each solution and their caustic geometries. The competing solutions show relatively small $\Delta\chi^{2}$ values compared to the best-fit solution (i.e., $s_{+}$ case): $8.53$, $5.70$, $8.78$, and $10.78$ for $s^{\prime}_{+}$, $s_{-}$, $s^{\prime}_{-}$, and $s^{\prime\prime}_{-}$ cases, respectively. For the $s_{\pm}$ and $s_{\pm}^{\prime}$ case, we obtain a best-fit value for $\rho_{\ast}$. However, as might be expected from the geometries, we find that the measurements are consistent with zero at 3$\sigma$. Thus, in these cases, we effectively have only an upper limit on $\rho_{\ast}$, so we will apply a $\rho_{\ast}$ weight function in the Bayesian analysis in Section \ref{sec:bayes_analysis}. For the $s_{-}^{\prime\prime}$ case, $\rho_{\ast}$ is measured from the caustic--crossing. However, the $s_{-}^{\prime\prime}$ solution does not satisfy our $\chi^{2}$ criterion (i.e., $\Delta\chi^{2} < 10.0$). Thus, we remove the $s_{-}^{\prime\prime}$ case from our fiducial solutions for determining lens properties of this event. However, because the $\Delta\chi^{2}$ of this case is very close to the $\chi^{2}$ criterion, we present parameters and figures of this solution for completeness. Lastly, because of the short timescale (i.e., $t_{\rm E} \sim 10$ days), we do not conduct the APRX modeling for this event.

\subsection{\eighteenfiftyfive} % KMT-2016-BLG-1855
\label{sec:lc_analysis_1855}

\begin{figure}[htb!]
\epsscale{1.00}
\plotone{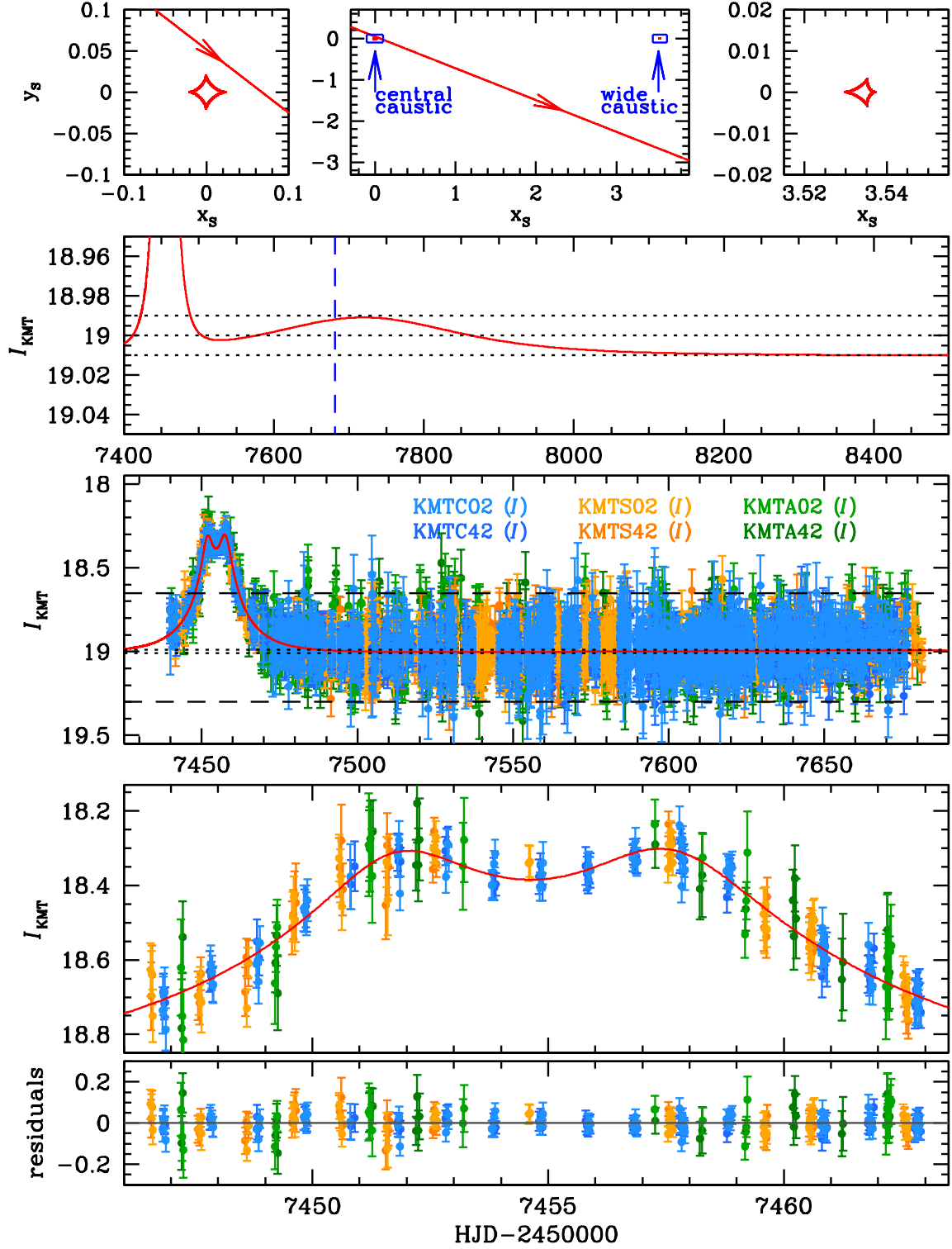}
\caption{Light curve of \eighteenfiftyfive\ with the best--fit model.
\label{fig:lc_1855_best}}
\end{figure}

\begin{figure}[htb!]
\epsscale{1.00}
\plotone{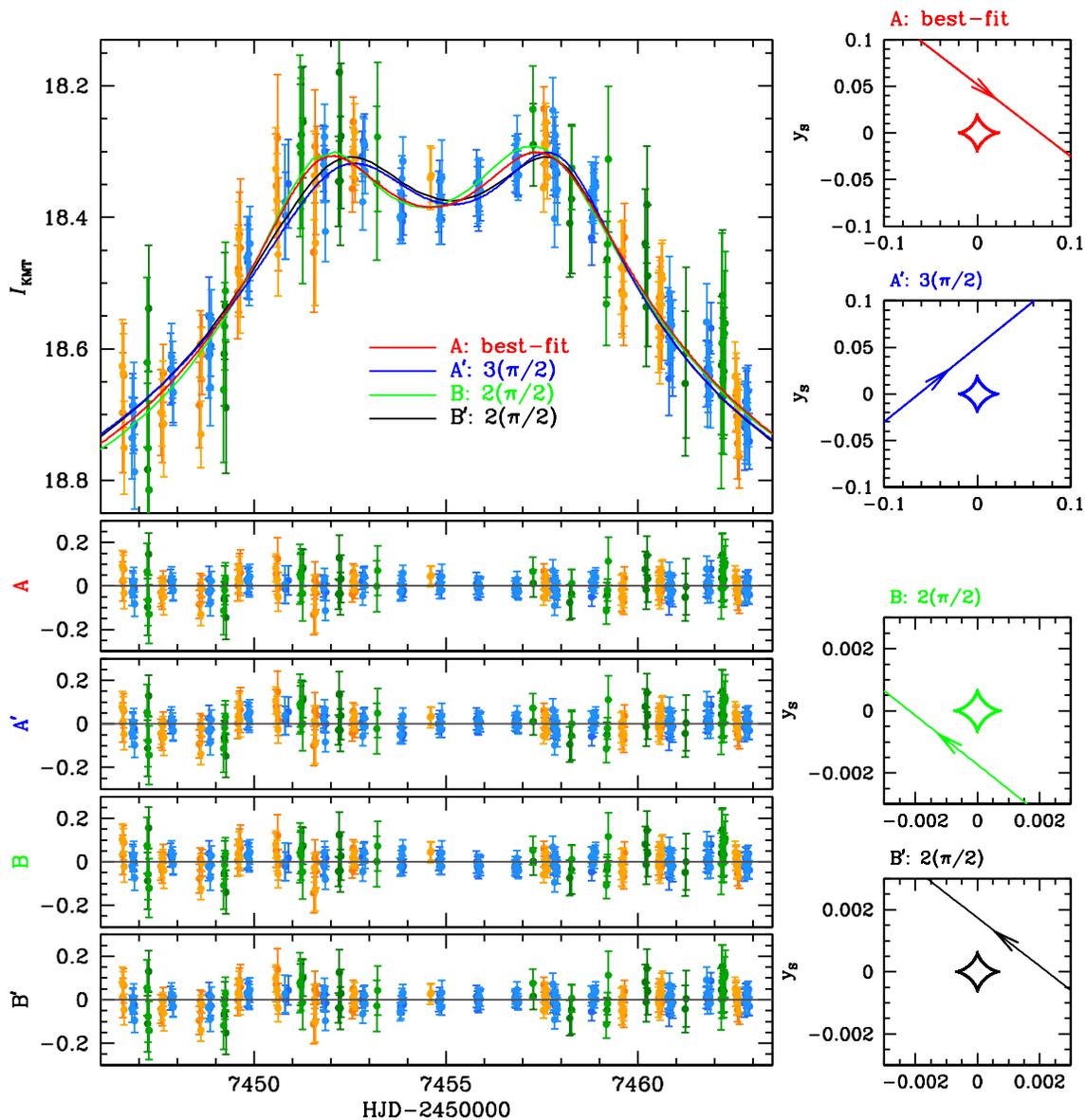}
\caption{Light curve of \eighteenfiftyfive\ with the best--fit (A) and degenerate models of A$^{\prime}$, B, and B$^{\prime}$.
\label{fig:lc_1855_degen_1}}
\end{figure}

\begin{figure}[htb!]
\epsscale{1.00}
\plotone{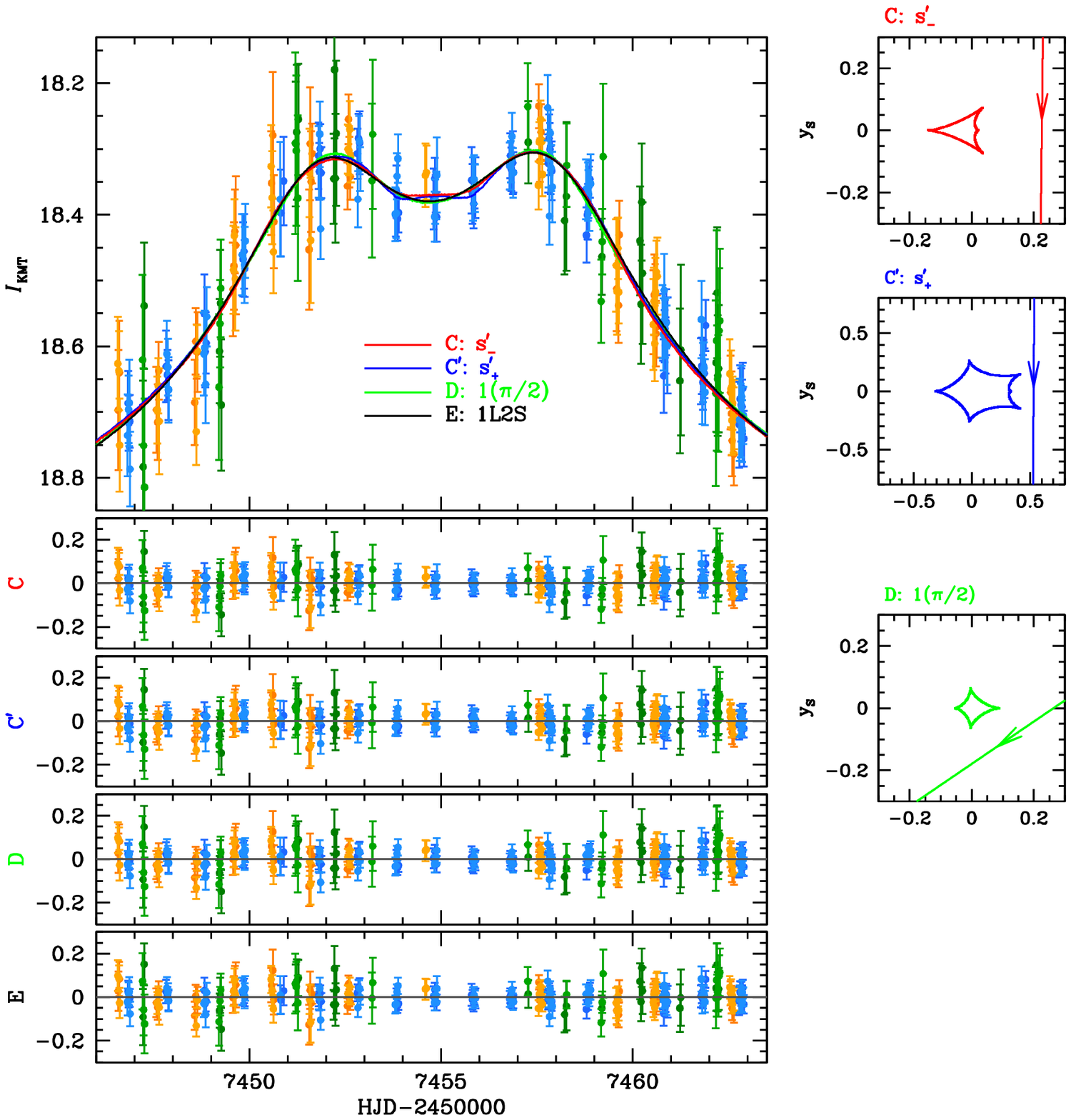}
\caption{Light curve of \eighteenfiftyfive\ with degenerate models of C, C$^{\prime}$, D, and E.
\label{fig:lc_1855_degen_2}}
\end{figure}

In Figure \ref{fig:lc_1855_best}, we present the observed light curve of \eighteenfiftyfive\ with the best--fit model curve and caustic geometry. The observed light curve exhibits anomalies at the peak. We find that the anomaly can be described by a source approaching a diamond--shaped central caustic, which is in the regime of a Chang \& Refsdal lens \citep[C--R:][]{chang79}. The best--fit model shows $\frac{1}{q} = 0.023\pm0.012$ which satisfies our mass ratio criterion to claim planet detection. However, we find that there exist possible solutions caused by the close/wide degeneracy \citep{griest98}, the offset degeneracy \citep{zhang22a, zhang22b}, and 2L1S/1L2S degeneracy \citep{gaudi98}. We also check for the $\alpha$--degeneracy (i.e., degeneracy caused by the angle of source trajectory), which can occur for C--R lenses. These solutions are denoted $n(\pi/2)$, where $n = (1,2,3)$.
In Tables \ref{table:model_1855_1} and \ref{table:model_1855_2}, we present model parameters of the best--fit and degenerate models. In Figures \ref{fig:lc_1855_degen_1} and \ref{fig:lc_1855_degen_2}, we present all possible solutions with their caustic geometries and residuals for comparison. 

We find total $7$ degenerate solutions including the 1L2S case. For the 2L1S cases, we find the initial parameters of A, B, C, and D solutions based on the grid search. We also find initial parameters for their paired offset-degeneracy solutions (i.e., A$^{\prime}$, B$^{\prime}$, C$^{\prime}$, and D$^{\prime}$) using heuristic analysis: $s^{\prime} = {(s_{\pm}^{\dagger})}^{2}/s_{\pm}$ where subscripts ${+}$ and ${-}$ indicates the $s > 1$ and $s < 1$ cases, respectively. Note that we transform our coordinates system from ``secondary'' to ``primary'' components to conduct the heuristic analysis because the analysis is valid for the ``primary'' coordinates system. Then, we refine the model parameters to check the degeneracy (note that we restore the coordinates for direct comparison). 

For the A and A$^{\prime}$ pair, the heuristic analysis predicts $s^{\prime} = 3.680$. The paired offset-degeneracy solution of the A case (i.e., A$^{\prime}$) is consistent with the $3(\pi/2)$ C--R case, which has an empirical value of $s = 3.780\pm0.088$. This A family degeneracy can be resolved (see below).
For the B and B$^{\prime}$ pair, the heuristic analysis predicts $s^{\prime} = 3.600$, which is consistent with the empirical $s = 3.708\pm0.121$ from the B$^{\prime}$ case. This B family is a C--R lensing case, which shows large uncertainties in the set of $(t_{\rm E}, s, q)$ parameters. 
For the C and C$^{\prime}$ pair, the heuristic analysis predicts $s^{\prime} = 1.162$, which is consistent with the empirical value of $s = 1.161\pm0.042$ from the C$^{\prime}$ case. Indeed, the C family is caused by close/wide degeneracy. 
For the D case, the heuristic analysis expects $s^{\prime} = 2.951$. However, we find that the paired offset solution evolves toward the B case. Indeed, the caustic geometry of the D case is asymmetric, which is different from the C--R lens case. Thus, because the source trajectory is not perpendicular the binary axis, the paired solution from the heuristic analysis cannot describe the peak of the light curve, and we would not necessarily expect it to \citep{gaudi97}. 
In all cases, the $\rho_{\ast}$ measurements are uncertain and give only upper limits of $\rho_{\ast}$ values as expected from non--crossing caustic geometries.

All the 2L1S models nominally have long timescales ($t_{\rm E}$), but they also have $q > 1.0$ (i.e., they approach the secondary, less-massive, lens component). For these cases, the actual timescale ($t_{\rm E}^{\prime}$) of the event should be scaled by $t_{\rm E}^{\prime} = t_{\rm E}\sqrt{q}$ as shown in Tables \ref{table:model_1855_1} and \ref{table:model_1855_2}. Hence, given that $t_{\rm E}^{\prime} \sim 15$ days, it is not surprising that we do not detect the APRX effect (i.e., $\Delta\chi^{2}_{\rm STD -APRX} = 2.7$). Thus, we conclude that the STD models are the fiducial solutions for this event.

As shown in Figures \ref{fig:lc_1855_degen_1} and \ref{fig:lc_1855_degen_2}, all cases describe the peak anomaly well. Although they nominally have $\Delta\chi^{2} > 10$ compared to the best--fit case, we find that $\chi^{2}$ differences mostly come from the baseline part (${\rm HJD^{\prime}} > 7600$). The best-fit case has a wide caustic, which creates a very shallow bump peaking at ${\rm HJD^{\prime} \sim 7717}$, $\Delta I \sim 0.01$ magnitudes above the baseline observations (see the blue dashed line).
However, systematics may exist in the baseline data at this level, especially considering the dispersion of the baseline data (i.e., $\Delta I \sim 0.65$ magnitudes). 
Thus, we compute $\Delta\chi^{2}$ without the baseline data of ${\rm HJD^{\prime} > 7600}$ (which $t_0 + \sim 1.5\,t_{\rm E} \sim 7595.25$) because the $\chi^{2}$ contributions at the baseline cannot be considered reliable. After this cut, the $\Delta\chi^{2}$ values for all cases (except the A$^{\prime}$ case) are less than $9$ as shown in Tables \ref{table:model_1855_1} and \ref{table:model_1855_2}. Hence, we cannot claim to resolve most degenerate solutions. 

The 1L2S model is completely degenerate with the 2L1S models and cannot be excluded based on physical considerations. First, the finite source effect is not measured; the $\rho_{\ast}$ distributions of both sources reach zero within $3\sigma$. Moreover, because of the severe extinction \citep[$A_{I} = 5.97$;][]{gonzalez12}, additional information to conclusively resolve the degeneracy, such as the source color (see Section \ref{sec:cmd_analysis}), is not available for this event. 

Thus, we treat \eighteenfiftyfive\ as a planet candidate, and we strongly counsel against cataloging it as a ``planet''.

\section{CMD Analysis} \label{sec:cmd_analysis}

For the five planetary events, we measure the angular source radius ($\theta_{\ast}$) using the conventional method described in \citet{yoo04}, i.e., the color-magnitude diagram (CMD) analysis. The $\theta_{\ast}$ measurement is important. If we measure $\rho_{\ast}$ from the finite--source effect, we can determine $\theta_{\rm E} = \theta_{\ast}/\rho_{\ast}$. Furthermore, even if we cannot measure the $\rho_{\ast}$, $\theta_{\ast}$ is required to apply the $\rho_{\ast}$ distributions as constraints on the Bayesian analysis.

\begin{figure}[htb!]
\epsscale{1.00}
\plotone{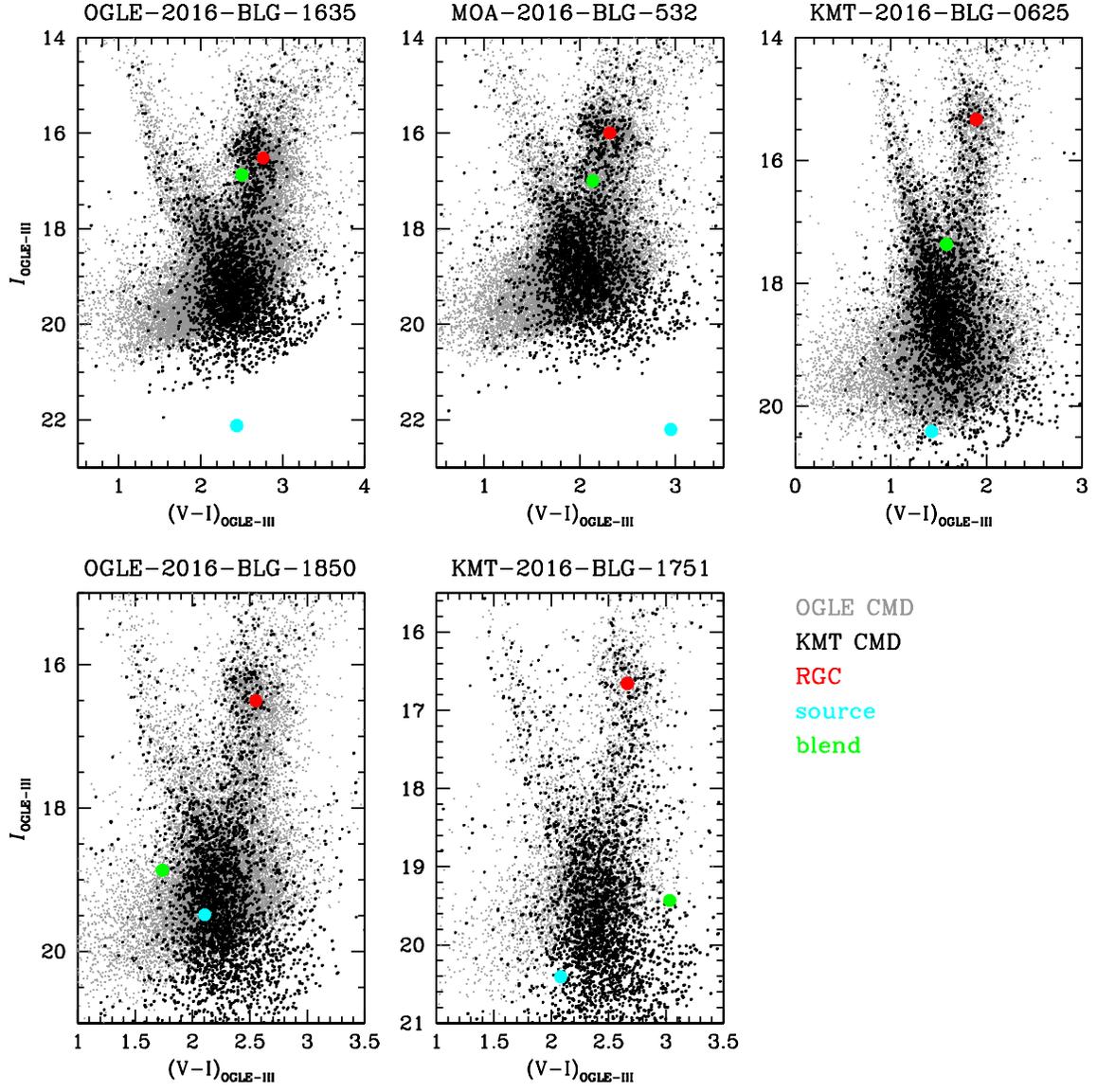}
\caption{Color--magnitude diagrams (CMD) of five planetary events.
\label{fig:CMDs}}
\end{figure}

We proceed with this analysis based on multi--band observations ($I$-- and $V$--bands) taken from KMTNet survey (i.e., KMTC). We align the KMTNet instrumental color and magnitudes to the OGLE--III scales using cross--matching of field stars. We note that the position of the red giant clump centroid (RGC) is determined based on the OGLE--III CMD \citep{szymanski11}. In Figure \ref{fig:CMDs}, we present CMDs of the five planetary events for the best--fit cases with the positions of RGC, source, and blend. We also present all information from the CMD analysis, including $\theta_{\ast}$, $\theta_{\rm E}$ and $\mu_{\rm rel}$, in Table \ref{table:cmd}. We note that the intrinsic color of the RGC is adopted from \citet{bensby11}. The de-reddened magnitude of the RGC is adopted from \citet{nataf13}. The de-reddened colors and magnitudes of source and blend are determined by assuming they experienced same amount of stellar extinction of the RGC. Lastly, we determine the $\theta_{\ast}$ using the surface brightness--color relation adopted from \citet{kervella04}.

Note that we proceed differently for the special case of the putative second source in the 1L2S solution for \sixtwentyfive. We find $I_{\rm S,0} = 19.345 \pm 0.010$ using the method of \citet{yoo04}. Then, we derive $I_{\rm S_{2},0} = 25.112 \pm 0.231$ based on the $q_{\rm flux}$ value of the 1L2S model. We convert the de-reddened $I$--band magnitude of the second source to absolute $I$--band magnitude ($M_{I}$) for the second source by adopting $M_{I,{\rm RGC}} = -0.12 \pm 0.09$ and $I_{\rm RGC,0} = 14.335$ from \citet{nataf13}: $M_{I,{\rm S_{2}}} = 10.656 \pm 0.248$. We can estimate the radius of the second source, $R_{\rm S_{2}} \sim 0.208\, R_{\odot}$, based on studies for stellar properties \citep{pecaut12,pecaut13}. Thus, we find the angular radius of the second source is $\theta_{\ast,{S_{2}}} \sim 0.128\, \mu{\rm as}$, which yields $\mu_{\rm rel, S_{2}} = 0.83 \pm 0.22\, {\rm mas\, yr^{-1}}$. Note that we adopt the distance to the second source ($D_{\rm S_{2}} \sim 7.59\, {\rm kpc}$) from \citet{nataf13}.

For \eighteenfiftyfive, the field is highly extincted \citep[$A_I = 5.97$;][]{gonzalez12}, so is not possible to measure the source color
in $V$-band from the KMTNet data. We construct an $I$-$H$ CMD for this event by cross-matching the OGLE-III catalog \citep{szymanski11} to VVV DR2 \citep{VVV_DR2}, and we convert the KMT pyDIA $I$ magnitude of the source to the OGLE-III
system. This that suggests the source is a red clump giant. However, the clump is extended in both color and magnitude in the CMD. Both the
lack of a color measurement and the uncertainty in the clump magnitude would make $\theta_{\star}$ highly uncertain. However, there are no
meaningful constraints on $\rho_{\star}$, so we do not calculate a value for $\theta_{\star}$ because it has no bearing on the analysis.

We also measure astrometric offsets between baseline objects and sources to check whether or not the blend light can be used as constraints. For all planetary events, we find that the blend is separated by $> 0.3^{\prime\prime}$, so it is dominated by a star that is not the lens.

\section{Planet Properties} \label{sec:bayes_analysis}

\subsection{Bayesian Formalism}

To determine the lens properties, two additional observables are simultaneously required. They are the angular Einstein ring radius ($\theta_{\rm E}$) and the amplitude of the microlens parallax vector ($|\pi_{\rm E}|$), which are measured from the effects of the finite source and microlens parallax, respectively. However, the events for which both observables are simultaneously measured are relatively rare. Indeed, we can measure only one of these observables out of five planetary events presented in this work. Thus, we estimate the lens properties using the Bayesian analysis. We follow the Bayesian formalism described in \citet{shin23} to generate the Galactic prior. Then, we apply the measured observable as a constraint on the Galactic prior. In Table \ref{table:lens}, we present the applied constraints and the lens properties for each event. For the notation of the constraint, $t_{\rm E}$ indicates a Gaussian weight function constructed based on the best--fit value of the $t_{\rm E}$ parameter and its uncertainty. $\theta_{\rm E}$ indicates the Gaussian weight adopted from the measured $\theta_{\rm E}$ if $\rho_{\ast}$ is certainly measured. $\rho_{\ast}$ indicates a weight function built based on the $\Delta\chi^{2}$ distribution as a function of $\rho_{\ast}$ if the $\rho_{\ast}$ measurement is uncertain. Lastly, $\pivec_{\rm E}$ indicates a constraint using the 2D APRX distributions described in \citet{ryu19}. In Table \ref{table:lens}, we present various lens properties for each event because each event has degenerate solutions, which yield different lens properties. For ease of cataloging, we present ``adopted" values for each property by adopting the method described in \citet{jung23}, i.e., weighted average values.

\subsection{\twosixtynine}

For the Bayesian analysis of this event, we apply constraints obtained from $t_{\rm E}$ (i.e., the Gaussian weight) and $\rho_{\ast}$ weight functions on the Galactic prior, because the $\rho_{\ast}$ measurements are uncertain and the APRX is not measured. Note that we can evaluate the effect of the $\rho_{\ast}$ weight on the posterior before conducting the Bayesian analysis. If the lower limit on the relative lens--source proper motion, $\mu_{\rm rel, +3\sigma} \equiv \theta_{\ast}/(\rho_{\ast, +3\sigma}t_{\rm E}) \lesssim 1\, {\rm mas\, yr^{-1}}$, the effect is minor. As expected (see the $\mu_{\rm rel}$ column of Table \ref{table:cmd}), the effects of $\rho_{\ast}$ are minor for both solutions.

The Bayesian results indicate that the lens system of this event consists of an M dwarf host star ($M_{\rm host} \sim 0.4\, M_{\odot}$) and a super--Jupiter--mass planet with a mass of $M_{\rm planet} \sim 11.5\, M_{\rm J}$, which is close to the limit of planetary objects. The planet orbits the host with a projected separation $a_{\perp} \sim 1.31$ au or $\sim 3.82$ au, which is beyond its snow line. The planetary system is located in the Galactic bulge with the distance $\sim 6.6$ kpc from us. Hence, this event is caused by a typical microlensing planetary system, which is a giant planet orbiting an M dwarf host beyond snow line \citep{ida05,kennedy08}.

\subsection{\fiveosix}

For this event, the $\rho_{\ast}$ values are measured. Thus, we apply $t_{\rm E}$ and $\theta_{\rm E}$ constraints on the Bayesian analyses. The lens system of this event consists of a late--type M--dwarf host star ($M_{\rm host} \sim 0.1\, M_{\odot}$) and a super--Neptune--mass planet ($M_{\rm planet} \sim 7.2\, M_{\rm N}$) orbiting with a projected separation, $a_{\perp} \sim 0.56$ au or $\sim 1.36$ au. The planetary system is located at the distance with $D_{\rm L} \sim 7.4$ kpc from us. Similarly to \twosixtynine\, this event is also caused by a typical microlensing planet.

\subsection{\sixtwentyfive}
Despite the non-optimal coverage, we can measure $\rho_{\ast}$ for this event. Thus, we apply $t_{\rm E}$ and $\theta_{\rm E}$ constraints on the Bayesian analyses. For this event, The Bayesian results for the lens system spans a wide range of properties because of the degenerate solutions (i.e., due to different $q$ and $\rho_{\ast}$ for each solution, see Table \ref{table:model_0625}). The host star is an M dwarf with the mass range of $M_{\rm host} \sim 0.2$--$0.3\, M_{\odot}$. For the $s_{-}$ case (the best--fit solution), the planet could be a Neptune--mass planet with a mass of $M_{\rm planet} \sim 1.36\, M_{\rm N}$ orbiting the host with a projected separation of $a_{\perp} \sim 1.3$ au. For the remaining cases, the planet could be a super--Earth--mass planet with a mass range of $M_{\rm planet} \sim 2.0$--$9.0\, M_{\oplus}$ orbiting the host with a projected separation range of $a_{\perp} \sim 0.9$--$1.9$ au. The planetary system belongs to the Galactic bulge with a distance range of $D_{\rm L} \sim 6.1$--$6.7$ kpc.

\subsection{\thirteenseven}

For this event, we find the APRX effect on the light curve. However, the $\rho_{\ast}$ measurements are uncertain. Thus, we apply $t_{\rm E}$, $\rho_{\ast}$ weights, and $\pivec_{\rm E}$ constraints on the Bayesian analyses. The $\pivec_{\rm E}$ constraints have major effects on the posteriors, while the $\rho_{\ast}$ constraints have only minor effects as expected from $\mu_{\rm rel, +3\sigma} \lesssim 1\, {\rm mas\, yr^{-1}}$ (see Table \ref{table:cmd}).

The planetary system of this event consists of an M--dwarf host star ($M_{\rm host} \sim 0.2$--$0.3\, M_{\odot}$) and a super--Earth--mass planet. We find that the planet mass of the inner cases ($M_{\rm planet} \sim 9\, M_{\oplus}$) is smaller than those of the outer cases ($M_{\rm planet} \sim 11\, M_{\oplus}$). The planet orbits the host with a projected separation of $a_{\perp} \sim 1.4$--$1.5$ au beyond its snow line. The system is located at the distance of $D_{\rm L} \sim 2$ kpc from us, i.e., in the disk, which is expected by considering the strong microlens parallax effect.

\subsection{\seventeenfiftyone}

For this event, we conduct Bayesian analyses for $s_{\pm}$ and $s_{\pm}^{\prime}$ solutions by applying $t_{\rm E}$ and $\rho_{\ast}$ weights as constraints. We find that the lens system consists of an M--dwarf host ($M_{\rm host} \sim 0.18\, M_{\odot}$) and a Jupiter--class planet ($M_{\rm planet} \sim 0.7$--$1.2\, M_{\rm J}$), which is located at the distance of $\sim 7.05$ kpc from us. The planet orbits the host with the projected separation of $a_{\perp} \sim 1.2$--$1.4$ au.    

We note that, as mentioned in Section \ref{sec:lc_analysis_1751}, the $s_{-}^{\prime\prime}$ case is removed from our fiducial solutions. Thus, although we conduct the Bayesian analysis for this case, we do not include the lens properties of this cease in Table \ref{table:lens}. However, for completeness, we present the lens properties of this case. The Bayesian analysis applied $t_{\rm E}$ constraint indicates that the lens system consists of a M--dwarf host star ($M_{\rm host} = 0.18_{-0.11}^{+0.28}\, M_{\odot}$) and a Super Neptune--mass planet ($M_{\rm planet} = 2.5_{-1.57}^{+4.25}\, M_{\rm N}$) with a projected separation of $1.28_{-0.44}^{+0.54}$ au. The system is located at the distance of $7.04_{-1.38}^{+0.54}$ kpc and the relative lens--source proper motion is $\mu_{\rm rel} = 7.56_{-2.68}^{+3.45}\, {\rm mas\, yr^{-1}}$. The results are similar to the lens properties of our fiducial solutions because the $t_{\rm E}$ value of the $s_{-}^{\prime\prime}$ is similar to them, except for the planet mass, which is caused by the smallest $q$ value of the $s_{-}^{\prime\prime}$ model. On the other hand, the Bayesian analysis applying $t_{\rm E}$ and $\rho_{\ast}$ weights shows an extreme lens system caused by the unusually small $\theta_{\rm E}$. That is, the lens system could consist of very--low mass host ($M_{\rm host} = 0.02_{-0.01}^{+0.05}\, M_{\odot}$ and sub Neptune--mass planet ($M_{\rm planet} = 0.31_{-0.19}^{0.71}\, M_{\rm N}$) with the projected separation of $0.31_{-0.07}^{+0.06}$. The system could be located at the distance of $8.17_{-1.04}^{+1.05}$ kpc. The relative lens--source proper motion is $\mu_{\rm rel} = 1.59_{-0.34}^{+0.21}\, {\rm mas\, yr^{-1}}$, which is inconsistent with the typical value of bulge--lens/bulge--source microlensing event ($5$--$10\, {\rm mas\, yr^{-1}}$). If the lens star were resolved by future adaptive optics imaging, this could definitively rule out the $s_{-}^{\prime\prime}$ solution.

\section{Summary} \label{sec:summary}

We found $5$ new planetary systems and one planet candidate through a systematic anomaly search for $2016$ prime--fields of the KMTNet data archive. These ``buried" planets have various properties. For \twosixtynine, the planetary system consists of an M dwarf host ($M_{\rm host} \sim 0.4\, M_{\odot}$) and a super--Jupiter--mass planet ($M_{\rm planet} \sim 11.5\, M_{\rm J}$), which orbits the host with a projected separation of $1.3$ or $3.8$ au. The system is located at a distance of $\sim 6.6$ kpc from us. For \fiveosix, the lens system indicates that a super--Neptune--mass planet ($M_{\rm planet} \sim 7.2\, M_{\rm N}$) orbits a late M--dwarf host star ($M_{\rm host} \sim 0.1\, M_{\odot}$) with a projected separation of $0.6$ or $1.4$ au. The planetary system is located at a distance of $\sim 7.4$ kpc from us. For \sixtwentyfive, because of the degenerate solutions, the planetary system consists of an M dwarf host star with mass in the range $0.1$--$0.3\, M_{\odot}$ and a planet with mass in the range $2.0\, M_{\oplus}$--$1.4\, M_{\rm N}$. The system is located at a distance in the range $6.1$--$6.7$ kpc. For \thirteenseven, the planetary system consists of an M dwarf host star ($M_{\rm host} \sim 0.3\, M_{\odot}$) and a Super--Earth-mass planet ($M_{\rm planet} = 9$--$11\, M_{\oplus}$) with a projected separation of $\sim 1.5$ au. The system is located at a distance of $2$ kpc. For \seventeenfiftyone, we adopt the lens properties of $s_{\pm}$ and $s_{\pm}^{\dagger}$ cases, which indicate that a Jupiter--class planet ($M_{\rm planet} = 0.7$--$1.2\, M_{\rm J}$) orbits an M--dwarf host ($M_{\rm host} \sim 0.18\, M_{\odot}$). The system is located at a distance of $\sim 7.05$ kpc. 
\begin{comment}
{\bf For \eighteenfiftyfive, the event is a planet candidate because the 2L1S/1L2S and planet/binary degeneracies are not resolved. For planetary cases, the lens system would consist of a planet with the mass range of $M_{\rm planet} = 2\, M_{\oplus}$--$7.6\, M_{\rm J}$ and an M-dwarf host star ($M_{\rm host} \sim 0.3\, M_{\odot}$). For binary cases, the lens system would consist of a secondary ($M_{\rm secondary} \sim 0.02\, M_{\odot}$ or $\sim 0.10\, M_{\odot}$) and an M-dwarf primary ($M_{\rm primary} \sim 0.3\, M_{\odot}$). The lens system would be located in the bulge ($D_{\rm L} \sim 6.8$ kpc). Note that there exists a possibility of no companion (i.e., 1L2S case).}
\end{comment}

Our goal in the series including this work is to build a complete planet sample discovered by the microlensing method for the $2016$--$2021$ KMTNet data archive. In Table \ref{table:2016_prime_planets}, we present all microlensing planets discovered on the KMTNet prime fields in $2016$, which are published planets that are recovered by the AnomalyFinder and five newly discovered in this work. The horizontal line separates planets expected to be part of the final statistical sample and those whose mass ratios are likely too uncertain or too large to be included.

As discussed in \citet{clanton14a,clanton14b} and \citet{shin19}, each planet detection method has different detection sensitivity, which provides complementary planet samples for studying planet demographics and planet frequency in our Galaxy. Our works are important for a complete microlensing planet sample. Indeed, although the sample size of microlensing planets is relatively small compared to other methods such as radial velocity and transits, the microlensing planet sample is less biased to host mass because, in principle, the microlensing method can detect any foreground objects regardless of their host brightness. Thus, a complete microlensing planet sample can help us to obtain a better understanding of planet demographics in our Galaxy.

\hbox{}
% =================================
% Acknowledgments
% =================================
%\begin{acknowledgments}
% KMTNet
This research has made use of the KMTNet system operated by the Korea Astronomy and Space Science Institute (KASI), and the data were obtained at three host sites of CTIO in Chile, SAAO in South Africa, and SSO in Australia.  
% Jennifer C. Yee
I.-G.S., S.-J.C., and J.C.Y. acknowledge support from N.S.F Grant No. AST--2108414.
% Han, Cheongho
Work by C.H. was supported by grants of the National Research Foundation of Korea (2017R1A4A1015178 and 2019R1A2C2085965).
% Yossi Shvartzvald
Y.S. acknowledges support from BSF Grant No. 2020740.
% Weicheng & Hongjing
W.Z. and H.Y. acknowledge support by the National Science Foundation of China (grant No. 12133005).
% MOA
The MOA project is supported by JSPS KAKENHI Grant Number JP24253004, JP26247023, JP23340064, JP15H00781, JP16H06287, JP17H02871 and JP22H00153.
% Hydra
The computations in this paper were conducted on the Smithsonian High Performance Cluster (SI/HPC), Smithsonian Institution (\url{https://doi.org/10.25572/SIHPC}).
% Exoplanet Archive
%This research has made use of the NASA Exoplanet Archive, which is operated by the California Institute of Technology, under contract with the National Aeronautics and Space Administration under the Exoplanet Exploration Program.
%\end{acknowledgments}
% =================================

\appendix 

\section{Non--planetary events} \label{sec:appendixA}

We report on the analysis of binary--lens events, that are found by AnomalyFinder as candidate planetary events. From the initial analyses, we find that the light curves of these events could be described both binary-lens and planet-lens interpretations. However, based on analyses using TLC reductions, we find that these events disfavor planetary solutions ($q < 0.03$) by $\Delta\chi^{2} > 15$. Thus, we cannot claim certain detection of the planets. In Table \ref{table:obs_binary}, we present observational information of these events.

\subsection{\doubleOtwenty} % KMT-2016-BLG-0020
The light curve of \doubleOtwenty\ (which we identified as KMT-2016-BLG-0020) shows deviations from the 1L1S interpretation ($\Delta\chi^{2}_{\rm 1L1S - 2L1S} = 159.68$). From the 2L1S modeling, we find several 2L1S solutions that can explain the deviations. Among the solutions, four are binary--lens cases and three are planet--lens cases. The best-fit solution is the binary--lens case with $(s,q) = (0.492\pm0.013, 0.108\pm0.005)$. However, the lowest $\chi^{2}$ planetary solution $((s, q(\times10^{-4}) = (0.702\pm0.081, 59.198\pm49.265))$ shows $\Delta\chi^{2} = 26.60$ compared to the best-fit solution. The $\Delta\chi^{2}$ cannot satisfy our criterion to claim the planet detection. Also, the planetary solutions cannot describe the subtle bump feature at HJD$^{\prime} \sim 7528$. Thus, we conclude \doubleOtwenty\ should be removed from the planet sample. We note that, although this event is not likely to be caused by a planetary lens system, the best-fit solution indicates that the companion could be a low-mass object such as a brown dwarf.

\subsection{\oneOsix} % KMT-2016-BLG-0106
For this event (which we identified as KMT-2016-BLG-0106), we find seven local solutions based on analysis using the TLC reductions. However, not all local minima satisfy our $q$ criterion (i.e., $q < 0.03$) for claiming a planet detection. The best-fit solution indicates that the event was caused by a binary lens system, i.e., $(s,\, q) = (2.671 \pm 0.089, 1.113 \pm 0.099)$. Among the local minima, the model showing the lowest $q$ value ($q = 0.052 \pm 0.003$) is disfavored by $\Delta\chi^{2} = 122.70$ compared to the best-fit solution. Thus, we conclude this event should be removed from the planet sample.

\subsection{\onefiftyseven} % KMT-2016-BLG-0157 
For this event (which we identified as KMT-2016-BLG-0157), we found two solutions that showed planet-like mass ratio ($q \sim 0.03$) from the initial analysis. We, therefore, refine the solutions based on the TLC reductions. The analysis using the TLC reductions clearly shows localized solutions (i.e., $s_{\pm}$ cases) with $(s,\,q) = (0.580 \pm 0.009, 0.048 \pm 0.003)$ and $(2.158 \pm 0.027, 0.057 \pm 0.003)$ for the $s_{-}$ and $s_{+}$ cases, respectively. However, the mass ratios do not satisfy our criterion ($q < 0.03$) to claim the planet detection although the companion is likely to be a low--mass object such as a brown dwarf. Hence, we remove this event from the planet sample.

\subsection{\threeseventyfour} % KMT-2016-BLG-0374
We find plausible solutions within the planetary regime ($q < 0.03$) from the initial analysis. However, based on the analysis using TLC reductions, we find that the best-fit solutions are binary--lens cases with $(s, q) = (6.89, 0.55 )$ and $(0.20, 0.22)$ for the $s_{+}$ and $s_{-}$ cases, respectively. We also find that the planet--like models are disfavored by $\Delta\chi^{2} = 19.74$ and $18.60$ for the $s_{+}$ and $s_{-}$ cases, respectively. The planet--like solutions cannot satisfy the criterion for the planet detection. Thus, we conclude to remove this event from our planet samples.

\subsection{\fourfortysix} % KMT-2016-BLG-0446
The AnomalyFinder detects subtle a deviation on the light curve at HJD$^{\prime} \sim 7631.0$--$7636.0$ based on the pipeline data. The anomaly can be explained by planetary models. However, the TLC reductions reveal that the anomaly is a false--positive. Then, we find that the light curve can be explained by the 1L1S interpretation rather than any 2L1S interpretations. Thus, we remove \fourfortysix\ from our sample.

\subsection{\seventeensixteen} % KMT-2016-BLG-1716
We find that the best-fit solution of \seventeensixteen\ (which we identified as KMT-2016-BLG-1716) is caused by a binary lens with the mass ratio, $q = 1.247 \pm 0.204$ (i.e., $q \sim 0.80$) for the $s_{-}$ case (the competing $s_{+}$ solution also exists). The best--fit light curves are caused by approaching a diamond-shaped caustic. Thus, a four-fold degeneracy exists (i.e., four solutions with different source trajectories for different $\alpha$ values).

We also find alternative planetary solutions with the mass ratio, $q = (26.095 \pm 8.013)\times10^{-4}$ for the $s_{-}$ case. However, $\Delta\chi^{2}_{\rm (planet - binary)} = 29.23$. The $\Delta\chi^{2}$ cannot satisfy our criterion ($\Delta\chi^{2} < 10$) to claim a planet detection. Indeed, these planetary models clearly show worse fits in their residuals. Hence, we decide to remove \seventeensixteen\ from our planet candidate sample for full analysis.

\subsection{\eighteensixtythree} % KMT-2016-BLG-1863
The best-fit solutions of \eighteensixtythree\ (which we identified as KMT-2016-BLG-1863) is a binary--lens model with $(s, q) = (0.277\pm0.005, 0.306\pm0.031)$. There exists a $s_{+}$ solution, $(s, q) = (5.626\pm0.215 , 0.782\pm0.133)$, with $\Delta\chi^{2} = 6.96$ caused by the close/wide degeneracy. We find that the solutions having the lowest $\chi^{2}$ in the planetary regime ($q < 0.03$) are disfavored by $\Delta\chi^{2} = 78.89$ and $73.52$ for the $s_{-}:[s,\, q]= [0.576\pm0.007, (161.679\pm8.552)\times10^{-4}]$ and $s_{+}:[s,\, q]= [1.657\pm0.025, (176.374\pm10.013)\times10^{-4}]$ cases, respectively. Thus, we conclude that \eighteensixtythree\ is caused by a binary lens system rather than a planetary lens system.

\section{\fourtwentyfive} \label{sec:appendixB}

We also present the analysis of \fourtwentyfive, which was identified by eye as a planet candidate but not selected as anomalous in the AnomalyFinder process. We conduct a detailed analysis based on TLC reductions for this event. We find that the best--fit solution is a binary lens case with $(s,q) = (4.834\pm0.201, 3.480\pm1.110)$. This is equivalent to $\frac{1}{q} = 0.287\pm0.085$, which clearly implies a binary lens origin. We also search for
a planetary model. The best  planetary model that satisfies our mass-ratio criterion has $(s,q) = (0.724\pm0.037, 0.013\pm0.004)$, but is disfavored by $\Delta\chi^{2} = 46$ compared to the best--fit model. Thus, we conclude that this event was caused by a binary lens system.  

\begin{figure}[htb!]
\epsscale{1.00}
\plotone{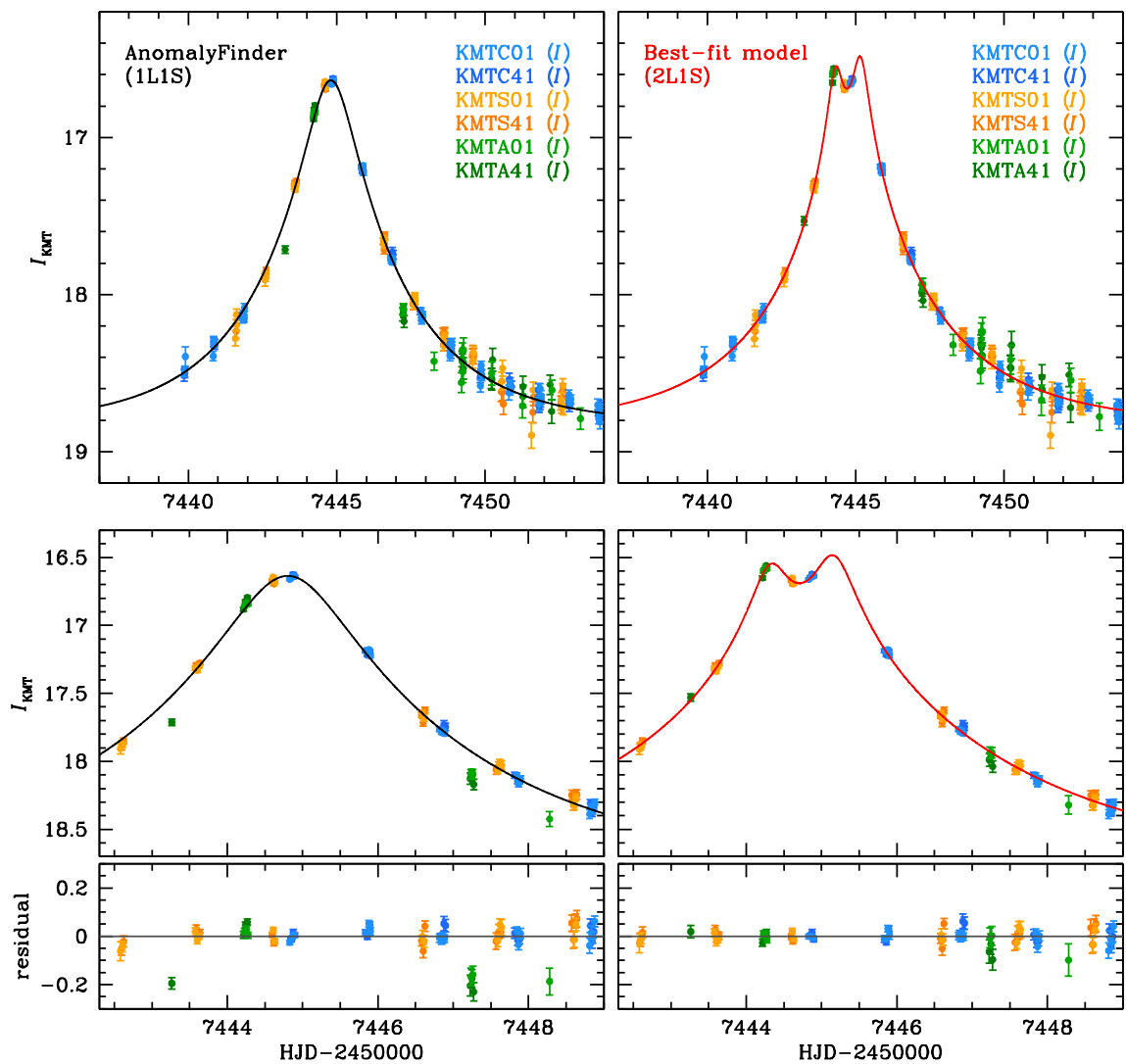}
\caption{Comparison of light curves between AnomalyFinder (1L1S) and the best--fit (2L1S) for 
\fourtwentyfive. Note that the data sets shown in this figure are produced by the KMTNet pipeline, 
which is actually used for the AnomalyFinder process. 
\label{fig:lc_0425_AF}}
\end{figure}

Although \fourtwentyfive\, turned out to be a binary lens event, it is still an important test case for verifying the AnomalyFinder process and assessing possible failure modes. In fact, the AnomalyFinder algorithm did identify a series of possible anomalies in this event, but the human operator rejected them as ``fake." In \fourtwentyfive, the anomaly occurs over the peak of the event, but because the event occurs early in the microlensing season, it is only sparsely covered, and the primary deviation from a point lens occurs in only the KMTA datasets. In addition, the event has a short timescale. Hence, due to the $\chi^2$ likelihood estimation, a point lens fit is biased toward the points at the peak (which have the smallest errorbars) and the baseline points (which dominate the numbers), and so it normalized the flux levels of the KMTA data so that the peak points (due to the anomaly) lay on the point lens light curve. As a result, the ``anomalies" identified by the AnomalyFinder were in the rising and falling parts of the light curve and caused by the bad flux normalization rather than the actual anomaly (see Figure \ref{fig:lc_0425_AF}). 

\fourtwentyfive\, is qualitatively similar to \twotwoninefour, which was also missed by the AnomalyFinder process \citep{shin23}. They are both short timescale events ($t_{\rm E} = 10.8\pm1.3$ and $7.1\pm0.3$ days, respectively) and had anomalies that occurred at the peak of the events. On the other hand, the reasons the anomalies were missed are distinctly different: in \fourtwentyfive, the wrong anomaly was identified, but in \twotwoninefour, the anomaly did not meet the detection threshold. The latter case is acceptable from the perspective of constructing a statistical sample of events. However, the failure for \fourtwentyfive\, is more concerning, but could be compensated for by adding an additional criterion to the AnomalyFinder algorithm to check for outliers in flux normalization.

% References ----------------------------------------------------------------     

% ---------------------------------------------------------------------     

% Tables & Figures 

% Table X (Observation info: planetary events) ------------------------------
\begin{deluxetable}{ccc|rrr|rc}
\tablecaption{Observations of $2016$ planetary events \label{table:obs_planet}}
\tablewidth{0pt}
\tablehead{
% ---------------------------------------------------------------------------
\multicolumn{3}{c|}{Event} &
\multicolumn{3}{c|}{Location} &
\multicolumn{2}{c}{obs. info.} \\
% --------------------------------------------
\multicolumn{1}{c}{KMTNet} &
\multicolumn{1}{c}{OGLE} &
\multicolumn{1}{c|}{MOA} &
% --------------------------------------------
\multicolumn{1}{c}{R.A. (J2000)} &
\multicolumn{1}{c}{Dec (J2000)} &
\multicolumn{1}{c|}{$(\ell, b)$} &
% --------------------------------------------
\multicolumn{1}{c}{$A_{I}$} &
\multicolumn{1}{c}{$\Gamma$ (${\rm hr}^{-1}$)} 
% ------------------------------------
}
\startdata
% ---------------------------------------------------------------------------------------------------------------------------------------------
    0269  & \bf{1635} & \nodata  & $17^{h} 54^{m} 01^{s}.22$ & $-30^{\circ} 46{'} 38{''}.10$ & $(-0^{\circ}.65, -2^{\circ}.51)$ & 1.82 & 2.0 \\
    0506  &     1749  & \bf{532} & $17^{h} 57^{m} 44^{s}.18$ & $-29^{\circ} 06{'} 25{''}.60$ & $(+1^{\circ}.20, -2^{\circ}.37)$ & 1.56 & 4.0 \\
\bf{0625} & \nodata   & \nodata  & $18^{h} 05^{m} 39^{s}.66$ & $-27^{\circ} 13{'} 36{''}.70$ & $(+3^{\circ}.70, -2^{\circ}.96)$ & 0.95 & 4.0 \\
    1307  & \bf{1850} & \nodata  & $17^{h} 52^{m} 00^{s}.18$ & $-32^{\circ} 12{'} 38{''}.20$ & $(-2^{\circ}.10, -2^{\circ}.86)$ & 2.01 & 4.0 \\
\bf{1751} & \nodata   & avail.   & $17^{h} 53^{m} 28^{s}.62$ & $-32^{\circ} 09{'} 06{''}.52$ & $(-1^{\circ}.89, -3^{\circ}.10)$ & 2.11 & 4.0 \\
\bf{1855} & \nodata   & \nodata  & $17^{h} 50^{m} 13^{s}.25$ & $-29{^\circ} 12{'} 26{''}.39$ & $(+0^{\circ}.29, -1^{\circ}.00)$ & 5.97 & 4.0 \\
% ---------------------------------------------------------------------------------------------------------------------------------------------
\enddata
\tablecomments{
Boldface indicates the ``discovery" name of each event.
}
\end{deluxetable}
% ---------------------------------------------------------------------------

% Table X (Observation info: binary events described in Appendix) ------------------------------
\begin{deluxetable}{ccc|rrr|rc}
\tablecaption{Observations of $2016$ non--planetary events \label{table:obs_binary}}
\tablewidth{0pt}
\tablehead{
% ---------------------------------------------------------------------------
\multicolumn{3}{c|}{Event} &
\multicolumn{3}{c|}{Location} &
\multicolumn{2}{c}{obs. info.} \\
% --------------------------------------------
\multicolumn{1}{c}{KMTNet} &
\multicolumn{1}{c}{OGLE} &
\multicolumn{1}{c|}{MOA} &
% --------------------------------------------
\multicolumn{1}{c}{R.A. (J2000)} &
\multicolumn{1}{c}{Dec (J2000)} &
\multicolumn{1}{c|}{$(\ell, b)$} &
% --------------------------------------------
\multicolumn{1}{c}{$A_{I}$} &
\multicolumn{1}{c}{$\Gamma$ (${\rm hr}^{-1}$)} 
% ------------------------------------
}
\startdata
% -------------------------------------------------------------------------------------------------------------------------------------
    0020  & \bf{0987} & \nodata  & $17^{h} 56^{m} 34^{s}.37$ & $-27^{\circ} 59{'} 31{''}.99$ & $(+2^{\circ}.04, -1^{\circ}.59)$ & 1.78 & 4.0  \\
    0106  & \nodata   & \bf{123} & $17^{h} 54^{m} 17^{s}.90$ & $-28{^\circ} 55{'} 15{''}.74$ & $(+0^{\circ}.99, -1^{\circ}.62)$ & 1.72 & 1.0  \\
    0157  & \bf{0558} & \nodata  & $17^{h} 57^{m} 45^{s}.40$ & $-28{^\circ} 20{'} 07{''}.40$ & $(+1^{\circ}.88, -1^{\circ}.98)$ & 1.61 & 4.0  \\
\bf{0374} & \nodata   & avail.   & $17^{h} 54^{m} 48^{s}.45$ & $-30{^\circ} 59{'} 04{''}.42$ & $(-0^{\circ}.74, -2^{\circ}.76)$ & 1.43 & 4.0  \\
    0425  & \bf{0185} & \nodata  & $17^{h} 52^{m} 35^{s}.43$ & $-31^{\circ} 19{'} 51{''}.60$ & $(-1^{\circ}.28, -2^{\circ}.52)$ & 2.24 & 4.0  \\
\bf{0446} & \nodata   & \nodata  & $17^{h} 51^{m} 52^{s}.02$ & $-28^{\circ} 50{'} 00{''}.17$ & $(+0^{\circ}.79, -1^{\circ}.12)$ & 3.13 & 4.0  \\
    0641  & \bf{0779} & \nodata  & $18^{h} 05^{m} 59^{s}.40$ & $-28^{\circ} 06{'} 13{''}.21$ & $(+2^{\circ}.97, -3^{\circ}.45)$ & 0.84 & 1.0  \\
    1716  & \bf{1722} &     555  & $17^{h} 55^{m} 21^{s}.82$ & $-30^{\circ} 42{'} 36{''}.90$ & $(-0^{\circ}.44, -2^{\circ}.72)$ & 1.73 & 4.0  \\
    1863  & \bf{0974} &     351  & $18^{h} 01^{m} 23^{s}.99$ & $-27{^\circ} 33{'} 20{''}.30$ & $(+2^{\circ}.95, -2^{\circ}.29)$ & 1.41 & 4.0  \\
% -------------------------------------------------------------------------------------------------------------------------------------
\enddata
\tablecomments{
Boldface indicates the ``discovery" name of each event.
}
\end{deluxetable}
% ---------------------------------------------------------------------------

% Table X (KB-16-0269: model parameters) ------------------------------------
\begin{deluxetable}{lrr}
\tablecaption{Model Parameters of \twosixtynine \label{table:model_0269}}
\tablewidth{0pt}
\tablehead{
% ---------------------------------------------------------------------------
\multicolumn{1}{c}{Parameter} &
\multicolumn{1}{c}{$s_{-}$} & 
\multicolumn{1}{c}{$s_{+}$} 
% ------------------------------------
}
\startdata
% ----------------------------------------------------------------------------------------------
$\chi^{2} / {\rm N}_{\rm data}$           & $ 5434.309 / 5436     $ & $ 5442.023 / 5436     $ \\
$t_0$ [${\rm HJD'}$]                      & $ 7624.124 \pm  0.013 $ & $ 7624.118 \pm  0.014 $ \\
$u_0$                                     & $    0.028 \pm  0.004 $ & $    0.028 \pm  0.003 $ \\
$t_{\rm E}$ [days]                        & $   20.908 \pm  2.239 $ & $   21.997 \pm  2.252 $ \\
$s$                                       & $    0.587 \pm  0.014 $ & $    1.684 \pm  0.043 $ \\
$q$ ($\times 10^{-4}$)                    & $  254.587 \pm 33.644 $ & $  247.586 \pm 30.858 $ \\
$\langle\log_{10} q\rangle$               & $   -1.594 \pm  0.058 $ & $   -1.620 \pm 0.056  $ \\
$\alpha$ [rad]                            & $    5.056 \pm  0.028 $ & $    5.047 \pm  0.030 $ \\
$\rho_{\ast, {\rm limit}}$                & $ < 0.0037            $ & $ < 0.0041            $ \\
% ----------------------------------------------------------------------------------------------
\enddata
\tablecomments{
${\rm HJD' = HJD - 2450000.0}$. 
}
\end{deluxetable}
% ---------------------------------------------------------------------------

% Figure X (KB-16-0269: Light curves) --------------------------------------------------------------
% --------------------------------------------------------------------------------------------------

% Table X (KB-16-0506: model parameters) ------------------------------------
\begin{deluxetable}{lrrlrr}
\tablecaption{Model Parameters of \fiveosix \label{table:model_0506}}
\tablewidth{0pt}
\tablehead{
% ---------------------------------------------------------------------------
\multicolumn{1}{c}{Cases} &
\multicolumn{2}{c}{\bf{Planet}} &
\multicolumn{3}{c}{Binary} \\
% ------------------------------------
\multicolumn{1}{c}{Parameter} &
\multicolumn{1}{c}{$\mathbf{s_{-}}$} & 
\multicolumn{1}{c}{$\mathbf{s_{+}}$} &
\multicolumn{1}{c}{Parameter} &
\multicolumn{1}{c}{$s_{-}$} & 
\multicolumn{1}{c}{$s_{+}$} 
% ------------------------------------
}
\startdata
% ------------------------------------------------------------------------------------------------------------------------------------------------------------------------
$\chi^{2} / {\rm N}_{\rm data}$  & $\mathbf{ 12638.981 / 12647  }$ & $\mathbf{ 12637.673 / 12647  }$ & $\Delta\chi^{2}_{\rm binary-planet}$ & $ 22.322             $ & $ 20.619             $ \\
$t_0$ [${\rm HJD'}$]             & $\mathbf{ 7636.877 \pm 0.003 }$ & $\mathbf{ 7636.885 \pm 0.003 }$ & $t_0$ [${\rm HJD'}$]                 & $ 7636.934 \pm 0.003 $ & $ 7636.938 \pm 0.003 $ \\
$u_0$                            & $\mathbf{    0.009 \pm 0.001 }$ & $\mathbf{    0.009 \pm 0.001 }$ & $u_0$                                & $    0.009 \pm 0.001 $ & $    0.007 \pm 0.001 $ \\
$t_{\rm E}$ [days]               & $\mathbf{   20.786 \pm 1.356 }$ & $\mathbf{   20.547 \pm 1.491 }$ & $t_{\rm E}$ [days]                   & $   21.606 \pm 1.392 $ & $   25.921 \pm 1.427 $ \\
$s$                              & $\mathbf{    0.653 \pm 0.018 }$ & $\mathbf{    1.584 \pm 0.048 }$ & $s$                                  & $    0.278 \pm 0.013 $ & $    4.742 \pm 0.362 $ \\
$q$ ($\times 10^{-4}$)           & $\mathbf{   40.382 \pm 5.534 }$ & $\mathbf{   40.441 \pm 6.491 }$ & $q$                                  & $    0.156 \pm 0.029 $ & $    0.232 \pm 0.062 $ \\ 
$\langle\log_{10} q\rangle$      & $\mathbf{   -2.387 \pm 0.058 }$ & $\mathbf{   -2.403 \pm 0.070 }$ & $\langle\log_{10} q\rangle$          & $   -0.765 \pm 0.073 $ & $   -0.595 \pm 0.100 $ \\
$\alpha$ [rad]                   & $\mathbf{    2.938 \pm 0.008 }$ & $\mathbf{    2.949 \pm 0.008 }$ & $\alpha$ [rad]                       & $    3.026 \pm 0.010 $ & $    3.036 \pm 0.008 $ \\
$\rho_{\ast}$ ($\times 10^{-4}$) & $\mathbf{   46.956 \pm 3.995 }$ & $\mathbf{   47.591 \pm 5.055 }$ & $\rho_{\ast}$ ($\times 10^{-4}$)     & $   73.386 \pm 5.591 $ & $   59.520 \pm 3.267 $ \\
% ------------------------------------------------------------------------------------------------------------------------------------------------------------------------
\enddata
\tablecomments{
${\rm HJD' = HJD - 2450000.0}$. 
}
\end{deluxetable}
% ---------------------------------------------------------------------------

% Figure X (KB-16-0506: Light curves) --------------------------------------------------------------
% --------------------------------------------------------------------------------------------------

% Table X (KB-16-0625: model parameters) ------------------------------------
\begin{deluxetable}{lrrrr|lr}
\tablecaption{Model Parameters of \sixtwentyfive \label{table:model_0625}}
\tablewidth{0pt}
\tablehead{
% ---------------------------------------------------------------------------
\multicolumn{1}{c}{Parameter} &
\multicolumn{1}{c}{$s_{-}$} & 
\multicolumn{1}{c}{$s^{\prime}_{-}$} &
\multicolumn{1}{c}{$s_{+}$} & 
\multicolumn{1}{c}{$s^{\prime}_{+}$} &
\multicolumn{1}{|c}{Parameter} &
\multicolumn{1}{c}{1L2S}
% ------------------------------------
}
\startdata
% -----------------------------------------------------------------------------------------------------------------------------------------------------------------------------------------------------------------
$\chi^{2} / {\rm N}_{\rm data}$           & $ 8014.084 / 8021    $ & $ 8015.097 / 8021    $ & $ 8015.065 / 8021    $ & $ 8017.379 / 8021    $ & $\chi^{2} / {\rm N}_{\rm data}$        & $ 8021.432 / 8021     $ \\
$t_0$ [${\rm HJD'}$]                      & $ 7655.951 \pm 0.008 $ & $ 7655.948 \pm 0.008 $ & $ 7655.951 \pm 0.008 $ & $ 7655.950 \pm 0.008 $ & $t_{0,S_{1}}$ [${\rm HJD'}$]           & $ 7655.953 \pm  0.008 $ \\
$u_0$                                     & $    0.073 \pm 0.004 $ & $    0.072 \pm 0.004 $ & $    0.075 \pm 0.005 $ & $    0.076 \pm 0.004 $ & $u_{0,S_{1}}$                          & $    0.078 \pm  0.007 $ \\
$t_{\rm E}$ [days]                        & $   11.494 \pm 0.466 $ & $   11.576 \pm 0.466 $ & $   11.335 \pm 0.508 $ & $   11.217 \pm 0.405 $ & $t_{\rm E}$ [days]                     & $   10.946 \pm  0.491 $ \\
$s$                                       & $    0.741 \pm 0.009 $ & $    0.741 \pm 0.009 $ & $    1.367 \pm 0.018 $ & $    1.358 \pm 0.015 $ & $t_{0,S_{2}}$ [${\rm HJD'}$]           & $ 7662.943 \pm  0.010 $ \\
$q$ ($\times 10^{-4}$)                    & $    2.357 \pm 1.123 $ & $    1.793 \pm 1.048 $ & $    0.727 \pm 0.254 $ & $    0.317 \pm 0.173 $ & $u_{0,S_{2}}$ ($\times 10^{-4}$)       & $    3.751 \pm 19.751 $ \\
$\langle\log_{10} q\rangle$               & $   -3.451 \pm 0.130 $ & $   -3.498 \pm 0.136 $ & $   -4.154 \pm 0.159 $ & $   -4.321 \pm 0.159 $ & \nodata                                & \nodata                 \\
$\alpha$ [rad]                            & $    3.217 \pm 0.008 $ & $    3.220 \pm 0.008 $ & $    0.122 \pm 0.003 $ & $    0.121 \pm 0.002 $ & $q_{\rm flux}$                         & $    0.005 \pm  0.001 $ \\
$\rho_{\ast}$ ($\times 10^{-4}$)          & $   12.256 \pm 6.613 $ & $   20.969 \pm 7.334 $ & $   17.498 \pm 7.796 $ & $   17.656 \pm 7.383 $ & $\rho_{\ast,S_{2}}$ ($\times 10^{-4}$) & $   51.309 \pm 13.447 $ \\
% -----------------------------------------------------------------------------------------------------------------------------------------------------------------------------------------------------------------
\enddata
\tablecomments{
${\rm HJD' = HJD - 2450000.0}$. We note that the $\rho_{\ast, S_{1}}$ is not measured for the 1L2S case. The upper limit ($3\sigma$ range) only can be measured: $\rho_{\ast, S_{1}} < 0.117$.  
}
\end{deluxetable}
% ---------------------------------------------------------------------------

% Figure X (KB-16-0625: Light curves) --------------------------------------------------------------
\begin{figure}[htb!]
\epsscale{1.00}
\plotone{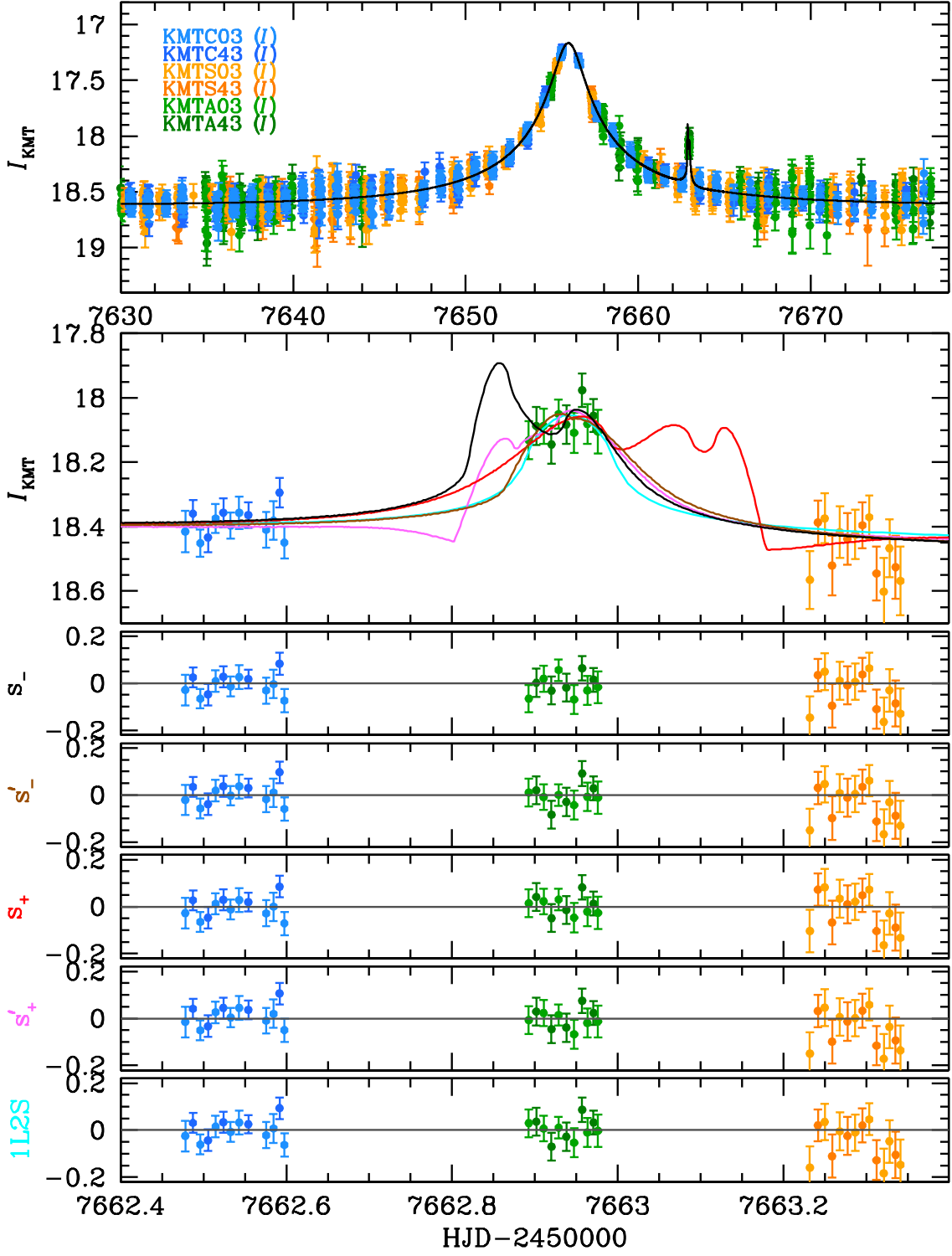}
\caption{Light curve of \sixtwentyfive\ with degenerate models.  
\label{fig:lc_0625}}
\end{figure}
% --------------------------------------------------------------------------------------------------

% Figure X (KB-16-0625: caustic geometry) ----------------------------------------------------------
% --------------------------------------------------------------------------------------------------

% Table X (KB-16-1307: model parameters) ------------------------------------
\begin{deluxetable}{lrr|rrrr}
\tablecaption{Model Parameters of \thirteenseven \label{table:model_1307}}
\tablewidth{0pt}
\tablehead{
% ---------------------------------------------------------------------------
\multicolumn{1}{c}{} &
\multicolumn{2}{c}{STD} &
\multicolumn{4}{|c}{APRX} \\
% ---------------------------------------------------------------------------
\multicolumn{1}{c}{Parameter} &
\multicolumn{1}{c}{inner} & 
\multicolumn{1}{c}{outer} &
\multicolumn{1}{|c}{inner ($u_{0} > 0$)} & 
\multicolumn{1}{c}{outer ($u_{0} > 0$)} & 
\multicolumn{1}{c}{inner ($u_{0} < 0$)} & 
\multicolumn{1}{c}{outer ($u_{0} < 0$)} 
% ------------------------------------
}
\startdata
% ---------------------------------------------------------------------------------------------------------------------------------------------------------------------------------------------------
$\chi^{2} / {\rm N}_{\rm data}$           & $ 8108.940 / 7995     $ & $ 8108.800 / 7995     $ & $ 8004.449 / 7995    $ & $ 8006.818 / 7995    $ & $ 8007.775 / 7995    $ & $ 8009.766 / 7995    $ \\
$t_0$ [${\rm HJD'}$]                      & $ 7654.235 \pm  0.192 $ & $ 7654.229 \pm  0.192 $ & $ 7655.221 \pm 0.175 $ & $ 7655.137 \pm 0.177 $ & $ 7655.258 \pm 0.177 $ & $ 7655.195 \pm 0.178 $ \\
$u_0$                                     & $    0.106 \pm  0.009 $ & $    0.105 \pm  0.010 $ & $    0.401 \pm 0.023 $ & $    0.397 \pm 0.020 $ & $   -0.397 \pm 0.020 $ & $   -0.398 \pm 0.031 $ \\
$t_{\rm E}$ [days]                        & $  209.677 \pm 16.086 $ & $  211.183 \pm 18.672 $ & $   62.803 \pm 3.843 $ & $   63.190 \pm 3.114 $ & $   59.885 \pm 2.917 $ & $   60.581 \pm 5.596 $ \\
$s$                                       & $    0.929 \pm  0.005 $ & $    0.961 \pm  0.006 $ & $    0.801 \pm 0.010 $ & $    0.826 \pm 0.009 $ & $    0.802 \pm 0.008 $ & $    0.825 \pm 0.014 $ \\
$q$ ($\times 10^{-4}$)                    & $    0.416 \pm  0.066 $ & $    0.443 \pm  0.068 $ & $    1.009 \pm 0.141 $ & $    1.258 \pm 0.161 $ & $    1.072 \pm 0.140 $ & $    1.334 \pm 0.180 $ \\ 
$\langle\log_{10} q\rangle$               & $   -4.378 \pm 0.068  $ & $    -4.366 \pm 0.068 $ & $   -3.983 \pm 0.057 $ & $   -3.923 \pm 0.058 $ & $   -3.945 \pm 0.052 $ & $   -3.890 \pm 0.060 $ \\
$\alpha$ [rad]                            & $    4.332 \pm  0.011 $ & $     4.329 \pm 0.011 $ & $    4.401 \pm 0.012 $ & $    4.394 \pm 0.012 $ & $   -4.385 \pm 0.012 $ & $   -4.384 \pm 0.012 $ \\
$\rho_{\ast, {\rm limit}}$                & $  < 0.003            $ & $  < 0.003            $ & $  < 0.007           $ & $  < 0.009           $ & $  < 0.007           $ & $    < 0.009         $ \\
$\pi_{{\rm E}, N}$                        & \nodata                 & \nodata                 & $    0.075 \pm 0.059 $ & $    0.070 \pm 0.060 $ & $    0.095 \pm 0.080 $ & $    0.074 \pm 0.084 $ \\
$\pi_{{\rm E}, E}$                        & \nodata                 & \nodata                 & $    0.455 \pm 0.037 $ & $    0.441 \pm 0.034 $ & $    0.465 \pm 0.035 $ & $    0.449 \pm 0.048 $ \\
% ---------------------------------------------------------------------------------------------------------------------------------------------------------------------------------------------------
\enddata
\tablecomments{
${\rm HJD' = HJD - 2450000.0}$. 
}
\end{deluxetable}
% ---------------------------------------------------------------------------

% Figure X (KB-16-1307: Light curve Whole) ---------------------------------------------------------
% --------------------------------------------------------------------------------------------------

% Figure X (KB-16-1307: Light curve Zoom) ---------------------------------------------------------
% --------------------------------------------------------------------------------------------------

% Figure X (KB-16-1307: Caustic Geometries) ---------------------------------------------------------
% --------------------------------------------------------------------------------------------------

% Figure X (KB-16-1307: APRX Contours) ---------------------------------------------------------
% --------------------------------------------------------------------------------------------------

% Figure X (KB-16-1307: Period -- chi2 plot) ---------------------------------------------------------
% --------------------------------------------------------------------------------------------------

% Table X (KB-16-1751: model parameters) ------------------------------------
\begin{deluxetable}{lrrrrr}
\tablecaption{Model Parameters of \seventeenfiftyone \label{table:model_1751}}
\tablewidth{0pt}
\tablehead{
% ---------------------------------------------------------------------------
\multicolumn{1}{c}{Parameter} &
\multicolumn{1}{c}{$\mathbf{s_{+}}$} & 
\multicolumn{1}{c}{$\mathbf{s^{\prime}_{+}}$} &
\multicolumn{1}{c}{$\mathbf{s_{-}}$} & 
\multicolumn{1}{c}{$\mathbf{s^{\prime}_{-}}$} &
\multicolumn{1}{c}{$s^{\prime\prime}_{-}$}
% ------------------------------------
}
\startdata
% -----------------------------------------------------------------------------------------------------------------------------------------------------------------------
$\chi^{2} / {\rm N}_{\rm data}$           & $\mathbf{ 8995.960 / 8994   } $ & $\mathbf{ 9004.487 / 8994    }$ & $\mathbf{ 9001.658 / 8994    }$ & $\mathbf{ 9004.738 / 8994    }$ & $ 9006.742 / 8994    $ \\
$t_0$ [${\rm HJD'}$]                      & $\mathbf{ 7501.132 \pm 0.016} $ & $\mathbf{ 7501.187 \pm 0.021 }$ & $\mathbf{ 7501.106 \pm 0.017 }$ & $\mathbf{ 7501.176 \pm 0.019 }$ & $ 7501.214 \pm 0.016 $ \\
$u_0$                                     & $\mathbf{    0.113 \pm 0.005} $ & $\mathbf{    0.111 \pm 0.005 }$ & $\mathbf{    0.107 \pm 0.005 }$ & $\mathbf{    0.110 \pm 0.005 }$ & $    0.103 \pm 0.006 $ \\
$t_{\rm E}$ [days]                        & $\mathbf{    9.625 \pm 0.293} $ & $\mathbf{    9.475 \pm 0.292 }$ & $\mathbf{    9.997 \pm 0.301 }$ & $\mathbf{    9.469 \pm 0.284 }$ & $    9.591 \pm 0.298 $ \\
$s$                                       & $\mathbf{    1.050 \pm 0.014} $ & $\mathbf{    1.027 \pm 0.017 }$ & $\mathbf{    0.848 \pm 0.011 }$ & $\mathbf{    0.873 \pm 0.015 }$ & $    0.950 \pm 0.007 $ \\
$q$ ($\times 10^{-4}$)                    & $\mathbf{   64.992 \pm 4.479} $ & $\mathbf{   39.995 \pm 8.312 }$ & $\mathbf{   61.147 \pm 3.556 }$ & $\mathbf{   38.161 \pm 7.169 }$ & $    7.207 \pm 2.902 $ \\ 
$\langle\log_{10} q\rangle$               & $\mathbf{   -2.200 \pm 0.031} $ & $\mathbf{   -2.447 \pm 0.113 }$ & $\mathbf{   -2.212 \pm 0.025 }$ & $\mathbf{   -2.485 \pm 0.104 }$ & $   -3.036 \pm 0.130 $ \\
$\alpha$ [rad]                            & $\mathbf{    4.339 \pm 0.022} $ & $\mathbf{    4.470 \pm 0.049 }$ & $\mathbf{    4.318 \pm 0.021 }$ & $\mathbf{    4.467 \pm 0.044 }$ & $    4.643 \pm 0.035 $ \\
$\rho_{\ast,{\rm limit}}$                 & $\mathbf{   < 0.019         } $ & $\mathbf{  < 0.020           }$ & $\mathbf{  < 0.021           }$ & $\mathbf{  < 0.020           }$ & \nodata                \\
$\rho_{\ast}$                             & \nodata                         & \nodata                         & \nodata                         & \nodata                         & $    0.019 \pm 0.002 $ \\
%$\rho_{\ast}$                            & $\mathbf{    0.014 \pm 0.003} $ & $    0.014 \pm 0.004 $ & $    0.013 \pm 0.004 $ & $    0.012 \pm 0.004 $ & $    0.019 \pm 0.002 $ \\
% -----------------------------------------------------------------------------------------------------------------------------------------------------------------------
\enddata
\tablecomments{
${\rm HJD' = HJD - 2450000.0}$. 
}
\end{deluxetable}
% ---------------------------------------------------------------------------

% Figure X (KB-16-1751: Light curves) --------------------------------------------------------------
% --------------------------------------------------------------------------------------------------

% Figure X (KB-16-1751: s-q space) -----------------------------------------------------------------
% --------------------------------------------------------------------------------------------------

% Figure X (KB-16-1751: caustic geometry) ----------------------------------------------------------
% --------------------------------------------------------------------------------------------------

% Figure X (KB-16-11855: Light curves best-fit) --------------------------------------------------------------
% --------------------------------------------------------------------------------------------------

% Figure X (KB-16-1855: Light curves all degenerate mdoels) --------------------------------------------------------------
% --------------------------------------------------------------------------------------------------

% Figure X (KB-16-1855: Light curves all degenerate mdoels) --------------------------------------------------------------
% --------------------------------------------------------------------------------------------------

% Table X (KB-16-1855: model parameters) ------------------------------------
\begin{deluxetable}{lrrrr}
\tablecaption{Model Parameters of \eighteenfiftyfive \label{table:model_1855_1}}
\tablewidth{0pt}
\tablehead{
% ---------------------------------------------------------------------------
\multicolumn{1}{c}{Parameter} &
\multicolumn{1}{c}{A: best--fit} & 
\multicolumn{1}{c}{A$^{\prime}$: $3(\pi/2)$} & 
\multicolumn{1}{c}{B: $2(\pi/2)$} & 
\multicolumn{1}{c}{B$^{\prime}$: $2(\pi/2)$}
% ------------------------------------
}
\startdata
% --------------------------------------------------------------------------------------------------------------------------------------
$\chi^{2} / {\rm N}_{\rm data}$                & $ \mathbf{7734.217 / 7735    } $ & $ 7749.828 / 7735     $ & $  7758.514 / 7735                  $ & $ 7757.884 / 7735                    $ \\
$\Delta\chi^{2}$ (full data)                   & \nodata                          & $   15.611            $ & $    24.297                         $ & $   23.667                           $ \\
$\Delta\chi^{2}$ (${\rm HJD^{\prime} < 7600}$) & \nodata                          & $   13.196            $ & $     8.055                         $ & $    8.821                           $ \\
$t_0$ [${\rm HJD'}$]                           & $ \mathbf{7454.015 \pm  0.080} $ & $ 7455.619 \pm  0.078 $ & $  7453.981 \pm 0.104               $ & $ 7455.694 \pm   0.086               $ \\
$u_0$                                          & $ \mathbf{   0.041 \pm  0.007} $ & $    0.040 \pm  0.006 $ & $     0.001 \pm 0.001               $ & $   -0.001 \pm   0.001               $ \\
$t_{\rm E}$ [days]                             & $ \mathbf{  94.156 \pm 17.078} $ & $  103.244 \pm 18.364 $ & $  2889.644 \pm 493.547             $ & $ 2924.318 \pm 120.251               $ \\
$t_{\rm E}^{\prime}$ [days]                    & $ \mathbf{  14.424 \pm  0.908} $ & $   15.331 \pm  0.904 $ & $    13.282 \pm 0.984               $ & $   13.321 \pm   0.945               $ \\
$s$                                            & $ \mathbf{   3.798 \pm  0.108} $ & $    3.780 \pm  0.088 $ & $     3.598 \pm 0.132               $ & $    3.708 \pm   0.121               $ \\
$q$                                            & $ \mathbf{  42.614 \pm 21.193} $ & $   45.352 \pm 24.063 $ & $   (47.334 \pm 11.658)\times10^{3} $ & $  (48.189 \pm   5.978)\times10^{3}  $ \\ 
$\frac{1}{q}$                                  & $ \mathbf{   0.023 \pm  0.012} $ & $    0.022 \pm  0.009 $ & $    (0.211 \pm 0.419)\times10^{-4} $ & $   (0.208 \pm   0.042)\times10^{-4} $ \\
$\langle\log_{10} \frac{1}{q}\rangle$          & $ \mathbf{  -1.605 \pm  0.196} $ & $   -1.666 \pm  0.173 $ & $    -4.249 \pm 0.248               $ & $   -4.574 \pm   0.068               $ \\ 
$\alpha$ [rad]                                 & $ \mathbf{   0.657 \pm  0.012} $ & $    5.603 \pm  0.016 $ & $     3.812 \pm 0.016               $ & $    3.803 \pm   0.015               $ \\
$\rho_{\ast,{\rm limit}}$                      & $ \mathbf{ < 0.023           } $ & $  < 0.024            $ & $   < 0.002                         $ & $    < 0.0006                        $ \\
% --------------------------------------------------------------------------------------------------------------------------------------
\enddata
\tablecomments{
${\rm HJD' = HJD - 2450000.0}$. 
}
\end{deluxetable}
% ---------------------------------------------------------------------------

% Table X (KB-16-1855: model parameters) ------------------------------------
\begin{deluxetable}{lrrr|lr}
\tablecaption{Model Parameters of \eighteenfiftyfive\ (continue) \label{table:model_1855_2}}
\tablewidth{0pt}
\tablehead{
% ---------------------------------------------------------------------------
\multicolumn{1}{c}{Parameter} &
\multicolumn{1}{c}{C: $s^{\prime}_{-}$} & 
\multicolumn{1}{c}{C$^{\prime}$: $s^{\prime}_{+}$} & 
\multicolumn{1}{c}{D: $1(\pi/2)$} & 
\multicolumn{1}{|c}{Parameter} &
\multicolumn{1}{c}{E: 1L2S} 
% ------------------------------------
}
\startdata
% --------------------------------------------------------------------------------------------------------------------------------------
$\chi^{2} / {\rm N}_{\rm data}$                & $ 7751.738 / 7735    $ & $ 7750.972 / 7735    $ & $ 7755.555 / 7735    $ & $\chi^{2} / {\rm N}_{\rm data}$                & $ 7752.925 / 7735    $ \\
$\Delta\chi^{2}$ (full data)                   & $   17.521           $ & $   16.775           $ & $   21.338           $ & $\Delta\chi^{2}$ (full data)                   & $   18.708           $ \\
$\Delta\chi^{2}$ (${\rm HJD^{\prime} < 7600}$) & $    2.269           $ & $    1.414           $ & $    5.727           $ & $\Delta\chi^{2}$ (${\rm HJD^{\prime} < 7600}$) & $    3.060           $ \\
$t_0$ [${\rm HJD'}$]                           & $ 7454.864 \pm 0.061 $ & $ 7454.893 \pm 0.090 $ & $ 7455.816 \pm 0.102 $ & $t_{0,S_{1}}$ [${\rm HJD'}$]                   & $ 7451.870 \pm 0.112 $ \\
$u_0$                                          & $    0.226 \pm 0.049 $ & $    0.531 \pm 0.066 $ & $    0.148 \pm 0.024 $ & $u_{0,S_{1}}$                                  & $    0.124 \pm 0.015 $ \\
$t_{\rm E}$ [days]                             & $   15.382 \pm 1.392 $ & $   15.408 \pm 1.392 $ & $   24.015 \pm 4.242 $ & $t_{\rm E}$ [days]                             & $   17.409 \pm 1.362 $ \\
$t_{\rm E}^{\prime}$ [days]                    & \nodata                & \nodata                & $   14.176 \pm 1.999 $ & $t_{0,S_{2}}$ [${\rm HJD'}$]                   & $ 7457.697 \pm 0.106 $ \\
$s$                                            & $    0.687 \pm 0.025 $ & $    1.161 \pm 0.042 $ & $    3.392 \pm 0.162 $ & $u_{0,S_{1}}$                                  & $    0.133 \pm 0.018 $ \\
$q$                                            & $   18.911 \pm 3.028 $ & $   16.223 \pm 2.143 $ & $    2.870 \pm 1.418 $ & $q_{\rm flux}$                                 & $    1.146 \pm 0.181 $ \\ 
$\frac{1}{q}$                                  & $    0.053 \pm 0.006 $ & $    0.062 \pm 0.007 $ & $    0.348 \pm 0.215 $ & \nodata                                        & \nodata                \\
$\langle\log_{10} \frac{1}{q}\rangle$          & $   -1.328 \pm 0.059 $ & $   -1.229 \pm 0.053 $ & $   -0.389 \pm 0.203 $ & \nodata                                        & \nodata                \\ 
$\alpha$ [rad]                                 & $    1.582 \pm 0.013 $ & $    1.578 \pm 0.012 $ & $    2.544 \pm 0.014 $ & \nodata                                        & \nodata                \\
$\rho_{\ast,{\rm limit}}$                      & $  < 0.168           $ & $  < 0.178           $ & $  < 0.101           $ & \nodata                                        & \nodata                \\
% --------------------------------------------------------------------------------------------------------------------------------------
\enddata
\tablecomments{
${\rm HJD' = HJD - 2450000.0}$. 
}
\end{deluxetable}
% ---------------------------------------------------------------------------

% Table X (source color) ----------------------------------------------------
\begin{longrotatetable}
\begin{deluxetable}{crrrrrrrr}
\tablecaption{CMD analyses of Planetary Events \label{table:cmd}}
\tablewidth{0pt}
%\tabletypesize{\scriptsize}
\tablehead{
% ---------------------------------------------------------------------------
\multicolumn{1}{c}{Event} & 
\multicolumn{1}{c}{} & 
\multicolumn{1}{c}{} & 
\multicolumn{1}{c}{} & 
\multicolumn{1}{c}{} & 
\multicolumn{1}{c}{} & 
\multicolumn{1}{c}{} &
\multicolumn{1}{c}{} &
\multicolumn{1}{c}{} \\
% ----------------------------------------------
\multicolumn{1}{c}{Case} &
\multicolumn{1}{r}{$(V-I)_{\rm RGC}$} & 
\multicolumn{1}{r}{$(V-I)_{0, \rm RGC}$} &
\multicolumn{1}{r}{$(V-I)_{\rm S}$} &
\multicolumn{1}{r}{$(V-I)_{0, \rm S}$} &
\multicolumn{1}{r}{$(V-I)_{\rm B}$} &
\multicolumn{1}{c}{$\theta_{\ast}$} &
\multicolumn{1}{c}{$\theta_{\rm E}$} &
\multicolumn{1}{c}{$\mu_{\rm rel}$} \\
% ----------------------------------------------
\multicolumn{1}{c}{} &
\multicolumn{1}{r}{$I_{\rm RGC}$} & 
\multicolumn{1}{r}{$I_{0, \rm RGC}$} &
\multicolumn{1}{r}{$I_{\rm S}$} &
\multicolumn{1}{r}{$I_{0, \rm S}$} &
\multicolumn{1}{r}{$I_{\rm B}$} &
\multicolumn{1}{c}{($\mu{\rm as}$)} &
\multicolumn{1}{c}{(mas)} &
\multicolumn{1}{c}{($\rm mas\, yr^{-1}$)}
% ------------------------------------
}
\startdata
% -------------------------------------------------------------------------------------------------------------------------------------------
%KB160269 &            &            &                      &                      &                      &                     &             \\
OB161635 &            &            &                      &                      &                      &                     &             &          \\
$s_{-}$  & $  2.763 $ & $  1.060 $ & $  2.441 \pm 0.136 $ & $  0.738 \pm 0.145 $ & $  2.503 \pm 0.005 $ & $ 0.315 \pm 0.050 $ & $ > 0.085 $ & $> 1.49$ \\
         & $ 16.519 $ & $ 14.482 $ & $ 22.121 \pm 0.015 $ & $ 20.084 \pm 0.015 $ & $ 16.872 \pm 0.001 $ & $                 $ &             &          \\
$s_{+}$  & $  2.763 $ & $  1.060 $ & $  2.442 \pm 0.137 $ & $  0.739 \pm 0.145 $ & $  2.503 \pm 0.005 $ & $ 0.311 \pm 0.050 $ & $ > 0.076 $ & $> 1.26$ \\
         & $ 16.519 $ & $ 14.482 $ & $ 22.153 \pm 0.015 $ & $ 20.116 \pm 0.015 $ & $ 16.872 \pm 0.001 $ &                     &             &          \\ 
% -------------------------------------------------------------------------------------------------------------------------------------------
\hline      
%KB160506 &            &            &                      &                      &                      &                     &                     \\
MB16532 &            &            &                      &                      &                      &                     &                     &                   \\
$s_{-}$  & $  2.313 $ & $  1.060 $ & $  2.952 \pm 0.150 $ & $  1.699 \pm 0.158 $ & $  2.135 \pm 0.002 $ & $ 0.529 \pm 0.035 $ & $ 0.113 \pm 0.015 $ & $ 1.98 \pm 0.26 $ \\
         & $ 15.995 $ & $ 14.391 $ & $ 22.200 \pm 0.008 $ & $ 20.596 \pm 0.008 $ & $ 16.998 \pm 0.001 $ &                     &                     &                   \\
$s_{+}$  & $  2.313 $ & $  1.060 $ & $  2.958 \pm 0.150 $ & $  1.705 \pm 0.158 $ & $  2.135 \pm 0.002 $ & $ 0.534 \pm 0.036 $ & $ 0.112 \pm 0.016 $ & $ 1.99 \pm 0.29 $ \\
         & $ 15.995 $ & $ 14.391 $ & $ 22.188 \pm 0.008 $ & $ 20.584 \pm 0.008 $ & $ 16.997 \pm 0.001 $ &                     &                     &                   \\
% -------------------------------------------------------------------------------------------------------------------------------------------
\hline      
KB160625         &            &            &                      &                      &                      &                     &                     &                   \\
$s_{-}$          & $  1.892 $ & $  1.060 $ & $  1.427 \pm 0.044 $ & $  0.595 \pm 0.066 $ & $  1.584 \pm 0.005 $ & $ 0.364 \pm 0.025 $ & $ 0.297 \pm 0.029 $ & $ 9.44 \pm 0.93 $ \\
                 & $ 15.331 $ & $ 14.335 $ & $ 20.412 \pm 0.010 $ & $ 19.416 \pm 0.010 $ & $ 17.363 \pm 0.001 $ &                     &                     &                   \\
$s_{-}^{\prime}$ & $  1.892 $ & $  1.060 $ & $  1.429 \pm 0.044 $ & $  0.597 \pm 0.066 $ & $  1.583 \pm 0.005 $ & $ 0.363 \pm 0.025 $ & $ 0.173 \pm 0.017 $ & $ 5.46 \pm 0.54 $ \\
                 & $ 15.331 $ & $ 14.335 $ & $ 20.423 \pm 0.010 $ & $ 19.427 \pm 0.010 $ & $ 17.362 \pm 0.001 $ &                     &                     &                   \\
$s_{+}$          & $  1.892 $ & $  1.060 $ & $  1.429 \pm 0.043 $ & $  0.597 \pm 0.066 $ & $  1.584 \pm 0.005 $ & $ 0.368 \pm 0.026 $ & $ 0.210 \pm 0.021 $ & $ 6.78 \pm 0.67 $ \\
                 & $ 15.331 $ & $ 14.335 $ & $ 20.390 \pm 0.010 $ & $ 19.394 \pm 0.010 $ & $ 17.364 \pm 0.001 $ &                     &                     &                   \\
$s_{+}^{\prime}$ & $  1.892 $ & $  1.060 $ & $  1.429 \pm 0.043 $ & $  0.597 \pm 0.066 $ & $  1.584 \pm 0.005 $ & $ 0.372 \pm 0.026 $ & $ 0.210 \pm 0.021 $ & $ 6.85 \pm 0.68 $ \\
                 & $ 15.331 $ & $ 14.335 $ & $ 20.371 \pm 0.010 $ & $ 19.375 \pm 0.010 $ & $ 17.366 \pm 0.001 $ &                     &                     &                   \\
% -------------------------------------------------------------------------------------------------------------------------------------------
\hline      
%KB161307            &            &            &                      &                      &                    &                       &             \\
OB161850            &            &            &                      &                      &                    &                       &             &            \\
APRX inner $u_{0}+$ & $  2.552 $ & $  1.060 $ & $  2.106 \pm 0.048 $ & $  0.614 \pm 0.069 $ & $  1.738 \pm 0.031 $ & $ 0.880 \pm 0.064 $ & $ > 0.126 $ & $ > 0.73 $ \\
                    & $ 16.502 $ & $ 14.558 $ & $ 19.488 \pm 0.007 $ & $ 17.544 \pm 0.007 $ & $ 18.868 \pm 0.007 $ &                     &             &            \\
APRX outer $u_{0}+$ & $  2.552 $ & $  1.060 $ & $  2.108 \pm 0.048 $ & $  0.616 \pm 0.069 $ & $  1.740 \pm 0.030 $ & $ 0.874 \pm 0.064 $ & $ > 0.097 $ & $ > 0.56 $ \\
                    & $ 16.502 $ & $ 14.558 $ & $ 19.508 \pm 0.007 $ & $ 17.564 \pm 0.007 $ & $ 18.854 \pm 0.006 $ &                     &             &            \\
APRX inner $u_{0}-$ & $  2.552 $ & $  1.060 $ & $  2.106 \pm 0.048 $ & $  0.614 \pm 0.069 $ & $  1.743 \pm 0.030 $ & $ 0.871 \pm 0.064 $ & $ > 0.124 $ & $ > 0.76 $ \\
                    & $ 16.502 $ & $ 14.558 $ & $ 19.510 \pm 0.007 $ & $ 17.565 \pm 0.007 $ & $ 18.848 \pm 0.006 $ &                     &             &            \\
APRX outer $u_{0}-$ & $  2.552 $ & $  1.060 $ & $  2.108 \pm 0.048 $ & $  0.616 \pm 0.069 $ & $  1.743 \pm 0.030 $ & $ 0.873 \pm 0.064 $ & $ > 0.097 $ & $ > 0.58 $ \\
                    & $ 16.502 $ & $ 14.558 $ & $  9.511 \pm 0.007 $ & $ 17.566 \pm 0.007 $ & $ 18.847 \pm 0.006 $ &                     &             &            \\
% -------------------------------------------------------------------------------------------------------------------------------------------
\hline      
KB161751               &            &            &                      &                      &                      &                     &                     &            \\
$s_{+}$                & $  2.664 $ & $  1.060 $ & $  2.083 \pm 0.056 $ & $  0.479 \pm 0.075 $ & $  3.034 \pm 0.068 $ & $ 0.543 \pm 0.038 $ & $ 0.039 \pm 0.010 $ & $ > 1.06 $ \\
                       & $ 16.657 $ & $ 14.548 $ & $ 20.407 \pm 0.007 $ & $ 18.299 \pm 0.007 $ & $ 19.431 \pm 0.004 $ &                     &                     &            \\
$s_{+}^{\prime}$       & $  2.664 $ & $  1.060 $ & $  2.083 \pm 0.057 $ & $  0.479 \pm 0.076 $ & $  3.059 \pm 0.073 $ & $ 0.554 \pm 0.039 $ & $ 0.040 \pm 0.011 $ & $ > 1.06 $ \\
                       & $ 16.657 $ & $ 14.548 $ & $ 20.366 \pm 0.007 $ & $ 18.258 \pm 0.007 $ & $ 19.448 \pm 0.004 $ &                     &                     &            \\
$s_{-}$                & $  2.664 $ & $  1.060 $ & $  2.084 \pm 0.057 $ & $  0.480 \pm 0.076 $ & $  2.998 \pm 0.062 $ & $ 0.527 \pm 0.037 $ & $ 0.041 \pm 0.014 $ & $ > 0.94 $ \\
                       & $ 16.657 $ & $ 14.548 $ & $ 20.477 \pm 0.007 $ & $ 18.368 \pm 0.007 $ & $ 19.404 \pm 0.004 $ &                     &                     &            \\
$s_{-}^{\prime}$       & $  2.664 $ & $  1.060 $ & $  2.084 \pm 0.057 $ & $  0.480 \pm 0.076 $ & $  3.052 \pm 0.072 $ & $ 0.551 \pm 0.039 $ & $ 0.044 \pm 0.015 $ & $ > 1.05 $ \\
                       & $ 16.657 $ & $ 14.548 $ & $ 20.375 \pm 0.007 $ & $ 18.267 \pm 0.007 $ & $ 19.444 \pm 0.004 $ &                     &                     &            \\
$s_{-}^{\prime\prime}$ & $  2.664 $ & $  1.060 $ & $  2.084 \pm 0.057 $ & $  0.480 \pm 0.075 $ & $  3.047 \pm 0.071 $ & $ 0.549 \pm 0.039 $ & $ 0.029 \pm 0.004 $ & $ 1.09 \pm 0.17 $ \\
                       & $ 16.657 $ & $ 14.548 $ & $ 20.387 \pm 0.007 $ & $ 18.279 \pm 0.007 $ & $ 19.439 \pm 0.004 $ &                     &                     &            \\
% -------------------------------------------------------------------------------------------------------------------------------------------
\enddata
\tablecomments{We use the abbreviation for event names, e.g., \twosixtynine\ is abbreviated as OB161635.}
%\tabletypesize{\small}
\end{deluxetable}
\end{longrotatetable}
% ---------------------------------------------------------------------------

% Figure X (CMDs) ---------------------------------------------------------------------------------
% --------------------------------------------------------------------------------------------------

% Table X (Lens properties) -------------------------------------------------
\begin{longrotatetable}
\begin{deluxetable}{lccrrrrrcc}
\tablecaption{Lens Properties of Planetary Events \label{table:lens}}
%\tabletypesize{\scriptsize}
\tablewidth{0pt}
\tablehead{
% ---------------------------------------------------------------------------
\multicolumn{1}{c}{Event}                 &
\multicolumn{1}{c}{Constraints}           &
\multicolumn{1}{c}{Case}                  &
\multicolumn{1}{c}{$M_{\rm host}$}        &
\multicolumn{1}{c}{$M_{\rm planet}$}      &
\multicolumn{1}{c}{$D_{\rm L}$}           &
\multicolumn{1}{c}{$a_{\perp}$}           &
\multicolumn{1}{c}{$\mu_{\rm rel}$}       &
\multicolumn{1}{c}{Gal. Mod.}             &
\multicolumn{1}{c}{$\chi^{2}$}            \\
% ---------------------------------------------------------------------------
\multicolumn{1}{c}{}                      &
\multicolumn{1}{c}{}                      &
\multicolumn{1}{c}{}                      &
\multicolumn{1}{c}{($M_{\odot}$)}         &
\multicolumn{1}{c}{($M_{\rm J}$ / $M_{\rm N}$ / $M_{\oplus}$)}       &
\multicolumn{1}{c}{(kpc)}                 &
\multicolumn{1}{c}{(au)}                  &
\multicolumn{1}{c}{($\rm mas\, yr^{-1}$)} &
\multicolumn{1}{c}{}                      &
\multicolumn{1}{c}{}                      
% ------------------------------------
}
\startdata
% -------------------------------------------------------------------------------------------------------------------------------------
% KB160269
OB161635      & $t_{\rm E} + \rho_{\ast}$                & $s_{-}$              & $ 0.43_{-0.27}^{+0.33} $ & $ 11.49_{-7.25}^{+8.91}\, M_{\rm J}$ & $ 6.62_{-1.68}^{+1.15} $ & $ 1.31_{-0.47}^{+0.50} $ & $ 6.32_{-2.25}^{+2.89} $ & 1.000 & 1.000 \\  
              &                                          & $s_{+}$              & $ 0.44_{-0.27}^{+0.33} $ & $ 11.48_{-7.28}^{+8.57}\, M_{\rm J}$ & $ 6.60_{-1.71}^{+1.15} $ & $ 3.82_{-1.37}^{+1.43} $ & $ 6.14_{-2.20}^{+2.82} $ & 0.723 & 0.021 \\
{\bf Adopted} &                                          &                      & $ 0.43 \pm 0.29 $        & $ 11.49 \pm 7.96\, M_{\rm J}$        & $ 6.62 \pm 1.39 $        & $ 1.34 \pm 0.47 $        & $ 6.32 \pm 2.53 $        &       &       \\ 
% -------------------------------------------------------------------------------------------------------------
\hline
% KB160506
MB16532       & $t_{\rm E} + \theta_{\rm E}$             & $s_{-}$              & $ 0.09_{-0.05}^{+0.14} $ & $ 7.18_{-3.90}^{+11.35}\, M_{\rm N}$ & $ 7.38_{-1.02}^{+0.97} $ & $ 0.56_{-0.10}^{+0.10} $ & $ 2.10_{-0.28}^{+0.30} $ & 1.000 & 0.520 \\
              &                                          & $s_{+}$              & $ 0.09_{-0.05}^{+0.15} $ & $ 7.23_{-4.05}^{+11.46}\, M_{\rm N}$ & $ 7.37_{-1.02}^{+0.97} $ & $ 1.36_{-0.25}^{+0.26} $ & $ 2.14_{-0.31}^{+0.34} $ & 0.997 & 1.000 \\
{\bf Adopted} &                                          &                      & $ 0.09 \pm 0.07 $        & $ 7.21 \pm 5.73\, M_{\rm N}$         & $ 7.37 \pm 0.74 $        & $ 1.09 \pm 0.17 $        & $ 2.13 \pm 0.23 $        &       &       \\   
% ------------------------------------------------------------------------------------------------------------------------------------
\hline
KB160625 & $t_{\rm E} + \theta_{\rm E}$                  & $s_{-}$              & $ 0.30_{-0.16}^{+0.30} $ & $  1.36_{-0.73}^{+1.90} \, M_{\rm N}$  & $ 6.14_{-1.28}^{+0.95} $ & $ 1.31_{-0.28}^{+0.24} $ & $ 9.25_{-0.97}^{+0.98} $ & 1.000 & 1.000 \\
         &                                               & $s_{-}^{\prime}$     & $ 0.15_{-0.08}^{+0.21} $ & $  8.98_{-5.20}^{+17.84}\, M_{\oplus}$ & $ 6.68_{-1.04}^{+0.94} $ & $ 0.86_{-0.15}^{+0.15} $ & $ 5.53_{-0.57}^{+0.58} $ & 0.365 & 0.603 \\
         &                                               & $s_{+}$              & $ 0.19_{-0.10}^{+0.26} $ & $  4.68_{-2.92}^{+6.42} \, M_{\oplus}$ & $ 6.52_{-1.11}^{+0.94} $ & $ 1.85_{-0.35}^{+0.33} $ & $ 6.80_{-0.70}^{+0.72} $ & 0.556 & 0.612 \\
         &                                               & $s_{+}^{\prime}$     & $ 0.19_{-0.10}^{+0.26} $ & $  2.03_{-1.06}^{+3.53} \, M_{\oplus}$ & $ 6.51_{-1.11}^{+0.94} $ & $ 1.84_{-0.34}^{+0.33} $ & $ 6.85_{-0.69}^{+0.71} $ & 0.574 & 0.193 \\
{\bf Adopted} &                                          &                      & $ 0.25 \pm 0.14 $        & $  0.94 \pm 0.80\, M_{\rm N}$          & $ 6.31 \pm 0.71 $        & $ 1.40 \pm 0.17 $        & $ 8.10 \pm 0.61 $        &       &       \\
% ------------------------------------------------------------------------------------------------------------------------------------
\hline
% KB161307
OB161850 & $t_{\rm E} + \rho_{\ast} + \pivec_{\rm E}$    & APRX inner $u_{0}+$  & $ 0.26_{-0.12}^{+0.20} $ & $  8.85_{-4.19}^{+7.03}\, M_{\oplus} $ & $ 2.12_{-0.80}^{+1.17} $ & $ 1.46_{-0.26}^{+0.22} $ & $ 5.09_{-2.30}^{+3.94} $ & 0.905 & 1.000 \\
         &                                               & APRX outer $u_{0}+$  & $ 0.26_{-0.12}^{+0.20} $ & $ 10.73_{-5.36}^{+8.59}\, M_{\oplus} $ & $ 2.14_{-0.82}^{+1.19} $ & $ 1.49_{-0.27}^{+0.22} $ & $ 4.95_{-2.28}^{+3.94} $ & 0.914 & 0.306 \\
         &                                               & APRX inner $u_{0}-$  & $ 0.24_{-0.11}^{+0.20} $ & $  8.70_{-3.98}^{+7.20}\, M_{\oplus} $ & $ 2.09_{-0.79}^{+1.13} $ & $ 1.41_{-0.24}^{+0.20} $ & $ 5.17_{-2.32}^{+4.09} $ & 0.923 & 0.190 \\
         &                                               & APRX outer $u_{0}-$  & $ 0.26_{-0.12}^{+0.20} $ & $ 11.47_{-5.70}^{+9.05}\, M_{\oplus} $ & $ 2.22_{-0.83}^{+1.19} $ & $ 1.50_{-0.28}^{+0.26} $ & $ 5.17_{-2.37}^{+3.94} $ & 1.000 & 0.070 \\
{\bf Adopted} &                                          &                      & $ 0.26 \pm 0.11 $        & $ 9.33 \pm 3.88 \, M_{\oplus}$         & $ 2.12 \pm 0.67 $        & $ 1.46 \pm 0.16 $        & $ 5.07 \pm 2.11 $        &       &       \\ 
% ------------------------------------------------------------------------------------------------------------------------------------
\hline
KB161751      & $t_{\rm E} + \rho_{\ast}$                & $s_{+}$                & $ 0.18_{-0.11}^{+0.28} $ & $ 1.21_{-0.76}^{+1.88}\, M_{\rm J} $ & $ 7.05_{-1.38}^{+1.16} $ & $ 1.40_{-0.49}^{+0.60} $ & $ 7.49_{-2.70}^{+3.46} $ & 0.835 & 1.000 \\
              &                                          & $s_{+}^{\prime}$       & $ 0.17_{-0.11}^{+0.27} $ & $ 0.73_{-0.48}^{+1.14}\, M_{\rm J} $ & $ 7.05_{-1.38}^{+1.16} $ & $ 1.36_{-0.48}^{+0.58} $ & $ 7.51_{-2.71}^{+3.48} $ & 0.719 & 0.014 \\
              &                                          & $s_{-}$                & $ 0.18_{-0.12}^{+0.28} $ & $ 1.18_{-0.74}^{+1.82}\, M_{\rm J} $ & $ 7.05_{-1.38}^{+1.15} $ & $ 1.16_{-0.41}^{+0.50} $ & $ 7.37_{-2.66}^{+3.42} $ & 1.000 & 0.058 \\
              &                                          & $s_{-}^{\prime}$       & $ 0.17_{-0.11}^{+0.27} $ & $ 0.69_{-0.48}^{+1.08}\, M_{\rm J} $ & $ 7.05_{-1.37}^{+1.16} $ & $ 1.15_{-0.41}^{+0.49} $ & $ 7.49_{-2.70}^{+3.47} $ & 0.696 & 0.012 \\
{\bf Adopted} &                                          &                        & $ 0.18 \pm 0.18 $        & $ 1.20 \pm 1.21\, M_{\rm J}$         & $ 7.05 \pm 1.17 $        & $ 1.39 \pm 0.50 $        & $ 7.49 \pm 2.83 $        &  &  \\
\enddata
\tablecomments{For the planet mass, we present the value in Jupiter ($M_{\rm J}$), Neptune ($M_{\rm N}$), or Earth ($M_{\oplus}$) masses as appropriate.  
}
%\tabletypesize{\small}
\end{deluxetable}
\end{longrotatetable}
% --------------------------------------------------------------------------- 

% Table X (2016 Prime-filed planets) -------------------------------------------------
\begin{deluxetable}{llcccl}
\tablecaption{Planetary events discovered on KMTNet prime fields in $2016$ \label{table:2016_prime_planets}}
\tablewidth{0pt}
\tablehead{
% ---------------------------------------------------------------------------
\multicolumn{1}{c}{Event Name}            &
\multicolumn{1}{c}{KMT Name}              &
\multicolumn{1}{c}{$\log_{10}(q)$}        &
\multicolumn{1}{c}{$s$}                   &
\multicolumn{1}{c}{Degeneracy}            &
\multicolumn{1}{c}{Reference}             
% ---------------------------------------------------------------------------
}
\startdata
% -------------------------------------------------------------------------------------------------------------------------------------
OB160007                  & KB161991 & -5.17 & 2.83 &                             & \citet{zangyy}        \\
OB161195$^{\ast}$         & KB160372 & -4.34 & 0.99 & c/w, ecliptic               & \citet{gould23}       \\
OB161850                  & KB161307 & -4.00 & 0.80 & i/o, ecliptic               & This work             \\
MB16319                   & KB161816 & -2.41 & 0.82 & i/o                         & \citet{han18}         \\
MB16532                   & KB160506 & -2.39 & 0.65 & c/w                         & This work             \\
KB161836                  &          & -2.35 & 1.30 & c/w, ecliptic               & \citet{yang20}        \\
MB16227                   & KB160622 & -2.03 & 0.93 &                             & \citet{koshimoto17}   \\
OB160596                  & KB161677 & -1.93 & 1.08 &                             & \citet{mroz17}        \\
KB162605                  &          & -1.92 & 0.94 &                             & \citet{ryu21}         \\
OB161190                  & KB160113 & -1.84 & 0.60 & ecliptic                    & \citet{ryu18}         \\
OB161635                  & KB160269 & -1.59 & 0.59 & c/w                         & This work             \\
\hline                                                                            
KB160625                  &          & -3.63 & 0.74 & c/w                         & This work             \\
OB160613$^{\ast\ast}$     & KB160017 & -2.26 & 1.06 & c/w                         & \citet{han17}         \\
KB161751                  &          & -2.19 & 1.05 & c/w                         & This work             \\
KB161855$^{\ast\ast\ast}$ &          & -1.61 & 3.80 & c/w, $\alpha$, offset, 1L2S & This work             \\
KB160212                  &          & -1.43 & 0.83 & c/w                         & \citet{hwang18}       \\
KB161820                  &          & -0.95 & 1.40 &                             & \citet{jung18}        \\
KB162142$^{\ast\ast\ast}$ &          & -0.69 & 0.97 & c/w                         & \citet{jung18}        \\
% ------------------------------------------------------------------------------------------------------------------------------------
\enddata
\tablecomments{
$^{\ast}$For OB161195, the properties of this planetary system was reported by \citet{shvartzvald17} and \citet{bond17}. 
However, we adopt $\log_{10}(q)$ and $s$ values from \citet{gould23}, which re-analyze the event and measure 
a more precise mass ratio. 
$^{\ast\ast}$For OB160613, the event was caused by a lens system consisting of a planet and binary host stars.
$^{\ast\ast\ast}$For KB161855 and KB162142, these are planet candidates.
In the column of ``Degeneracy'', we present the type of degeneracies for the solutions: 
``c/w'', ``i/o'', ``ecliptic'', ``offset'', ``$\alpha$'', and ``1L2S'' indicate
the close/wide degeneracy, inner/outer degeneracy, ecliptic degeneracy of the microlens--parallax effect,
offset--degeneracy, $\alpha$--degeneracy (see Section \ref{sec:lc_analysis_1855}), and 2L1S/1L2S degeneracy, 
respectively.
}
\end{deluxetable}
% --------------------------------------------------------------------------- 

% Appendix part

% Figure X (KB-16-0425: AnomalyFinder) --------------------------------------------------------------
% --------------------------------------------------------------------------------------------------


\begin{thebibliography}{999}

\bibitem[Alard \& Lupton(1998)]{alard98} Alard, C. \& Lupton, R.~H.\ 1998, \apj, 503, 325. doi:10.1086/305984
\bibitem[Albrow et al.(2009)]{albrow09} Albrow, M.~D., Horne, K., Bramich, D.~M., et al.\ 2009, \mnras, 397, 2099. doi:10.1111/j.1365-2966.2009.15098.x
\bibitem[Albrow (2017)]{albrow17} Albrow, M.~D.\ 2017, MichaelDAlbrow/pyDIA: Initial release on github, V1.0.0, Zenodo, doi:10.5281/zenodo.268049
\bibitem[Bensby et al.(2011)]{bensby11} Bensby, T., Ad{\'e}n, D., Mel{\'e}ndez, J., et al.\ 2011, \aap, 533, A134. doi:10.1051/0004-6361/201117059
\bibitem[Bond et al.(2017)]{bond17} Bond, I.~A., Bennett, D.~P., Sumi, T., et al.\ 2017, \mnras, 469, 2434. doi:10.1093/mnras/stx1049
\bibitem[Bond et al.(2001)]{bond01} Bond, I.~A., Abe, F., Dodd, R.~J., et al.\ 2001, \mnras, 327, 868. doi:10.1046/j.1365-8711.2001.04776.x
\bibitem[Bramich et al.(2013)]{bramich13} Bramich, D.~M., Horne, K., Albrow, M.~D., et al.\ 2013, \mnras, 428, 2275, doi:10.1093/mnras/sts184
\bibitem[Chang \& Refsdal(1979)]{chang79} Chang, K. \& Refsdal, S.\ 1979, \nat, 282, 561. doi:10.1038/282561a0
\bibitem[Clanton \& Gaudi(2014a)]{clanton14a} Clanton, C. \& Gaudi, B.~S.\ 2014a, \apj, 791, 90. doi:10.1088/0004-637X/791/2/90
\bibitem[Clanton \& Gaudi(2014b)]{clanton14b} Clanton, C. \& Gaudi, B.~S.\ 2014b, \apj, 791, 91. doi:10.1088/0004-637X/791/2/91
\bibitem[Doran \& M{\"u}ller(2004)]{doran04} Doran, M. \& M{\"u}ller, C.~M.\ 2004, \jcap, 2004, 003. doi:10.1088/1475-7516/2004/09/003
\bibitem[Gaudi(1998)]{gaudi98} Gaudi, B.~S.\ 1998, \apj, 506, 533. doi:10.1086/306256
\bibitem[Gaudi \& Gould(1997)]{gaudi97} Gaudi, B.~S. \& Gould, A.\ 1997, \apj, 486, 85. doi:10.1086/304491
\bibitem[Gonzalez et al.(2012)]{gonzalez12} Gonzalez, O.~A., Rejkuba, M., Zoccali, M., et al.\ 2012, \aap, 543, A13. doi:10.1051/0004-6361/201219222
\bibitem[Gould(1992)]{gould92} Gould, A.\ 1992, \apj, 392, 442. doi:10.1086/171443
\bibitem[Gould \& Loeb(1992)]{gouldloeb92} Gould, A. \& Loeb, A.\ 1992, \apj, 396, 104. doi:10.1086/171700
\bibitem[Gould et al.(2022)]{gould22} Gould, A., Han, C., Zang, W., et al.\ 2022, \aap, 664, A13. doi:10.1051/0004-6361/202243744 % Systematic Search Paper V.
\bibitem[Gould(2022)]{masada} Gould, A.\ 2022, arXiv:2209.12501. doi:10.48550/arXiv.2209.12501
\bibitem[Gould et al.(2023)]{gould23} Gould, A., Shvartzvald, Y., Zhang, J., et al.\ 2023, arXiv:2303.08876. doi:10.48550/arXiv.2303.08876
\bibitem[Griest \& Safizadeh(1998)]{griest98} Griest, K. \& Safizadeh, N.\ 1998, \apj, 500, 37. doi:10.1086/305729
\bibitem[Griest \& Hu(1992)]{griest92} Griest, K. \& Hu, W.\ 1992, \apj, 397, 362. doi:10.1086/171793
\bibitem[Han \& Gould(1997)]{han97} Han, C. \& Gould, A.\ 1997, \apj, 480, 196. doi:10.1086/303944
\bibitem[Han et al.(2017)]{han17} Han, C., Udalski, A., Gould, A., et al.\ 2017, \aj, 154, 223. doi:10.3847/1538-3881/aa9179
\bibitem[Han et al.(2018)]{han18} Han, C., Bond, I.~A., Gould, A., et al.\ 2018, \aj, 156, 226. doi:10.3847/1538-3881/aae38e
\bibitem[Hwang et al.(2018)]{hwang18} Hwang, K.-H., Kim, H.-W., Kim, D.-J., et al.\ 2018, Journal of Korean Astronomical Society, 51, 197. doi:10.5303/JKAS.2018.51.6.197
\bibitem[Hwang et al.(2022)]{hwang22} Hwang, K.-H., Zang, W., Gould, A., et al.\ 2022, \aj, 163, 43. doi:10.3847/1538-3881/ac38ad % Systematic Search Paper II.
\bibitem[Ida \& Lin(2005)]{ida05} Ida, S. \& Lin, D.~N.~C.\ 2005, \apj, 626, 1045. doi:10.1086/429953 
\bibitem[Jung et al.(2018)]{jung18} Jung, Y.~K., Hwang, K.-H., Ryu, Y.-H., et al.\ 2018, \aj, 156, 208. doi:10.3847/1538-3881/aae319
\bibitem[Jung et al.(2022)]{jung22} Jung, Y.~K., Zang, W., Han, C., et al.\ 2022, \aj, 164, 262. doi:10.3847/1538-3881/ac9c5c % Systematic Search Paper VI.
\bibitem[Jung et al.(2023)]{jung23} Jung, Y.~K., Zang, W., Wang, H., et al.\ 2023, in prep.
\bibitem[Kennedy \& Kenyon(2008)]{kennedy08} Kennedy, G.~M. \& Kenyon, S.~J.\ 2008, \apj, 673, 502. doi:10.1086/524130 
\bibitem[Kervella et al.(2004)]{kervella04} Kervella, P., Th{\'e}venin, F., Di Folco, E., et al.\ 2004, \aap, 426, 297. doi:10.1051/0004-6361:20035930 
\bibitem[Kim et al.(2016)]{kim16} Kim, S.-L., Lee, C.-U., Park, B.-G., et al.\ 2016, Journal of Korean Astronomical Society, 49, 37. doi:10.5303/JKAS.2016.49.1.037 
\bibitem[Kim et al.(2018)]{kim18} Kim, D.-J., Kim, H.-W., Hwang, K.-H., et al.\ 2018, \aj, 155, 76. doi:10.3847/1538-3881/aaa47b
\bibitem[Koshimoto et al.(2017)]{koshimoto17} Koshimoto, N., Shvartzvald, Y., Bennett, D.~P., et al.\ 2017, \aj, 154, 3. doi:10.3847/1538-3881/aa72e0     
\bibitem[Minniti et al.(2017)]{VVV_DR2} Minniti, D., Lucas, P., \& VVV Team\ 2017, VizieR Online Data Catalog, II/348
\bibitem[Mr{\'o}z et al.(2017)]{mroz17} Mr{\'o}z, P., Han, C., and, et al.\ 2017, \aj, 153, 143. doi:10.3847/1538-3881/aa5da2
\bibitem[Nataf et al.(2013)]{nataf13} Nataf, D.~M., Gould, A., Fouqu{\'e}, P., et al.\ 2013, \apj, 769, 88. doi:10.1088/0004-637X/769/2/88
\bibitem[Paczynski(1997)]{paczynski97} Paczynski, B.\ 1997, astro-ph/9711007
\bibitem[Pecaut et al.(2012)]{pecaut12} Pecaut, M.~J., Mamajek, E.~E., \& Bubar, E.~J.\ 2012, \apj, 746, 154. doi:10.1088/0004-637X/746/2/154
\bibitem[Pecaut \& Mamajek(2013)]{pecaut13} Pecaut, M.~J. \& Mamajek, E.~E.\ 2013, \apjs, 208, 9. doi:10.1088/0067-0049/208/1/9
\bibitem[Poindexter et al.(2005)]{poindexter05} Poindexter, S., Afonso, C., Bennett, D.~P., et al.\ 2005, \apj, 633, 914. doi:10.1086/468182
\bibitem[Ryu et al.(2018)]{ryu18} Ryu, Y.-H., Yee, J.~C., Udalski, A., et al.\ 2018, \aj, 155, 40. doi:10.3847/1538-3881/aa9be4
\bibitem[Ryu et al.(2019)]{ryu19} Ryu, Y.-H., Hwang, K.-H., Gould, A., et al.\ 2019, \aj, 158, 151. doi:10.3847/1538-3881/ab3a34
\bibitem[Ryu et al.(2021)]{ryu21} Ryu, Y.-H., Hwang, K.-H., Gould, A., et al.\ 2021, \aj, 162, 96. doi:10.3847/1538-3881/ac062a      
\bibitem[Ryu et al.(2022)]{ryu22} Ryu, Y.-H., Kil Jung, Y., Yang, H., et al.\ 2022, \aj, 164, 180. doi:10.3847/1538-3881/ac8d6c
\bibitem[Shin et al.(2019)]{shin19} Shin, I.-G., Ryu, Y.-H., Yee, J.~C., et al.\ 2019, \aj, 157, 146. doi:10.3847/1538-3881/ab07c2
\bibitem[Shin et al.(2023)]{shin23} Shin, I.-G., Yee, J.~C., Gould, A., et al.\ 2023, \aj, 165, 8. doi:10.3847/1538-3881/ac9d93
\bibitem[Shvartzvald et al.(2017)]{shvartzvald17} Shvartzvald, Y., Yee, J.~C., Calchi Novati, S., et al.\ 2017, \apjl, 840, L3. doi:10.3847/2041-8213/aa6d09
\bibitem[Sumi et al.(2003)]{sumi03} Sumi, T., Abe, F., Bond, I.~A., et al.\ 2003, \apj, 591, 204. doi:10.1086/375212 
\bibitem[Szyma{\'n}ski et al.(2011)]{szymanski11} Szyma{\'n}ski, M.~K., Udalski, A., Soszy{\'n}ski, I., et al.\ 2011, \actaa, 61, 83
\bibitem[Tomaney \& Crotts(1996)]{tomaney96} Tomaney, A.~B. \& Crotts, A.~P.~S.\ 1996, \aj, 112, 2872. doi:10.1086/118228
\bibitem[Udalski(2003)]{udalski03} Udalski, A.\ 2003, \actaa, 53, 291
\bibitem[Udalski et al.(1994)]{udalski94} Udalski, A., Szymanski, M., Kaluzny, J., et al.\ 1994, \actaa, 44, 227
\bibitem[Udalski et al.(2015)]{udalski15} Udalski, A., Szyma{\'n}ski, M.~K., \& Szyma{\'n}ski, G.\ 2015, \actaa, 65, 1
\bibitem[Wang et al.(2022)]{wang22} Wang, H., Zang, W., Zhu, W., et al.\ 2022, \mnras, 510, 1778. doi:10.1093/mnras/stab3581 % Systematic Search Paper III.
\bibitem[Wozniak(2000)]{wozniak00} Wozniak, P.~R.\ 2000, \actaa, 50, 421
\bibitem[Yang et al.(2020)]{yang20} Yang, H., Zhang, X., Hwang, K.-H., et al.\ 2020, \aj, 159, 98. doi:10.3847/1538-3881/ab660e
\bibitem[Yee et al.(2012)]{yee12} Yee, J.~C., Shvartzvald, Y., Gal-Yam, A., et al.\ 2012, \apj, 755, 102. doi:10.1088/0004-637X/755/2/102
\bibitem[Yoo et al.(2004)]{yoo04} Yoo, J., DePoy, D.~L., Gal-Yam, A., et al.\ 2004, \apj, 603, 139. doi:10.1086/381241
\bibitem[Zang et al.(2021)]{zang21} Zang, W., Hwang, K.-H., Udalski, A., et al.\ 2021, \aj, 162, 163. doi:10.3847/1538-3881/ac12d4 % Systematic Search Paper I.
\bibitem[Zang et al.(2022a)]{zang22a} Zang, W., Yang, H., Han, C., et al.\ 2022a, \mnras, 515, 928. doi:10.1093/mnras/stac1883 % Systematic Search Paper IV.
\bibitem[Zang et al.(2023)]{zang23} Zang, W., Jung, Y.~K., Yang, H., et al.\ 2023, \aj, 165, 103. doi:10.3847/1538-3881/acb34b % Systematic Search Paper VII.
\bibitem[Zang et al.(in prep.)]{zangyy} Zang et al.\ in prep.
\bibitem[Zhang \& Gaudi(2022)]{zhang22a} Zhang, K. \& Gaudi, B.~S.\ 2022, \apjl, 936, L22. doi:10.3847/2041-8213/ac8c2b
\bibitem[Zhang et al.(2022)]{zhang22b} Zhang, K., Gaudi, B.~S., \& Bloom, J.~S.\ 2022, Nature Astronomy, 6, 782. doi:10.1038/s41550-022-01671-6
\end{thebibliography}
\end{document}